\documentclass[12pt,a4paper,twoside]{kkd}
\usepackage{epsfig}
\usepackage{amssymb}
\usepackage{setspace}
\usepackage[round]{natbib}

\makeatletter
\begin{document}

\pagenumbering{roman}
\addcontentsline{toc}{chapter}{Title page}
 
%\newpage
%\begin{singlespacing}
\setlength{\textheight}{285.9 mm}
\oddsidemargin 1.2 cm 

\thispagestyle{empty}
\def\maketitle{%
  \null
  \thispagestyle{empty}%
%  \vfill
  \begin{center}%\leavevmode
   % \normalfont
    {\Large {\bf \@title\par}}%
    \vskip 2.2 cm
    {\bf   {\it Thesis submitted to}}\\
    %\vskip .2 cm
    {\bf {\it Indian Institute of Technology, Kharagpur}}\\
    %\vskip .2 cm
    {\bf {\it for the award of the degree}}\\
    \vspace{1.0cm}
    {\bf {\it of}}\\
    \vspace{.35cm}
    {\Large {\bf Doctor of Philosophy }}\\
    \vspace{.25cm}   
    {\bf {\it by}}\\
    %\vspace{.25cm}
    {\Large {\bf\@author\par}}%
    \vspace{.25cm}
    {\bf {\it   under the guidance of}}\\
    \vspace{.6cm}
    {\Large {\bf Dr. Somnath Bharadwaj}}
    \end{center}%
  %\vfill
  \null
}
\setlength{\topmargin}{-15. mm}

\begin{singlespacing}
\title{PROBING COSMOLOGICAL REIONIZATION THROUGH \\ 
RADIO-INTERFEROMETRIC OBSERVATIONS \\
OF NEUTRAL HYDROGEN }
\author{Kanan Kumar Datta}
\maketitle
\vspace{.25cm}
\begin{figure}[h]
\centerline{\psfig{figure=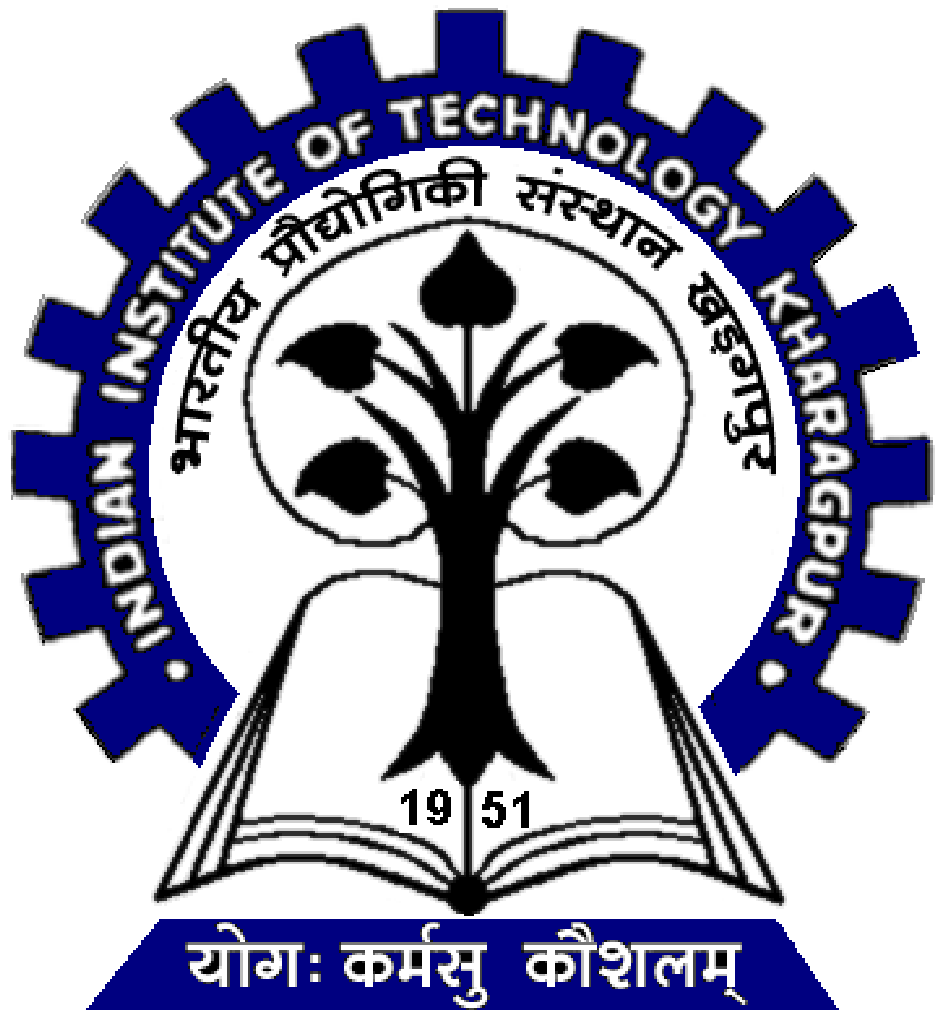,width=4.5cm,height=4.5cm}}
\end{figure}
\begin{center}
%\vspace{.3cm}
{\large {\bf DEPARTMENT OF PHYSICS \& METEOROLOGY}}\\
\vspace{.15cm}
{\large {\bf INDIAN INSTITUTE OF TECHNOLOGY, KHARAGPUR}}\\
\vspace{.15cm}
{\large {\bf MAY 2009}}

\vspace{.25cm}
{\bf $\copyright$ 2009, Kanan Kumar Datta. All rights reserved.}

\end{center}

\end{singlespacing}

\textheight 21.1 cm \textwidth 16.0 cm \pagestyle{headings}

\newcommand{\degr}{\ensuremath{^{\circ}}}

\renewcommand{\labelenumi}{(\roman{enumi})}
\renewcommand{\thefootnote}{\fnsymbol{footnote}}

\setlength{\topmargin}{-015. mm}
\setlength{\headheight}{10.0 mm}
\setlength{\headsep}{10.0 mm}
\setlength{\topskip}{0.00 mm}
\setlength{\textheight}{215.9 mm}
\setlength{\textwidth}{152.4 mm}

\setlength{\parskip}{.75 mm}
\setlength{\parindent}{6.00 mm}
\setlength{\floatsep}{4 mm}
\setlength{\textfloatsep}{4 mm} \setlength{\intextsep}{4 mm}
\setcounter{secnumdepth}{3} \setcounter{tocdepth}{2}
\newtheorem{theorem}{Theorem}[section]
\newtheorem{lemma}{Lemma}[chapter]
\newtheorem{Remark}{Remark}[chapter]
\newtheorem{corollary}{Corollary}
\newtheorem{definition}[theorem]{Definition}
\newtheorem{observation}[theorem]{Observation}
\newtheorem{fact}{Fact}[theorem]
\newtheorem{proposition}[theorem]{Proposition}
\newtheorem{rule-def}[theorem]{Rule}
\newcommand{\ccnt}[1]{\multicolumn{1}{|c|}{#1}}
\newcommand{\fns}[1]{\footnotesize{#1}}

%\singlespacing
%\doublespacing
% defineing short form------

\newcommand{\sce}{\setcounter{equation}}
\newcommand{\nn}{\nonumber}
\newcommand{\lb}{\label}
\newcounter{saveeqn}
\newcommand{\alpheqn}{\setcounter{saveeqn}{\value{equation}}
\stepcounter{saveeqn}\setcounter{equation}{0}%
\renewcommand{\theequation}{\mbox{\arabic{chapter}.\arabic{saveeqn}\alph{equation}}}}
\newcommand{\reseteqn}{\setcounter{equation}{\value{saveeqn}}%
\renewcommand{\theequation}{\arabic{chapter}.\arabic{equation}}}
\newcounter{savecite}
\newcommand{\alphcite}{\setcounter{savecite}{\value{cite}}
\stepcounter{savecite}\setcounter{cite}{0}%
\renewcommand{\thecite}{\mbox{\arabic{savecite}\alph{cite}}}}
\newcommand{\resetcite}{\setcounter{cite}{\value{savecite}}%
\renewcommand{\thecite}{\arabic{cite}}}

\newcommand{\be}{\begin{equation}}
\newcommand{\e}{\end{equation}}
\newcommand{\bear}{\begin{eqnarray}}
\newcommand{\ear}{\end{eqnarray}}
\newcommand{\nline}{\nonumber \\}
\newcommand{\f}{\frac}
\newcommand{\de}{{\rm d}}
\newcommand{\del}{\partial}
\newcommand{\R}{{\cal R}}
\def\apj{Astrophysical Journal}
\def\apjl{Astrophysical Journal Letters}
\def\apjs{Astrophysical Journal Supplement Series}
\def\mnras{Monthly Notices of Royal Astronomical Society}
\def\mnrasl{Monthly Notices of Royal Astronomical Society: Letters}
\def\prd{Physical Review D}
\def\prl{Physical Review Letters}
\def\phyrep{Physics Report}
\def\aj{Astronomical Journal}
\def\aa{Astronomy \& Astrophysics}
\def\japa{Journal of Astrophysics and Astronomy}
\def\jcap{Journal of Cosmology and Astroparticle Physics}
\def\u{{\vec U}}
\def\th{\vec{\theta}}
\def\rn{r_{\nu}}
\def\rnp{r'_{\nu}}
\def\newblock{}
\def\E{\hat{E}}
\def\k{{\bf k}}
\def\kp{k_\parallel}
\def\HI{\rm HI}

\def\newblock{}

\clearpage{\pagestyle{empty}\cleardoublepage} %%%%%%%%%%%%%%%%%%%%
\addcontentsline{toc}{chapter}{Certificate of approval}
%MAKE IT TOTALLY BLANK
%\thispagestyle{empty}
\vspace{2cm}
 %\draftstring{}
%\watermarkgraphic{water.png} \watermark
\begin{center}
{\Large \underline{\bf CERTIFICATE OF APPROVAL}}
\end{center}

\vspace{.5cm}
\hspace{12cm} Date: 14/05/09

\hspace{-.55cm}Certified that the thesis entitled {\bf PROBING
COSMOLOGICAL REIONIZATION THROUGH RADIO-INTERFEROMETRIC OBSERVATIONS
OF NEUTRAL HYDROGEN} submitted by {\bf KANAN KUMAR DATTA} to Indian 
Institute of Technology, Kharagpur, for the award of the degree of
Doctor of Philosophy has been accepted by the external examiners and
that the student has successfully defended the thesis in the viva-voce
examination held today.

\vspace{2.0cm}
\hspace{-.55cm}Signature\hspace{4.25cm}Signature\hspace{4.25cm}Signature\newline
\hspace{-.5cm}Name\hspace{4.95cm}Name\hspace{4.9cm}Name

\vspace{.6cm}
\hspace{-.55cm}(Member of the DSC)\hspace{2.cm}(Member of the
DSC)\hspace{2.cm}(Member of the DSC)

\vspace{2.0cm}
\hspace{-.55cm}Signature\hspace{4.25cm}Signature\hspace{4.25cm}Signature\newline
\hspace{-.5cm}Name\hspace{4.95cm}Name\hspace{4.9cm}Name

\vspace{.6cm}
\hspace{-.55cm}(Supervisor)\hspace{3.5cm}(External Examiner)\hspace{2.25cm}(Chairman)

\clearpage{\pagestyle{empty}\cleardoublepage} %%%%%%%%%%%%%%%%%%%%
\addcontentsline{toc}{chapter}{Declaration}

 \vspace*{2cm}
 %\draftstring{}
%\watermarkgraphic{water.png} \watermark
\begin{singlespacing}
{\Large {\bf DECLARATION}}

\vspace{1cm} I certify that 

\hspace{1cm} a. \hspace{.4cm}the work contained in this thesis is original and has been done by me

\hspace{1.9cm} under the guidance of my supervisor(s).

\hspace{1cm} b. \hspace{.4cm}the work has not been submitted to any other Institute for any degree or 

\hspace{1.9cm} diploma.

\hspace{1cm} c. \hspace{.4cm}I have followed the guidelines provided by the Institute in preparing the

\hspace{1.9cm} thesis.

\hspace{1cm} d. \hspace{.4cm}I have conformed to the norms and guidelines given in the Ethical Code 

\hspace{1.9cm} of Conduct of the Institute.

\hspace{1cm} e. \hspace{.4cm}whenever I have used materials (data, theoretical analysis, figures, and  

\hspace{1.9cm} text) from other sources, I have given due credit to them by citing them

\hspace{1.9cm}  in the text of the thesis and giving their details in the references. Further,

\hspace{1.9cm}  I have taken permission from the copyright owners of the  sources,

\hspace{1.9cm} whenever necessary.

\vspace{2.5cm}
\hspace{10.5cm} Signature of the Student

%%\vspace{5.65cm}
%%\hspace{10.5cm} iv

\end{singlespacing}

\clearpage{\pagestyle{empty}\cleardoublepage} %%%%%%%%%%%%%%%%%%%%
\addcontentsline{toc}{chapter}{Certificate by the Supervisor}
%MAKE IT TOTALLY BLANK
%\thispagestyle{empty}
\vspace{3cm}
 %\draftstring{}
%\watermarkgraphic{water.png} \watermark
\begin{center}
{\Large \underline{\bf CERTIFICATE}}
\end{center}

\vspace{.5cm}
\hspace{-.55cm}This is to certify that the thesis entitled {\bf
Probing Cosmological Reionization through Radio-Interferometric
Observations of Neutral Hydrogen}, submitted by {\bf Kanan Kumar
Datta} to Indian Institute of Technology, Kharagpur, is a record of
bona fide research work under my supervision and is worthy of
consideration for the award of the degree of Doctor of Philosophy of
the Institute.

\vspace{5.0cm}
\hspace{-.55cm}Date:\hspace{10cm}Supervisor

\clearpage{\pagestyle{empty}\cleardoublepage} %%%%%%%%%%%%%%%%%%%%
\addcontentsline{toc}{chapter}{Acknowledgement}
 \vspace*{-0.2cm}
\begin{center}
{\Large \bf Acknowledgment}
\end{center}
%{\setlength{\baselineskip}{14pt} \setlength{\parskip}{2pt}
This thesis is the end of my long journey in obtaining my PhD
degree. There are many people who made this journey easier with their proper
guidance, words of encouragement, generous help. It is a pleasant aspect that
I have now the opportunity to convey my gratitude to all of them.

First I bow down before the Almighty who has made everything possible.
I express my heart-felt gratitude to my parents  and other
family members (uncle, aunty, sisters and brothers), for their love,
support and encouragement. I dedicate my thesis to them. Special
thanks to Rituparna Ghosal for her constant inspiration and
support. She never let me down. 

It is great pleasure and proud privilege to express my sincere thanks
and gratitude to my supervisor, Prof. S. Bharadwaj, for all the guidance
and encouragement he has offered me throughout the period of this
research. I feel that I have been extremely lucky to get an adviser
like him. I am grateful to him
for his numerous ideas, suggestions, criticism of my research work. 
His style of thinking and his scientific interests
has always been  a source of inspiration for me. He always had enough
patience to clarify all my various stupid doubts. Working  with him
has been sheer joy and I hope to do more fruitful work with him in
future.  I honestly thank God that He gave you as my adviser.

It is a pleasure to pay tribute also to my collaborators. I gratefully
acknowledge Dr. Tirthankar Roychoudhury for his crucial 
contribution, advice, and supervision. He was always there when I
needed him. I would also acknowledge Suman Majumdar who generated
HI 21-cm maps. I am grateful for Tapomoy Guha Sarkar from
whom I learned a lot. It was a pleasure to work with them. I hope to
keep up our collaboration in the future.

I express my deep respectful thanks to Prof. S. Kar.  I was taught
relativity course by him. I learnt many things from his excellent
teaching. He also has read the synopsis of my thesis and provided me valuable
comments. I wish to thank Dr. S. Konar for her teaching on
``Introduction to Astrophysics'' course during my course work. I would
like to thank  Prof. A. Dasgupta, Prof. A. Taraphdar, Prof. S. P. Khastgir 
for useful discussion.  It was nice experience to discuss  scientific problems
with them.

My friends at the Centre for Theoretical studies (CTS) gave me the feeling
of being at home at work.  I would like to thank Prasun Dutta, Dr Saiyad
Ali, Prakash Sarkar, Abhik Ghosh, Dr Biswajit Pandey for scientific
discussions that 
helped me a lot in staying at the right track. I would like to 
thank Anupam Das, Sanjit Das, Suman Ghosh, Partha Dutta, Soma Dey,
Biswambhar Raksit, Dr Hemwati Nandan, Somnath Maity, Subhasis
Panda, Tatan Ghosh with whom 
I share many exciting memorable 
moments of laughter and pure joy. I wish to thank Ujal Haldar, Gopal,
Subhabrata, Venu  for 
helping me various ways.

I am also grateful to all the staff members of our department and
Centre for Theoretical Studies  and the
authorities of Indian Institute of Technology (IIT), Kharagpur, with
whom I had to interact many times during my period of work. I would
like to thank the CSIR, Govt. of India for financial support 
through a Research Fellowship. Finally I would like to acknowledge the
anonymous referee of the thesis for useful comments and suggetions
which have certainly improved the quality of the thesis.

%\clearpage{\pagestyle{empty}\cleardoublepage} %%%%%%%%%%%%%%%%%%%%

\addcontentsline{toc}{chapter}{List of Symbols}
%\chaptermark{List of Symbols}

%\thispagestyle{empty}
 %\vspace*{2cm}
 %\draftstring{}
%\watermarkgraphic{water.png} \watermark
\newpage
\begin{center}
{\Large \textbf{List of Symbols}}
\end{center}

\vspace{1.0cm}
\begin{center}
\begin{tabular}{ccl}\hline
{\bf Symbols}&\hspace{4cm} & {\bf Definition (unit)}\\ \hline\hline
$A$&&Amplitude for foreground\\
&&power spectrum ($mK^2$)\\
$A_{eff}$&&Effective antenna collecting \\
&&area ($m^2$)\\
$A({\vec \theta})$&&Antenna beam pattern (unit less)\\
$a_{lm}$&&Spherical Harmonics coefficient\\
$B$&&Frequency bandwidth (MHz)\\
$b_c$&& bias of the ionized sphere centers with \\
&&the dark matter (unit less)\\
$C_l(\Delta \nu)$&& Multifrequency angular power \\
&&spectrum (m$K^2$)\\
$C_l^{flat}(\Delta \nu)$&&$C_l(\Delta \nu)$ calculated using the flat
sky (m$K^2$)\\
&& approximation\\
$D(z)$&&Growth rate of dark matter density\\
&& contrast $\delta$ (unit less)\\
${\vec d}$&&Antenna separation vector projected \\
&&in the plane perpendicular \\
&&to the line of sight (m)\\
$(\de B/{\de T})_{\nu}$&&Conversion factor from temperature\\
&&to specific intensity (mJy/mK)\\
${\hat E}$&&Estimator for bubble detection ($mJy^2$)\\
$E_o$&&Observed value of the estimator ($mJy^2$)\\

\hline

\end{tabular}
\newpage

\begin{tabular}{lcl}\hline
{\bf Symbols}&\hspace{2cm} & {\bf Definition (unit)}\\ \hline\hline
$F_l(\Delta \nu)$&&Ratio of decrements of the 21-cm signal\\
&& to the foreground (unit less)\\

$F({\vec U},\nu)$&&Total foreground contribution (mJy)\\

$HF({\vec U},\nu))$&&Contribution from fluctuating HI (mJy)\\
$H_0$&&Hubble parameter at present({\rm Km/s/Mpc}) \\ 
$H(z)$&&Hubble parameter at redshift $z$ ({\rm Km/s/Mpc})\\
$h$&&$H_0/100$ \,{\rm km/s/Mpc} (unit less)\\
$I_l(\Delta \nu)$&& Foreground frequency decorrelation \\
&&function (unit less)\\
${\bar I_{\nu}}$&& Average specific intensity of the redshifted HI\\
&&21-cm from the EoR (mJy)\\
$I_{\nu}({\vec \theta})$&&Sky specific intensity pattern at position
${\vec \theta}$ (Jy)\\
$J_1(x)$&&First order Bessel function\\
$j_l$&&Spherical Bessel function \\
$j_l^{''}$&&$\frac{d^2}{dx^2}j_{l}(x)$\\
${\bf k}$&&Fourier mode ($Mpc^{-1})$\\
${\hat k}$&&Unit vector of ${\bf k}$\\
$k_B$&&Boltzmann constant (Joule/K)\\
$k_{\parallel}$&&Component of ${\bf k}$ along ${\bf \theta}$ (${\rm Mpc^{-1}}$)\\
$\ell$&&Angular mutipole (unit less)\\
${\bf m}$&&Component of ${\hat n}$ along ${\bf \theta}$\\
$N$&&Total number of antennae of a\\
&&radio experiment \\
$N_b$&&Total number of independent\\
&&baselines\\

\hline

\end{tabular}

\newpage

\begin{tabular}{lcl}\hline
{\bf Symbols}&\hspace{2cm} & {\bf Definition (unit)}\\ \hline\hline
$N({\vec U},\nu)$&&System noise (mJy)\\
$n$&&Power law index describes scaling\\
&&relation of SNR with redshift\\
${\hat n}$&&Unit vector along line of sight\\
${\tilde n_{HI}}$&&Mean comoving number density of \\
&&ionized spheres (${\rm Mpc}^{-3}$)\\
$n_s$&&spectral index in the matter power spectrum\\
$P_{\rm HI}(k)$&&HI 21-cm power spectrum in redshift space (${\rm Mpc}^{-3}$)\\
$P_{\Delta_{HI}}(z,k)$&&Cross-correlation power spectrum \\
&&between the above two(${\rm Mpc}^{-3}$)\\
$P_{\Delta^2_{HI}}(z,k)$&&HI density power spectrum(${\rm Mpc}^{-3}$) \\
$P(z,k)$&&Dark matter power spectrum (${\rm Mpc}^{-3}$)\\
$R$&&Comoving radius of the ionized sphere (Mpc, \\
&&in the Chapter 2 only)\\
$\R$&&${\bar x_{HII}}/{\bar x_{HI}}$ (unit less)\\
$R_b$&& Comoving radius of ionized bubble (Mpc)\\
$R_f$&&Comoving filter size (Mpc)\\
$R_{\nu}$&&Comoving radius of the planar section \\ 
&& the bubble at a frequency $\nu$ ({\rm Mpc})\\
$r_{\nu}$&&Comoving distance from present\\
&&to the redshift $z=1420/{\nu}-1$ ({\rm Mpc})\\
$r_{\nu}'$&&$\de r_{\nu}/{\de \nu}$ (Mpc/MHz)\\
$S_{center}({\vec U},\nu)$&&HI signal from ionized bubble\\
&&when it is at the center of \\
&&FoV (mJy)\\

\hline

\end{tabular}

\newpage

\begin{tabular}{lcl}\hline
{\bf Symbols}&\hspace{2cm} & {\bf Definition (unit)}\\ \hline\hline
$S_{cut}$&&A flux level in the image \\
&&above which all points sources can \\
&&be identified and removed (mJy)\\
$S_f({\vec U},\nu)$&&Filter to detect individual ionized bubbles (mJy)\\
$S({\vec U},\nu)$&&HI signal from ionized bubble (mJy)\\
$t_{obs}$&&Total observation time (sec)\\
$T_s$&& Spin temperature of hydrogen \\
&&gas (mK)\\
$T_{sky}$&& Sky brightness temperature (Kelvin)\\
$T_{\gamma}$&&CMB temperature (K)\\
$T(\nu, {\hat n})$&& Excess 21-cm brightness temperature \\
&&observed at a frequency $\nu$ along a \\
&&direction ${\hat n}$ (mK)\\
${\bar T(z)}$&& Spatially averaged Excess 21-cm brightness\\
&&temperature at redshift $z$ (mK)\\
${\vec U}$&&Baseline vector (unit less)\\
$u, v$&&Two components of ${\vec U}$ (unit less)\\
$V({\vec U},\nu)$&&Total Visibility measured at baseline ${\vec U}$\\
&&and frequency $\nu$ (Jy)\\
$v(z,{\hat n}r_{\nu})$&&peculiar velocity of HI gas (km/sec)\\
$W(y)$&& Spherical top hat window function\\
$x_{HI}$&& Neutral hydrogen fraction (unit less)\\
${\bar x_{HII}}$&&Average ionized hydrogen \\
&&fraction (unit less)\\

\hline

\end{tabular}

\newpage

\begin{tabular}{lcl}\hline
{\bf Symbols}&\hspace{2cm} & {\bf Definition (unit)}\\ \hline\hline
${\bar x_{HI}}(z)$&& Spatially averaged HI fraction (unit less)\\
$Y$&&Helium mass fraction (unit less)\\
$Y_{lm}$&&Spherical Harmonics\\
$z$ & & Redshift (unit less)\\
$z_c$&&Redshift of ionized bubble \\
&&center (unit less)\\
${\bar \alpha}$&&Mean spectral index (unit less)\\
$\alpha_{eff}$&& Effective spectral index (unit less)\\
$\beta$&&power law index for foreground \\
&&power spectrum (unit less)\\
$\Delta_{HI}(z,{\bf k})$&& 3D Fourier transform of the fluctuations\\
&&in the HI densities ($Mpc^{-3}$)\\
$\Delta_p$&& Poisson fluctuations ($Mpc^{-3}$)\\
$\Delta t$&&Integration time for a radio\\
&& experiments (sec)\\
$\Delta \nu$&&Frequency separation between two\\
&& channels $\nu_1$ and $\nu_2$(MHz, for chapter 2)\\
&&$\nu_c-\nu$(for other chapters)\\
$\Delta \nu_b$&&Bubble size in frequency (MHz)\\
$\Delta \nu_c$&&Frequency channel width (MHz)\\
$\Delta \nu_{1/2}$&&Frequency separation $\Delta \nu$ at which
the \\
&&frequency decorrelation function \\
&&$\kappa_l(\Delta \nu)$ becomes $1/2$ (MHz)\\
$\Delta(z,{\bf k})$&& 3D Fourier transform of the fluctuations \\
&&in the dark matter densities ($Mpc^{-3}$)\\

\hline

\end{tabular}

\newpage

\begin{tabular}{lcl}\hline
{\bf Symbols}&\hspace{2cm} & {\bf Definition (unit)}\\ \hline\hline

$\langle(\Delta E)^2\rangle$&&Variance of the estimator ($mJy^2$)\\
$\langle(\Delta E)^2\rangle_i$&&Contribution to the variance
$\langle(\Delta E)^2\rangle$\\
&&from $i^{th}$ component ($mJy^2$) \\
$\delta$&&Dark matter density contrast (unit less)\\
$\delta_D^2({\vec U})$&&Two dimensional Dirac delta function\\
$\delta_{HI}(z,{\hat n}r_{\nu})$&& Fluctuations in HI
density \\
&&field (unit less)\\
$\delta_{ij}$&&Kronekar delta\\
$\eta_{HI}(z,{\hat n}r_{\nu})$&&21-cm radiation efficiency in redshift
\\
&&space (unit less)\\
${\bar \eta_{HI}({\bf k})}$&& Fourier transform of $\eta_{HI}({\hat
n}r_{\nu})$ ($Mpc^{-3}$)\\
$\Theta(x)$&&Heaviside step function\\
${\bf \theta}$&&Two dimensional vector in \\
&&the sky plane (unit less)\\
${\vec \theta}_c$&&Ionized bubble center (rad)\\
$\theta_x,\theta_y$&&Two perpendicular components of ${\bf \theta}$ (rad)\\
$\theta_{\nu}$&&Angular radius of circular disk \\
&&of radius $R_{\nu}$ (rad)\\
$\theta_{\nu_c}$&&Angular size of ionized bubble (rad) \\
$\theta_0$&&$0.6\,\theta_{FWHM}$ (FWHM of the antenna beam pattern)\\
$\kappa_l(\Delta \nu)$&& Dimensionless frequency decorrelation \\
&&function (unit less)\\
$\lambda$&&Observing wavelength (m)\\

\hline

\end{tabular}

\newpage

\newpage

\begin{tabular}{lcl}\hline
{\bf Symbols}&\hspace{2cm} & {\bf Definition (unit)}\\ \hline\hline

$\lambda_c$&&Wavelength corresponding to frequency $\nu_c$ (m)\\
$\lambda_J$&&Comoving Jeans Length (Mpc)\\
$\mu$&&Angle between the line of sight\\
&& and the wave vector (${\hat k}.{\hat n}$)\\
$\nu_c$&&Redshifted frequency corresponding \\
&&to the redshift $z_c$ (MHz)\\
$\nu_f$&&Filter position along frequency \\
&&axis (MHz)\\
$\xi$&&Foreground frequency decorrelation \\
&&length (MHz)\\
$\rho_N({\vec U},\nu))$&&Normalized baseline distribution\\
&&function(${\rm MHz^{-1}}$)\\
$\rho_{ante}(r)$&&Antenna distribution function\\
$\sigma$&&Noise rms in image (mJy)\\
$\sigma_p$&&One dimensional pair velocity dispersion\\
&&in relative galaxy velocity (km/s)\\
$\sigma_8$&&RMS mass fluctuations on $8h^{-1}$ {\rm Mpc} scale (unit less)\\
$\tau_e$&& CMB electron scattering optical \\
&&depth (unit less)\\
$\Omega$&& Solid angle ($rad^2$)\\
$\Omega_b$&&Baryon density parameter (unit less)\\
$\Omega_m$&&Matter density parameter (unit less)\\
$<..>$ && Ensemble average\\
\hline

\end{tabular}

\end{center}

%\clearpage{\pagestyle{empty}\cleardoublepage} %%%%%%%%%%%%%%%%%%%%
\addcontentsline{toc}{chapter}{List of Tables}
\listoftables

%\clearpage{\pagestyle{empty}\cleardoublepage} %%%%%%%%%%%%%%%%%%%%
\addcontentsline{toc}{chapter}{List of Figures}

\listoffigures

\clearpage{\pagestyle{empty}\cleardoublepage} %%%%%%%%%%%%%%%%%%%%
\chaptermark{$\empty$}
\addcontentsline{toc}{chapter}{Abstract}

%\thispagestyle{empty}

%\pagenumbering{roman}

%\vspace{-0.2cm}
\begin{center}

{\Large \bf Abstract}
\end{center}
%\vspace{.2cm}

\begin{singlespace}
One of the major challenges in modern cosmology is to understand the 
reionization history of the Universe. This is directly related to
galaxy formation and the formation of the first luminous
objects. Observations of redshifted 21-cm radiation from neutral
hydrogen (HI) is probably the most promising future probe of
reionization. Several approaches have been proposed to extract  
information about the epoch of reionization from the data which is
expected to come in near future.

The most discussed approach has been to study the global statistical
properties of the reionization HI 21-cm. We develop the formalism to
calculate the Multi-frequency Angular Power Spectrum (MAPS) and
quantify the statistics  
of the HI signal  as a joint function of the angular multipole $l$ and 
frequency separation $\Delta\nu$. We adopt 
a simple model for the HI distribution which incorporates patchy 
reionization and use it to study the signatures of 
ionized bubbles on MAPS. We also study the implications of the
foreground subtraction.

This thesis also investigates the possibility of detecting 
ionized bubbles around individual sources through radio interferometric 
observations of redshifted HI 21-cm radiation. We present a visibility 
based matched filter technique to optimally combine the 
signal from an ionized bubble and minimize the noise and foreground 
contributions. The formalism makes definite predictions on the 
ability to detect an ionized bubble or 
conclusively rule out its  presence within a radio map. Results are
presented for the GMRT and the MWA. Using simulated
HI maps we analyzed the  
impact of HI fluctuations outside the bubble on its detectability. Various 
other issues 
such as (i) bubble size determination (ii) blind search for bubbles, 
(iii) optimum redshift for bubble detection are  also discussed.

\end{singlespace}
\newpage

%\clearpage{\pagestyle{empty}\cleardoublepage} %%%%%%%%%%%%%%%%%%%%
\addcontentsline{toc}{chapter}{Contents}
\tableofcontents
%\clearpage{\pagestyle{empty}\cleardoublepage} \pagenumbering{}%%%%%%%%%%%%%%%%%%%%

\setcounter{chapter}{0}

%KKD
\setcounter{section}{0}
\setcounter{subsection}{0}
\setcounter{subsubsection}{2}
\setcounter{equation}{0}
%\pagenumbering{arabic}
\pagenumbering{arabic} \setcounter{page}{0} 
\oddsidemargin 0.9 cm \evensidemargin -.4 cm
\setlength{\textwidth}{152.4 mm}

\chapter[Introduction]
{\bf \textbf{Introduction}}

Understanding the evolutionary history of the Universe is one of the
major goals in modern cosmology. Cosmic Microwave Background Radiation 
(CMBR) observations
(COBE\footnote{http://lambda.gsfc.nasa.gov/product/cobe/},
WMAP\footnote{http://map.gsfc.nasa.gov/}) give a picture of the
early Universe (only $\sim 370,000\,{\rm years}$ after the Big
Bang). During the  first $\sim
100000\,\, {\rm years}$, the universe was a fully ionized plasma with a
strong coupling caused by the Thomson scattering between photons and
electrons. Because of adiabatic expansion, the temperature of the Universe
dropped down to few thousand Kelvin ($\sim 3000 \,{\rm K}$) at redshit
$z \sim 1100$ and the protons and electrons  
combined for the first time to form neutral Hydrogen (HI) atoms. The
scattering of photons reduced and they decoupled from baryonic
matter.  After this the photons were mostly undisturbed except that
the expansion of the Universe redshifted them into the microwave at
present. This relic background radiation is known as the CMBR. After
the recombination, the Universe became almost neutral with $75\%$ and
$25\%$ in weight of total baryonic matter was in the form of HI and
neutral Helium respectively (neutral Helium formed earlier than the
HI). On the other hand observations of Ly alpha forest in quasar (QSO)
absorption spectra show the diffuse Hydrogen gas in the Universe to be 
completely ionized at redshifts $z\le 5$ (\cite{fan02})

\section{The Epoch of Reionization}
The above two observations suggest that the HI of the Universe was
ionized sometime in the redshift range $z=1000$ to $5$. The period
when the HI was ionized is called the Epoch of Reionization (EoR). The
EoR is one of the least known chapters in the evolutionary history of
the Universe. Its exact timing, duration, nature of the reionizing
sources, their relative contribution to the reionization, large scale
distribution and evolution of HI are highly unknown. Again, once the first 
reionizing sources  were formed, their various feedback mechanisms
such as mass deposition, energy injection and emitted radiation deeply
affect subsequent galaxy formation and influence the evolution of the 
Intergalactic Medium (IGM). The epoch of reionization, therefore,
can be considered as a complicated era which involves a large number of
interconnected processes (for a review see \cite{cf06}). The study
reionization  has been at the fore font of research over the last few
years (\cite{barkana01,barkana06}).

Though the EoR is yet to be observed in detail, theorists have
proposed possible pictures of how the reionization took place. The weak
density perturbations which were generated during 
inflation era grew through gravitational instability and lead to 
overdense regions. The first generation of galaxies formed at redshift
$z\sim20$ in these overdense regions. The gas 
in these galaxies cooled by molecular cooling and fragmented.  Then the 
first generation of stars which are believed to be massive ($\sim 100M_o$) and 
metal free were created. Enormous amount of radiation produced by
these stars ionized the surrounding IGM. Then
new generation of galaxies and stars formed. Ionized 
bubbles thus grow and  filled the entire space. In another scenario, black 
hole were created at the centers of galaxies. Enormous amount of  x-ray 
radiation ionized the IGM.

Currently two types of experiments give information about the EoR. First, the study of Lyman-$\alpha$ line absorption  
in the high redshift QSO spectra has been be used to probe the
ionization state and  
the HI distribution at high redshifts. The analysis of Gunn-Peterson
troughs (\cite{gunn}) in the high redshift QSO absorption spectra
suggests that reionization finished around redshift $z\sim 6$
(\citealt{becker,fan02,white03}). The  study of the dark gap
distribution and its evolution in QSO spectra has been shown as an
efficient and independent probe of the reionization. This analysis puts
an upper limit on the HI fraction $x_{\rm HI}<0.36$ at redshift
$z=6.3$ \citep{gallerani08}. Measurements of the sizes of HII regions
around high redshift QSOs are also consistent with the above results
(\cite{fan06}). Another independent constraint comes from CMBR
observations. The CMBR photons scatter off free electrons (produced
during reionization) which results in the suppression of 
the intrinsic temperature and polarization anisotropies on angular
scale below the horizon at EoR. At the same time a 
polarization signal is generated at large angular scale. The amplitude
and the position of the peak in the polarization angular power
spectrum depend on the reionization redshift.    
From recent WMAP measurements of the electron scattering optical depth
$\tau_e$ from the temperature-polarization and
polarization-polarization power spectrum
(\citealt{page06,dunkley08,komatsu08}) imply that 
reionization started before $z \sim 10$. 
It thus seems from these two experimental results
that the EoR is an extended and complex process
which occurred over a redshift range $6-15$
(\citealt{cf06a,fck06,alvarez}). However, there exist limitations of
using these observations  
to study the details of reionization. The Lyman-$\alpha$ absorption
feature (similarly for Lyman-$\beta$ and  Lyman-$\gamma$) in QSO  
spectra is not sensitive to the higher neutral fraction of hydrogen ($>0.1$). 
Therefore, these observations can not be used to probe the
reionization at its earlier stages. The CMBR experiments are  
sensitive only to the integrated history of the EoR  and it may not 
be useful to study the progress of reionization with redshift. 
In fact, it has been shown that the CMBR polarization power
spectrum is weakly dependent on the details of the reionization
history (\citealt{kaplinghat,hu,haiman03,colombo}), though
weak constraints could be obtained from  
upcoming experiments such as PLANCK\footnote{http://www.rssd.esa.int/Planck/}.

Several other probes have been discussed in the literature to unveil the 
reionization
history of the Universe. According to theory we expect QSOs and massive 
galaxies to form in highly 
overdense regions. Moreover first generation stars which are expected to be 
massive and short lived would produce gamma ray bursts. Observations of high 
redshift QSOs, galaxies and 
gamma ray bursts could in principle provide a great deal of information. 
Though currently 
available experimental sensitivity is not sufficient enough to detect those, 
future space based experiments like JWST\footnote{http://www.jwst.nasa.gov/} 
will have enough sensitivity to detect 
those objects.

\section{21-cm Tomography}
The interaction between the
spins of the proton and the electron in a hydrogen atom in its ground
state gives rise to two hyperfine
states, i.e., the triplet states of parallel spins and the singlet state 
of anti-parallel spins. The triplet state has higher energy than the singlet 
state. When a hydrogen atom jumps from the triplet to the singlet
state it emits a photon with the wavelength of 21-cm.

Observations of the redshifted 21 cm line from the EoR is 
perhaps one of the most promising tools for studying the EoR 
(for recent review see \cite{furlanetto4}). The advantage of this probe lies 
in the fact that the EoR can be probed at any desired redshift
by appropriately tuning  the observation frequency. Since HI is distributed 
all over space these observations have the potential  to probe the 
large scale distribution of HI. A wealth of information about the 
EoR can be extracted from these observations. Unlike the observations of 
QSO absorption spectra the redshifted
21 cm radiation does not suffer from
saturation  because the optical depth
for 21 cm radiation is much less than unity.

The possibility of observing 21 cm emission from the cosmological
structure formation was first recognized by \citet{suny} and later studied
by \citet{hogan}, \citet{scott} and \citet{madau97} considering both
emission and absorption against the CMBR. 
More recently, the effect of heating of the
HI gas and its reionization on 21 cm signal has drawn great deal of attention
and has been studied in detail 
(\citealt{gnedin,shaver,tozzi,isfm,iliev,ciardi,furlanetto,miralda,chen,cooray1,cooray2,mcquinn,sethi,salvaterra,carilli}).

There could be several approaches in interpreting the data which is
expected to come in coming years. Measurements of 21-cm signal in emission 
averaged over large area of sky would provide a direct probe of the 
evolution of neutral fraction with redshift(\cite{shaver,gnedin2}). Telescopes  
are being set up to measure this average signal
(e.g, Compact Reionization Experiments (CORE) at Australia Telescope National 
Facility, Experiments to Detect Global EOR Signature (EDGES) at MIT Haystack 
Observatory). Measurements of the HI power spectrum and individual ionized bubble detection are other two major approaches which are discussed in the following subsections.

\subsection{Statistical analysis of the 21-cm signal}
The most discussed
approach has been to study the global  statistical
properties of the HI distribution through quantities like the 
power spectrum. The precise measurements of the HI 21-cm fluctuations in terms of their multifrequency angular power spectrum would provide a wealth of information of the size, spatial distribution and evolution of the ionized regions. This would also help us  understand the effect of reionization on the structure formation, radiative feedback
mechanisms in star- forming zones, the physics of the first generation stars,  
galaxies etc. This approach has been considered in the context of lower 
redshifts (\citealt{bharad01a,bharad01b,bharad03,bharad04a}). A similar 
formalism can also be applied at high redshifts to probe reionization and also
the pre-reionization era (\citealt{zal,furlanetto2,
bharad04,bharad05a, bharad05b,ali1,ali2,lz,he}). 
It is expected that the radiation from first generation luminous
objects changes  the character of the 21 cm sky completely. During this
epoch, an unique signature of ionized regions will be imprinted on the 
redshifted 21 cm
signal that manifests the processes for the ionizing radiations and
that evolves with redshift as reionization proceeds. Chapter 2 presents the 
multifrequency angular power
spectrum (MAPS) of the epoch of reionization 21-cm signal as a joint 
function of the angular multipole $l$ and frequency separation $\Delta
\nu$. This studies the signature of ionized regions (bubbles) on the MAPS and 
its implication for separating foregrounds from the signal.

\subsection{Detecting individual ionized bubbles in 21-cm maps}
It is believed that
the ionizing radiation from QSOs and the stars within  galaxies 
reionize the surrounding neutral  IGM.  
The initial framework for the  growth of HII regions around
individual galaxies have been developed  by 
\citet{arons} and \citet{shapiro}. Later various types of 
sophisticated models for the growth of HII regions are prescribed and used to
sharpen our understanding about inhomogeneous reionizaton 
(\citealt{furlanetto04a,cohn,kramer}).
A different, complimentary 
approach would be to directly observe the individual ionized regions 
around luminous sources (stars/QSOs). The issue of  detecting  these  
bubbles in 
radio-interferometric observations of redshifted 
HI $21$ cm radiation has been drawing considerable  attention.
The  detection of individual ionized 
bubbles would be a direct probe
of the reionization process. It has also been proposed  that 
such observations  will probe the properties of the ionizing
sources and the evolution of the surrounding  IGM
\citep{wyithe04a,wyithe05,kohler05,maselli07,alvarez07,geil07,wyithe08,geil08}.
Observations of 
individual ionized bubbles would  complement the  study of
reionization through the power spectrum  of HI  brightness temperature
fluctuations.

Nearly  all the above mentioned  work on detecting ionized regions consider 
the contrast between the ionized regions and the neutral IGM in images
of redshifted HI 21 cm radiation. The HI signal is expected to be
only a small contribution buried deep in the emission  from other
astrophysical sources (foregrounds) and in the system noise.  Chapter 3 
introduces a matched filter  to   optimally combine the
entire signal of an ionized  bubble while  minimizing the noise and
foreground  contributions. This technique uses the visibilities which
are the fundamental quantity measured in radio-interferometric
observations. Using visibilities  has an advantage  over the image
based techniques because the system noise contribution in different
visibilities is independent whereas   the noise in different pixels of
a radio-interferometric images is not. Chapter 4 presents simulation
results for bubble detection in 21-cm maps and also studies the size and
position determination of ionized bubbles.

\section{The Radio- Interferometric Experiments}

On the experimental side several low frequency radio experiments are either 
functional or being set up. 
This motivated us to study the expected redshifted 21 cm background from
 the EoR and possibility to detect it. Two different observational
 strategies will be followed. The first approach is to measure the
 global 
evolution of mean HI signal with redshift, and second is to
measure large scale distribution of HI through the power spectrum
measurements and  
detect the HII regions. In principle both approaches would provide
a wealth of  
information about the reionization.  The global signature experiments
use a single, small antenna which provides a large field of view (FoV). These 
will measure the HI 21-cm signal in emission 
averaged over a large area of the sky. Two main 
experiments, namely  the Cosmological Reionization Experiments (CORE) at the 
Australian Telescope National Facility, and
the Experiment to Detect the Global EoR signature (EDGES), at the MIT Haystack 
Observatory \footnote{http://www.haystack.mit.edu} are underway in
this direction.

The majority of the recent or upcoming radio-interferometric
experiments  are aimed at 
measuring the HI 21-cm signal statistically. Individual ionized wholes
(bubbles)  
can also be detected using interferometric observations of HI. The Giant 
Metre-Wave Radio Telescope
 (GMRT\footnote{http://www.gmrt.ncra.tifr.res.in};  
\citealt{swarup}) is already
functioning at several bands in the frequency range 150-1420 MHz and can
potentially detect the 21 cm signal at high redshifts. In addition, 
construction of other low-frequency experiments such as the Murchison 
Widefield Array (MWA), LOw Frequency
ARray (LOFAR\footnote{http://www.lofar.org/}),  21 Centemeter Array
(21CMA\footnote{http://web.phys.cmu.edu/$\sim$past/}), Precision Array
to  Probe Epoch of
Reionization (PAPER), Square Kilometer
Array(SKA \footnote{http://www.skatelescope.org/}) has  
 raised the possibility to detect 21 cm signal from very high redshifts.  The 
above first generation experiments will probably start their operation at the
end of the decade.

\section{Challenges to Overcome}

Although the redshifted 21-cm line can provide enormous amount of
information, its detection is going to be a huge challenge. The signal
is expected to be highly
contaminated by foreground radio emission. Potential 
sources for these foregrounds include synchrotron and free-free
emission from our Galaxy and external galaxies, low-frequency radio
point sources and free-free emission from electrons in the IGM
(\citealt{shaver,dimat1,dimat,oh,gleser07,ali08}).   
Contributions from astrophysical foregrounds are expected to be several 
order of magnitude stronger than the HI signal (\cite{santos}).

However, there have been various proposals for tackling the
foregrounds, the most promising being the application of
multi-frequency observations. The foregrounds are expected to have a 
continuum spectra, and the
contribution at two different frequencies separated by $\Delta \nu
\sim 1 \,{\rm MHz}$ are expected to be highly correlated. The HI
signal, on the other hand, is expected to be  uncorrelated at such a frequency 
separation   and this holds the promise of allowing us
to separate the signal from the foregrounds. It has been 
proposed that multi-frequency analysis of the 
radio signal can be useful in separating out the 
foreground (e.g. \citealt{zal,santos}). 
 An alternate approach 
is to subtract  a best fit continuum spectra along each line of sight
\citep{wt06} and then determine the  power spectrum. 
 This is expected to be an  effective foreground subtraction method
 in data  with very  low noise levels.  \citet{morales06b} have
discussed the complementarity of different foreground removal
techniques and the implications for array design and the analysis of 
reionization data.

The system noise in all low frequency radio experiments relevant for the 
reionization is dominated by the sky contribution $T_{\rm sky}$. We also 
expect $T_{\rm sky}$ to vary depending on whether the source is in the 
galactic plane or away from it. This is expected to be an independent 
Gaussian random variable and can be reduced by increasing observation time. 
The observation time required to detect the HI signal is an important issue.

 Man-made radio frequency interference (RFI) is a growing problem in
 all earth-based radio astronomy. Signal from television, FM radio, 
satellites, mobile communication, electric spark etc. all fall in the same 
frequency band 
as the redshifted 21-cm reionization signal from the reionization. These are
expected to be much stronger than the expected $21 {\rm cm}$ signal, and
it is necessary to quantify and characterize the RFI. It has been suggested to 
construct EoR experiments at remote sites which are expected to have
 low RFI. Recently \citet{bowman07} have  characterized   the RFI for the 
 MWA  site on the frequency range $80$ to $300 \rm MHz$. They find an
 excellent  RFI environment except for a few channels which are
 dominated by satellite  communication signal. There are several methods and
 techniques being developed to identify, characterize, and
 ultimately subtract interfering signals (\citealt{fridman,elli}).
Telescope design also plays a role in mitigating interfering signals 
(\citealt{leshem}). The effect of polarization leakage is another
 issue which needs to be  
investigated in detail.  This could cause polarization structures on the sky to
appear as frequency dependent ripples in the foregrounds intensity
. This could be particularly severe for the MWA which has a wide field
 of view. Radio recombinations lines could be a
significant contaminant in the low frequency radio
 observations. Unfortunately due to  
lack of observations we have
little knowledge of the impact of these lines in the epoch of reionization 
observation (\citealt{morales04}). Refraction index of Earth's
 atmosphere varies significantly in space as well as in time  
in the frequency band relevant for EoR experiments. This creates significant 
calibration and imaging
problems that  must be solved in order to  reliably clean the strong 
foreground contamination (\citealt{thomp}).  This is another important
issue related to system noise. The FoV of the individual antenna
and  the baseline distribution change with observing wavelength and  
if these are neglected they could cause severe problem in extracting
 the signal.

\section{Outline of the Thesis}
We give an outline of the rest of the thesis. 
\vspace*{.2 cm}
%%%%%%%%%%%%%%%%%%%%%%%%%%%%%%%%%%%%%%%%%%%%%%%%%%%%%%%%%%%%%%%%%%%%%%%%

{\bf {\em  Chapter 2}}  calculates the Multi-frequency Angular Power Spectrum 
(MAPS) to quantify the statistics of the HI signal  as a joint 
function of the angular multipole $l$ and frequency separation $\Delta
\nu$.  Assuming a small portion of a spherical sky as a flat-sky
we develop formulae for MAPS, including the effect of peculiar velocities 
(\cite{bharad04}). The flat sky approximation is found to be a good 
representation over the angular
scales of interest. The final expression is very simple to
calculate and interpret in comparison to the formulae obtained using
full spherical sky. We adopt 
a simple model for the HI distribution which incorporates patchy 
reionization and use it to study the signatures of 
ionized bubbles on MAPS. We also study the implications of the
foreground subtraction. 
 
\vspace*{0.2 cm}

{\bf {\em  Chapter 3}} investigates  the possibility of detecting
individual ionized regions (bubbles) in 
radio-interferometric observations of HI $21$ cm radiation. We develop
a visibility based 
formalism that uses a matched filter to  optimally combine the entire signal
from a bubble while  minimizing the noise and foreground
contributions. The method makes definite predictions on the ability to
detect an ionized bubble or conclusively rule out its presence within
a radio map. We make predictions for the GMRT and the MWA at a
frequency of $150\, {\rm MHz}$ (corresponding to a redshift of $8.5$).

\vspace{0.2 cm}

{\bf {\em Chapter 4} } studies the impact of the HI fluctuations
outside the bubble that we are trying to detect on the detectibility
of the bubble in 21-cm maps. We use simulated HI maps which
incorporates the patchy reionization scenarios and investigate the
restrictions imposed by the HI fluctuations on bubble detection.
We validate the matched filter technique presented in Chapter 3
through simulation of bubble detection. We also use the simulations to
determine the accuracy to which the GMRT and the MWA will be able to
determine the size and position of an ionized bubble, and test if this
is limited due to the presence of HI fluctuations.

\vspace{0.2 cm}

In {\bf {\em Chapter 5} } we estimate the optimum redshift for detecting
ionized bubbles in 21-cm maps for different reionization scenarios. We
investigate the situations under which bubbles can be detected.  
Results are also presented in terms of the scaling relations.
\newpage

%\clearpage{\pagestyle{empty}\cleardoublepage} %%%%%%%%%%%%%%%%%%%%

 %\newpage
 \setcounter{section}{0}
 \setcounter{subsection}{0}
 \setcounter{subsubsection}{2}
 \setcounter{equation}{0}
 %\pagenumbering{arabic}

%-------------------------------------------
\chapter[Multi-frequency  Angular Power Spectrum of 21 cm 
    Signal]
{\bf \textbf {Multi-frequency  Angular Power Spectrum of 21 cm 
    Signal  \footnote{{\bf \em { This chapter is adapted from the paper  `` The multi-frequency  angular power spectrum of 
the epoch of reionization   21 cm   signal'' by \citet{kkd1}}.}}}}

\vspace{1.5cm}
\section{Introduction}

In this Chapter, we develop the formalism to calculate the
multi-frequency angular power spectrum (hereafter MAPS) which can be
used to analyse   the  21 cm   signal from HI  both in emission and 
absorption against the CMB. We restrict our attention
to HI emission which is the situation of interest for the epoch of
reionization.  In our formalism, we consider the  effect of
redshift-space distortions which has been  ignored in many of earlier
works. As noted by \cite{bharad04}, this is an important effect and can
enhance the mean signal by $50\%$ or more and the effect is expected
to be most pronounced in the multi-frequency analysis. 
We next use the flat sky
approximation to develop  a much simpler expression of MAPS which is
much easier to calculate and interpret than the   angular power
spectrum  written in terms of the spherical Bessel functions.   We
adopt a simple model for the HI distribution \citep{bharad05a}
which incorporates patchy reionization  and use it to predict the
expected signal and study its multi-frequency properties. The model
allows us to vary properties like the size of the ionized regions and
their bias relative to the dark matter. We use MAPS to analyze the
imprint of these features on the HI signal and discuss their
implication for future HI observations.

As noted earlier, the HI signal at two different frequencies separated
by $\Delta \nu$ is expected to become uncorrelated as $\Delta \nu$ is
increases. As noted in \cite{bharad05a},  the value of $\Delta
\nu$ beyond which  the signal ceases to be correlated depends on the
angular scales being observed and it is $< 1 \, {\rm MHz}$ in most
situations  of interest. A prior estimate of the multi-frequency
behavior is extremely important when planning HI observations. The
width of the individual frequency  channels sets the frequency
resolution over which the signal is averaged. This should be chosen 
sufficiently small so that the signal remains  correlated over  the
channel width. Choosing a frequency channel which is too wide would
end up averaging uncorrelated HI signal which would wash out various
important features in the signal, and also lead to a degradation in the
signal to noise ratio. In this context we also note that an earlier
work \citep{santos} assumed  individual frequency channels $1 \, {\rm
  MHz}$ wide and smoothed the signal with this before performing the
multi-frequency analysis. This, as we have already noted and shall
study in detail in this Chapter, is considerably larger than the $\Delta
\nu$ where the signal is uncorrelated and hence is  not the optimal
strategy for the analysis. We avoid such a pitfall by not
incorporating the finite frequency resolution of any realistic  HI
observations. It is assumed that the analysis be used to
determine the optimal  frequency channel width for future HI
observations. Further, it is quite straightforward to introduce a
finite frequency window into our result through a convolution.

The outline of this Chapter is as follows. In Section 2.2 we present the 
theoretical formalism for calculating MAPS of the expected 21 cm
signal considering the effect of HI peculiar velocity. The calculation
in the full-sky and the flat-sky approximation are both presented with
the details being given in separate Appendices. Section 2.2.1 defines
various components of the HI power spectrum and Section  2.2.4 presents
to models for the HI distribution. We use these models when making
predictions for the expected HI signal. We present our results in
Section 2.3 and also summarize our findings. In Section 2.4 we discuss
the implications  for extracting the signal from the
foregrounds. 
\section{Theoretical Formalism}

\subsection{The HI power spectrum}

The aim of this Section is to set up the notation and calculate
the angular correlation $C_l$ for the 21cm brightness temperature
fluctuations. It is now well known (e.g.. \citealt{bharad05a}) that the
excess brightness temperature 
observed at a frequency $\nu$ 
along a direction ${\bf \hat{n}}$ is given by
\be
T(\nu,{\bf \hat{n}}) 
= \bar{T}(z)~\eta_{\rm HI}\left(z, {\bf \hat{n}} r_{\nu}\right)
\label{eq:1_1}
\e
where the frequency of observation is related to the redshift by
$\nu = 1420/(1 + z)\, {\rm MHz}$. We consider a flat Universe ($k=0$)
in which the
comoving distance $r_{\nu}$ can be written as  
\be
r_{\nu} = \int_0^z \de z' \f{c}{H(z')}.
\e
The mean background
excess brightness temperature $\bar{T}(z)$ at redshift $z$ is written as
\be
\bar{T}(z) \approx 25 {\rm mK} \sqrt{\frac{0.15}{\Omega_m h^2}}
~\left(\frac{\Omega_b h^2}{0.022}\right)
\left(\frac{1 - Y}{0.76}\right) \sqrt{\frac{1+z}{10}}
\e
where $Y \approx 0.24$ is the helium mass fraction and 
all other symbols have usual meaning. In the above relation, it
has been assumed that the Hubble parameter 
$H(z) \approx H_0 \Omega_m^{1/2} (1+z)^{3/2}$, which is a good
approximation for most cosmological models at $z > 3$.
The quantity $\eta_{\rm HI}$ is known as 
the ``21 cm radiation efficiency in redshift space'' (\citealt{bharad05a}) and can be written 
in terms of the mean neutral hydrogen fraction $\bar{x}_{\rm HI}$ and the 
fluctuation in neutral hydrogen density field $\delta_{\rm HI}$ as
\bear
\eta_{\rm HI}(z,{\bf \hat{n}} r_{\nu}) &=& \bar{x}_{\rm HI}(z) 
[1 + \delta_{\rm HI}(z,{\bf \hat{n}} r_{\nu})]
\left(1 - \f{T_{\gamma}}{T_s}\right)
\nline
&\times&
\left[1 - \f{(1+z)}{H(z)} \f{\del v(z,{\bf \hat{n}} r_{\nu})}{\del
    r_{\nu}} \right] 
\ear
where $T_{\gamma}$ and $T_s$ are the temperature of the CMB and 
the spin temperature of the gas respectively. The term in the square 
bracket arises from the coherent components of the HI peculiar
velocities. In the above derivation it is assumed that the
term  $(1+z)\del v(z,{\bf \hat{n}} r_{\nu})/\del
    r_{\nu}$ is small compared with $H(z)$ which is a reasonable
    assumption for
    the scale of our interest.

At this stage, it is useful to make a set of assumptions which
will simplify our analysis: (i) We assume that $T_s \gg T_{\gamma}$,
which corresponds to the scenario where the spin temperature
$T_s$ and the gas kinetic temperature are 
strongly coupled either through strong $Ly\alpha$ scattering  or
collisional coupling (\citealt{madau97}).  
Though the couplings are expected to be patchy (\citealt{higgins09})
the assumption is reasonable throughout the IGM soon after the
formation of first sources of radiation. 
(ii) We assume that the HI peculiar velocity field is determined
by the dark matter fluctuations, which is reasonable as
the peculiar velocities mostly trace the dark matter potential
wells. This assumption is valid  for scales larger than the Jeans
length scale which are the scales of our interest.
We then have 
\be
\eta_{\rm HI}(z,{\bf \hat{n}} r_{\nu}) = 
\int \f{\de^3 k}{(2 \pi)^3} 
{\rm e}^{-{\rm i} k r_{\nu} ({\bf \hat{k} \cdot \hat{n}})} 
\tilde{\eta}_{\rm HI}\left(z, {\bf k}\right)
\label{eq:eta}
\e
where for ${\bf k}\neq 0$
\be
\tilde{\eta}_{\rm HI}\left(z, {\bf k}\right) =\bar{x}_{\rm HI}(z)
 \left[\Delta_{\rm HI}(z,{\bf k})
+   ({\bf \hat{k} \cdot \hat{n}})^2
\Delta(z, {\bf k}) \right] \,,
\label{eq:eta_tilde}
\e
and $\Delta_{\rm HI}(z,{\bf k})$ and $\Delta(z,{\bf  k})$ are 
the Fourier transform of the fluctuations in the HI and the
dark matter densities respectively. Note  that
$f(\Omega_m)$, which
relates peculiar velocities to the dark matter, has been assumed to
have a  value  $f(\Omega_m)=1$ which is reasonable  at the high  $z$
of interest here. 

For future use, we define the 
relevant three dimensional (3D) power spectra 
\bear
\langle \Delta(z, {\bf k}) \Delta^*(z, {\bf k'}) \rangle
&=& (2 \pi)^3 \delta_D({\bf k - k'}) P(z,k) 
\nline
\langle \Delta_{\rm HI}(z, {\bf k}) \Delta^*_{\rm HI}(z, {\bf k'}) \rangle
&=& (2 \pi)^3 \delta_D({\bf k - k'}) P_{\Delta^2_{\rm HI}}(z,k) 
\nline
\langle \Delta(z, {\bf k}) \Delta^*_{\rm HI}(z, {\bf k'}) \rangle
&=& (2 \pi)^3 \delta_D({\bf k - k'}) P_{ \Delta_{\rm HI}}(z,k) 
\ear
where $P(z,k)$ and $P_{\Delta^2_{\rm HI}}(z,k) $ are the power spectra
of the fluctuations in the dark matter and the HI densities
respectively, while  $P_{ \Delta_{\rm HI}}(z,k) $ is the
cross-correlation between the two.

\subsection{The multi-frequency angular power spectrum (MAPS)} 
The multi-frequency angular power spectrum of 21 cm  
brightness temperature fluctuations at two different frequencies
$\nu_1$ and $\nu_2$ is defined as
\be
C_l(\nu_1,\nu_2) \equiv
\langle a_{lm}(\nu_1) ~ a^*_{lm}(\nu_2) \rangle \,.
\e
In our entire analysis  $\nu_1$ and $\nu_2$ are assumed to differ by
only a small amount $\Delta \nu \ll \nu_1$, and it is convenient to
introduce the notation 
\be
C_l(\Delta \nu) \equiv C_l(\nu, \nu + \Delta \nu)
\e
where we do not explicitly show the frequency $\nu$ whose value will
be clear from the context. Further, wherever possible, we shall
not explicitly show the $z$ dependence  of various quantities like 
$\bar{T}$, $\bar{x}_{\rm HI}$, $P(k)$ etc., and it is
to be understood that these are to be evaluated at the appropriate 
redshift determined by $\nu$.

The spherical harmonic moment of $T(\nu,{\bf \hat{n}})$ are 
defined as
\bear
a_{lm}(\nu) &=& \int \de \Omega ~ Y^*_{lm}({\bf \hat{n}}) ~ 
T(\nu,{\bf \hat{n}}) 
\nline
&=&
\bar{T}
\int \de \Omega ~ Y^*_{lm}({\bf \hat{n}}) ~ 
\int \f{\de^3 k}{(2 \pi)^3} \tilde{\eta}_{\rm HI}\left({\bf k}\right)
{\rm e}^{-{\rm i} k r_{\nu} ({\bf \hat{k} \cdot \hat{n}})} \,.
\nline 
\label{eq:a_lm}
\ear
Putting the expression (\ref{eq:eta_tilde})
for $\tilde{\eta}_{\rm HI}\left({\bf k}\right)$ 
in the above equation, one can explicitly calculate the 
MAPS
 in terms of the three dimensional power spectra defined
earlier. We give the details of the calculation 
in Appendix \ref{sec:cl} and present only the final expression  for
the angular power spectrum at a frequency $\nu$ 
\bear
C_l(\Delta\nu)\!\!\!\!\!&=&\!\!\!\!\!
\f{2\bar{T}^2 ~ \bar{x}^2_{\rm HI}}{\pi}
\int_0^{\infty} k^2 \de k \,
\left[j_l(k r_{\nu}) j_l(k r_{\nu_2})
P_{\Delta^2_{\rm HI}}(k)
\right.
\nline
&-&\!\!\!\!\!
 \{j_l(k r_{\nu}) j''_l(k r_{\nu_2}) 
+ j_l(k r_{\nu_2}) j''_l(k r_{\nu})\}
P_{\Delta_{\rm HI}}(k)
\nline
&+&\!\!\!\!\! 
\left. j''_l(k r_{\nu}) j''_l(k r_{\nu_2}) P(k)
\right]
\label{eq:cl_delta_nu}
\nline
\ear
Here 
%$j''_l(x)= \de^2 j_l(x)/ \de x^2$ 
$j''_l(x)=\frac{d^2}{dx^2} j_l(x)$ 
and 
we have used the notation 
$r_{\nu_2} = r_{\nu} + r'_{\nu}~\Delta \nu$ 
with
\be
r'_{\nu} \equiv \f{\del r_{\nu}}{\del \nu} = -\f{c}{\nu_0} \f{(1+z)^2}{H(z)}.
\e
Note that equation (\ref{eq:cl_delta_nu}) predicts $C_l(\Delta \nu)$ from
the cosmological 21 cm  HI signal to be real. 

With increasing $\Delta \nu$, we expect the two spherical Bessel
functions $j_l(k r_{\nu_1})$ and  $j_l(k r_{\nu_2})$ 
to oscillate out  of phase. As a consequence the value of
$C_l(\Delta \nu)$ is expected to fall increasing $\Delta \nu$. 
We quantify this through  a dimensionless frequency
decorrelation  function defined as the ratio 
\be
\kappa_l(\Delta \nu) \equiv \f{C_l(\Delta \nu)}{C_l(0)} \,.
\label{eq:kappa_l}
\e
For a fixed multipole $l$, this fall in this function with increasing
$\Delta \nu$  essentially measures how quickly features at the  angular
scale $\theta \sim \pi/l$ in the 21 cm HI maps  at two different
frequencies become uncorrelated.  Note that  $0 \le
|\kappa_l(\Delta \nu)| \le 1$.

\subsection{Flat-sky approximation}
\label{sec:flat-sky}
Radio interferometers have a finite field of view which is determined  by
the parameters of the individual elements in the array. For example, 
at $150 \, {\rm MHz}$  this is around $3^{\circ}$ for the GMRT.
In most cases of interest it suffices to consider only   small
angular scales which correspond to $l \gg 1$. 
For the currently favored set of flat 
$\Lambda$CDM models, a 
comoving length scale  $R$ at redshift $z > 5$ would 
roughly correspond to a multipole
\be
l \approx 3 \times 10^4 \left(1 - \f{1.1}{\sqrt{1+z}}\right)
\left(\f{R}{h^{-1}\mbox{Mpc}}\right)^{-1}
\e
Thus, for length scales of $R < 100 h^{-1}$ Mpc
at $z \approx 10$, one would
be interested in multipoles $l > 200$.
For such high values of $l$ one can
work in the flat-sky approximation. 

A small portion of the sky can  be well
approximated by a plane. The unit vector 
${\bf \hat{n}}$ towards the direction of observation
can be decomposed as
\be
{\bf \hat{n}} = {\bf m} + {\boldmath{ \theta}}; ~~
{\bf m} \cdot {\boldmath{ \theta}} = 0 ; ~~ \mid {\boldmath {\theta}}
\mid \ll 1
\e
where ${\bf m}$ is a vector towards the center of the 
field of view and ${\bf \theta}$ is a two-dimensional
vector in the plane of the sky.
It is then natural to define
the two-dimensional Fourier transform of $T(\nu,{\bf \hat{n}})$
in the flat-sky as
\bear
\tilde{T}(\nu, {\bf U}) &\equiv& \int \de^2 {\boldmath{\theta}} ~  
{\rm e}^{-2 \pi {\rm i} {\bf U \cdot} {\boldmath{\theta}}} ~
T(\nu,{\bf \hat{n}})  
\label{eq:T_four}
\ear
where ${\bf U}$, which corresponds to an  inverse angular scale, is the
Fourier space counterpart of {\boldmath{$\theta$}}. Using
equations  (\ref{eq:1_1}) and (\ref{eq:eta}),  and the fact that for the
flat-sky   we can approximate ${\bf k \cdot \hat{n}
  \approx  k \cdot \hat{m}}\equiv k_{\parallel}$ we have 
\be
\tilde{T}(\nu, {\bf U}) =\f{\bar{T}}{2 \pi \, r^2_{\nu}}  \int \de   
k_{\parallel} \, e^{-i k_{\parallel}  r_{\nu}} \, \tilde{\eta}_{\rm
  HI}(k_{\parallel} {\bf \hat{m}}+ 2 \pi {\bf U}/r_{\nu}) \,.  
\e
It is useful to introduce  the $\tilde{\eta}_{\rm HI}$
power spectrum  $P_{\rm HI}$ defined as 
\be
\langle \tilde{\eta}_{\rm HI}({\bf k}) \tilde{\eta}_{\rm HI}({\bf
  k'}) \rangle = (2 \pi)^3 \delta^3_D({\bf k - k'}) P_{\rm HI}({\bf
  k}) \,.
\e
This is related to the other three power spectra introduced earlier
through 
\be
 P_{\rm HI}({\bf k})=\bar{x}^2_{\rm HI}(z) [ P_{\Delta^2_{\rm
       HI}}(k) 
+ 2  \mu^2  P_{\rm \Delta_{\rm HI}}(k) + \mu^4 
 P(k)]
\label{eq:PHI}
\e
where $\mu={\bf  \hat{m} \cdot \hat{k}}=k_{\parallel}/k$ (\citealt{barkana05}).
Note that the anisotropy of  $ P_{\rm HI}({\bf k})$ {\it i.e.,} 
its $\mu$-dependence arises  from  the peculiar velocities.

The quantities calculated  in the flat-sky approximation 
can be expressed in terms of their all-sky counterparts.
The correspondence between the all-sky angular power spectra and its 
flat-sky approximation is given by
\be
\langle \tilde{T}(\nu_1, {\bf U}) \tilde{T}^*(\nu_2, {\bf U'}) \rangle
= 
C_{2 \pi U}(\nu_1, \nu_2) ~ \delta^{(2)}_D({\bf U - U'})
\label{eq:T_Cl}
\e
where $\delta^{(2)}_D({\bf U - U'})$ is the two-dimensional 
Dirac-delta function. The details of the above calculation
are presented in Appendix \ref{sec:flat-all}.
Thus allows us to estimate the angular power spectrum
$C_l$ under the flat-sky approximation which has 
a much simpler expression
\be
C^{\rm flat}_l(\Delta \nu)
=
\f{\bar{T}^2~ }{\pi r_{\nu}^2}
\int_{0}^{\infty} \de k_{\parallel} \, 
\cos (k_{\parallel} r'_{\nu} \Delta \nu) \, P_{\rm HI}({\bf k})
\label{eq:fsa} 
\e
where the vector ${\bf k}$ has magnitude  $k = \sqrt{k_{\parallel}^2 +
  l^2/r^2_{\nu}}$ {\it ie.} ${\bf k}$ has  components $k_{\parallel}$
and $l/r_{\nu}$  along the line of sight and  in the plane of the sky
respectively. 
It is clear that the angular power spectrum $C_l(\Delta \nu)$
is calculated by summing over all Fourier modes ${\bf k}$ whose
projection in the plane of the sky has a magnitude $l/r_{\nu}$. We
also see that $C_{l}$ is  determined by the power spectra only for
modes $k \ge l/r_{\nu}$. 

The flat-sky angular power spectrum $C^{\rm flat}_l(0)$ is
essentially the 2D power spectrum of the HI distribution on a plane 
at the distance $r_{\nu}$ from the observer, and for $\Delta \nu=0$
equation (\ref{eq:fsa}) is just the relation between the 2D power spectrum
and its 3D counterpart (\cite{peacock}). For $\Delta \nu \neq 0$ it is
the cross-correlation of the 2D Fourier components of the HI
distribution on two different planes, one at $r_{\nu}$ and another at
$r_{\nu+\Delta \nu}$. Any 2D Fourier mode is calculated from its full  3D
counterparts by projecting the 3D modes onto the  plane where the 2D
Fourier mode is being evaluated. The same set of 3D modes contribute
with different phases when they are projected onto two different
planes. This gives rise to  $\cos(k_{\parallel} r'_{\nu} \Delta \nu)
$ in equation (\ref{eq:fsa}) when the same 2D mode on two different planes
are cross-correlated and this in turn causes the  decorrelation of 
$C^{\rm   flat}_l(\Delta \nu)$  with increasing $\Delta \nu$. 

Testing the range of $l$ over which the flat-sky approximation is
valid, we find that for the typical HI power spectra 
$C^{\rm flat}_l(\Delta \nu)$ is in agreement with  the full-sky
$C_l(\Delta \nu)$  calculated using  equation (\ref{eq:cl_delta_nu})
at a level better than 1 per cent  for  angular modes $l > 10$.
Since the integral in equation (\ref{eq:fsa}) is much simpler
to compute, and   more straightforward to interpret, we use 
the flat-sky approximation of $C_l$ for our calculations in the rest
of this Chapter.

Note that equation (\ref{eq:fsa}) is very
similar to the expression for the visibility correlations 
[equation (16) of \citealt{bharad05a}] expected in radio
interferometric observations of redshifted 21 cm HI emission. 
The two relations differ only in a  proportionality factor which 
incorporates the parameters of the telescope being used for the
observation. This reflects the close relation between the visibility
correlations, which are the directly measurable quantities in radio
interferometry, and the $C_l$s considered  here.

\subsection{Modeling the  HI distribution}

The crucial quantities in calculating the angular correlation function
are the three dimensional power spectra $P(k)$,
$P_{\Delta_{\rm HI}}(k)$ and 
$P_{\Delta^2_{\rm HI}}(k)$.
The form of the 
dark matter power spectrum $P(k)$ is relatively
well-established, particularly within the linear theory. We shall be
using the standard expression given by (\cite{bunn}).

The power spectrum of HI density fluctuations $P_{\Delta^2_{\rm
    HI}}(k)$ and its cross-correlation with the dark matter 
fluctuations $P_{\Delta_{\rm HI}}(k)$  are both largely unknown,
and determining these is one of the most important aims of the future  
redshifted 21 cm observations.  A possible approach
could be to implement some specific model for reionization, then  attempt
to predict the expected patchiness in the HI  distribution and 
calculate the  power spectra.  Such an exercise is somewhat beyond the
scope of this paper.  The objective here is to quantify  
the  angular power spectrum  in terms of the physical attributes
characteristic of the HI distribution at the epoch of reionization. To this end  we adopt  two
simple models with a few parameters which capture the  salient
features of the HI distribution.

The first model, which we shall denote as DM,  assumes homogeneous
reionization where the HI 
traces the dark matter , i.e., $\Delta_{\rm HI} = \Delta$.
This model does not introduce  a characteristic length-scale in the
HI distribution, and hence it serves as the fiducial model against
which we can compare the predictions for patchy reionization.   
Under the standard scenario of reionization by UV sources, 
this is a valid assumption in very early stages of reionization when
most of the IGM is neutral. However, this assumption could have
another range of validity. This has to do with the scenarios where
the dominant source of reionization are the exotic decaying particles,
like neutrinos. In such case, there would be no bubbles associated
with individual galaxies, rather the reionization proceeds in 
a homogeneous manner. In this model we have

\be 
P_{\Delta^2_{\rm HI}}(k)=P_{\Delta_{\rm HI}}(k) =
P(k)
\e
which we use in equation (\ref{eq:cl_delta_nu}) to
calculate $C_l(\Delta \nu)$. Alternately, we have 
\be 
P_{\rm HI}(k)=\bar{x}^2_{\rm HI} \,
(1 + \mu^2)^2 \,P(k)
\label{eq:P_HI_DM}
\e
which we can use in equation (\ref{eq:fsa}) to
calculate $C_l(\Delta \nu)$ in the flat-sky approximation. 
This model has only one free parameter namely the mean neutral
fraction $\bar{x}_{\rm HI}$.

The second model, denoted as PR, 
incorporates  patchy reionization. It is assumed that
reionization occurs through the growth of completely ionized regions 
(bubbles) in the hydrogen distribution. The bubbles are assumed to be
spheres,  all with the same   
comoving radius $R$,  their centers tracing the dark matter
distribution with a possible bias $b_c$.  While in reality there will
be a spread  in the  shapes and sizes of the ionized patches, we
can consider $R$ as being the  characteristic size at any particular
epoch. The  distribution of the centers of the ionized regions
basically incorporates the fact that the ionizing sources are expected
to reside  at the peaks of the dark matter density distribution and
these are expected to be strongly clustered. 
For non-overlapping  spheres  the fraction of ionized volume is given
by 
\be
\bar{x}_{\rm HII} \equiv 1 - \bar{x}_{\rm HI}
= \f{4 \pi R^3}{3} \tilde{n}_{\rm HII}
\e
where $\tilde{n}_{\rm HII}$ is the mean comoving number density of
ionized spheres and we use $\R$ to denote the ratio 
$\R=\bar{x}_{\rm HII}/\bar{x}_{\rm HI}$. 
This model has been discussed in  detail 
in Bharadwaj and Ali (2005), and we have 
\be
\Delta_{\rm HI}({\bf k}) =
\left[1 -   b_c \R 
 W(k R)\right] 
\Delta({\bf k})
- \R
W(k R) \Delta_P({\bf k})\,.
\e
The HI fluctuation is a sum of  two parts, one which is correlated
with the dark matter distribution and an uncorrelated Poisson 
fluctuation $\Delta_P$. The latter  arises from the discrete nature of
the HII regions and has a  power spectrum $\tilde{n}_{\rm HII}^{-1}$. 
Also,  $W(y) = (3/y^3) [\sin y - y \cos y]$ is the 
spherical top hat window function arising from the Fourier transform
of the spherical  bubbles. This gives

\be 
P_{\Delta^2_{\rm HI}}(k)=
\left[1 -  \R b_c 
 W(k R)\right]^2 P(k)+ 
\f{[\R W(k R)]^2}{\tilde{n}_{\rm    HII}}
\e
and 
\be
P_{\Delta_{\rm HI}}(k) = \left[1 -  \R b_c  W(k R)\right] P(k)
\e
which we use in equation (\ref{eq:cl_delta_nu})  to
calculate $C_l(\Delta \nu)$. Alternately, we have 
the HI power spectrum (Bharadwaj and Ali, 2005) 
\be
P_{\rm HI}({\bf k})=\bar{x}^2_{\rm HI} \left\{[1-\R b_c W(k R)  +
  \mu^2]^2 
P(k) + \f{[\R W(k R)]^2}{\tilde{n}_{\rm    HII}} \right\}
\label{eq:P_HI_patchy}
\e
which we can use in equation (\ref{eq:fsa}) to calculate $C_l(\Delta \nu)$
in the flat-sky approximation. 

This model has three independent parameters, namely the average neutral
fraction $\bar{x}_{\rm HI}$, the comoving radius of the ionized
bubbles $R$ and the bias of the bubble centers with respect to
the dark matter $b_c$. Our analysis assumes non-overlapping spheres and
hence it is valid only when a small fraction of the HI is ionized
 and the bias is not very large. As a consequence we restrict these
 parameters to the range $\bar{x}_{\rm   HI} \ge 0.5$ and
 $b_c \leq 1.5$. We note that in the early stages of reionization 
({\it ie.} $\bar{x}_{\rm HII} \ll  1$) equation (\ref{eq:P_HI_patchy})
matches the  HI power spectrum  
 calculated by \cite{wh05}, 
though their  method of arriving at the final result is
somewhat  different and  is quite a bit  more involved.

\begin{figure}
\includegraphics[width=150mm]{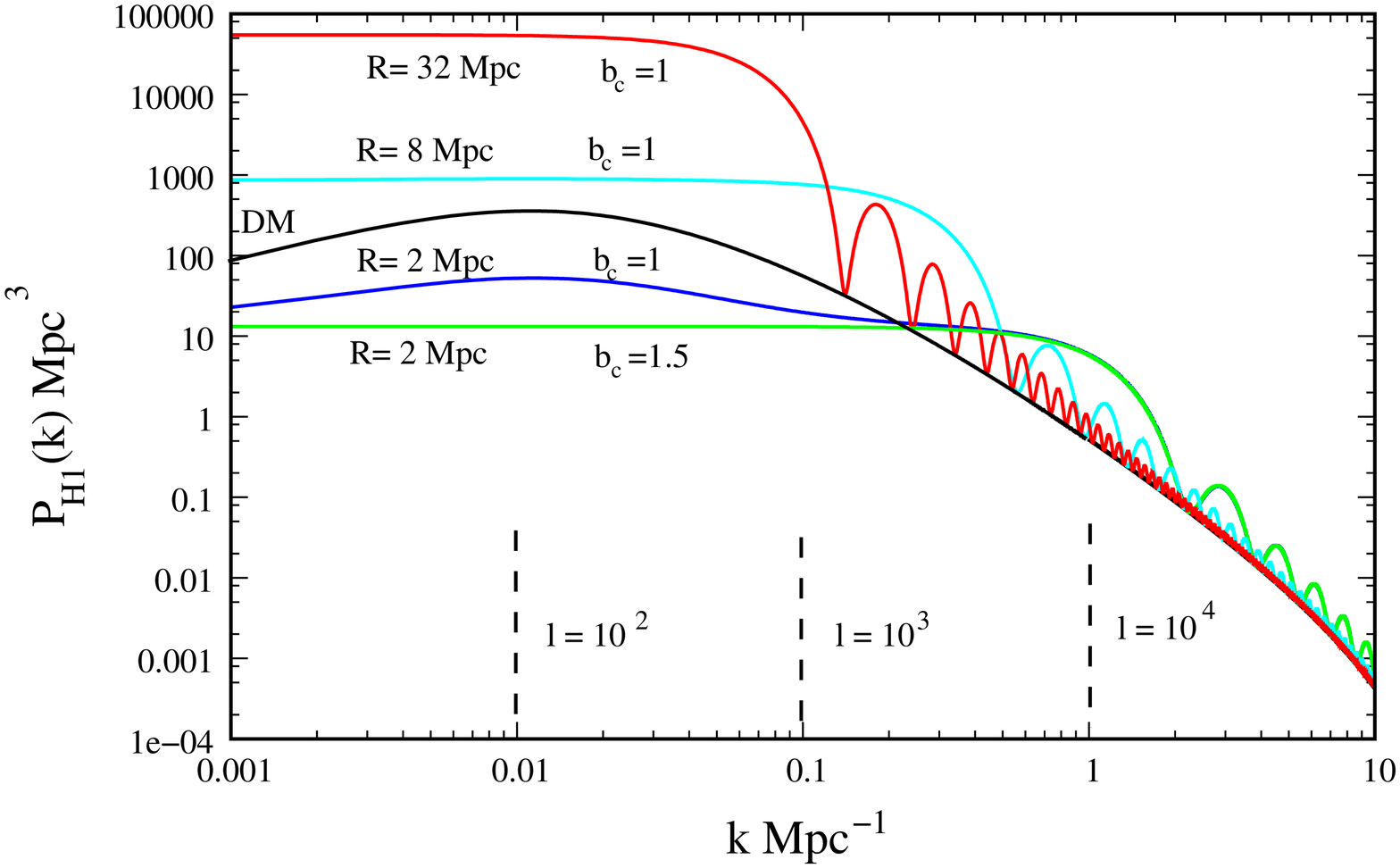}
\caption{The shows the HI  power spectrum $P_{\rm HI}({\bf k})$
for the different models of the HI distribution considered here. Other
than the one labeled DM, the curves are all for the PR model with
different values of $R$ and $b_c$ as indicated in the figure. 
Here the $\mu$ dependence  has been  incorporated  by using 
the  average value $\langle\mu^2\rangle = 1/3$.  
}
\label{fig:pk}
\end{figure}

Figure \ref{fig:pk} shows the behaviour of $P_{\rm HI}({\bf k})$ for
the two different models 
considered here. The cosmological parameters used
throughout this paper  are  those determined as the best-fit values by
WMAP 3-year data release,  i.e.,  
$\Omega_m=0.23, \Omega_b h^2 = 0.022, n_s = 0.96, h = 0.74, \sigma_8
= 0.76$ (\citealt{spergel}). Further, without any loss of generality, 
we have restricted our analysis to a single  redshift $z = 10$ which
corresponds to a frequency $\nu = 129$ MHz,  and have assumed 
$\bar{x}_{\rm HI} = 0.6$ (implying $\R = 2/3$) which is consistent
with currently favoured reionization models.  

The reason for choosing this particular redshift is that the
effects of patchy reionization are expected to be most prominent
around $z \approx 10$ in currently favoured reionization models.
At higher redshifts, the reionization is in its preliminary stages
($\bar{x}_{\rm HII} \ll 1$) and
the characteristic bubble size $R$ is quite  small. 
This implies that the effects of patchy reionization are not
substantial and hence the  HI distribution essentially traces the dark
matter.  Thus our results for the DM model are representative 
of what is expected at higher redshifts, just that the overall
normalization of $P_{\rm HI}({\bf k})$ would possibly be $z$ 
dependent through the values of the growing mode of
density perturbations  and the neutral fraction.  
At lower redshifts the HI signal
is expected to be drastically diminished because most of the
hydrogen would be ionized.
Given this, it is optimum to study the HI signal properties 
at some intermediate redshift where $\bar{x}_{\rm HII} \sim 0.5$
and $R$ is reasonably large. For the currently favoured
reionization scenarios, it seems that these properties
are satisfied at $z \approx 10$ (\citealt{cf06a}), 
which we shall be studying in the
rest of this paper.

The curve labeled  DM in Figure \ref{fig:pk}  shows $P_{\rm HI}(k)$
when the HI traces the dark matter. The characteristic scale in this
power spectrum is set by the Fourier mode entering the horizon at the
epoch of matter radiation equality. The imprint of the acoustic
oscillations in the dark matter power spectrum holds interesting
possibilities for determining cosmological parameters using high-$z$
HI observations, we do not consider this here.  The other curves in
Figure \ref{fig:pk}  all show  the PR model for  different values of
$R$ and $b_c$. 
The point to note is that for
large values of the bubble size ($R \ge 8$ Mpc) the 
power spectrum is essentially determined by the Poisson 
fluctuation term $P_{\rm HI}({\bf k}) \approx 
\bar{x}^2_{\rm HI} \R^2 W^2(k R)/\tilde{n}_{\rm HII}
= \bar{x}_{\rm HII} W^2(kR) (4 \pi R^3/3)$, which scales 
as $R^3$ and is independent of the bias parameter $b_c$. 
For length-scales larger than the bubble size  ($k < \pi/R$) we have
$W^2(k R) \approx 1$, and hence the power spectrum
is practically constant 
$P_{\rm HI}({\bf k}) \approx \bar{x}_{\rm HII} (4 \pi R^3/3)$. 
Around scales corresponding to the characteristic bubble
size $k \approx \pi/R$, the window function $W(kR)$ starts
decreasing which introduces a prominent drop in $P_{\rm HI}({\bf k})$.
For smaller length-scales ($k > \pi/R$),
the power spectrum shows oscillations arising from the nature of
the window function $W(k R)$. 
At these scales, the amplitude of $W^2(kR)$ decreases as $(kR)^{-4}$
which is more rapid than $P(k)$. Hence, at sufficiently large $k$ 
the power spectrum  $P_{\rm HI}({\bf k})$  is  dominated by the dark
matter fluctuations and it approaches the DM model,  with the approach
being faster for  large $R$. We should mention  here that the
oscillations in $P_{\rm HI}({\bf k})$ are a consequence of the fact
that we have chosen the ionized bubbles to be spheres, all of the same
size.   In reality the ionized regions will have a spread in the
bubble shapes and sizes, and it is quite likely that such
oscillation will  be washed out (\citealt{wh05})
but we expect the other features of
the PR model discussed above to hold if the characteristic  bubble size
is large $(R \ge 8 \, {\rm Mpc})$. 

For smaller values of $R$, the power spectrum $P_{\rm HI}({\bf k})$ 
could be dominated either by the term containing the dark matter
power spectrum $P(k)$ or by the Poisson fluctuation
term, depending on the value of $\R b_c = 2 b_c/3$. For values
of $b_c \approx 3/2$, the coefficient of $P(k)$ in equation 
(\ref{eq:P_HI_patchy}) tends to vanish, and hence the Poisson fluctuations 
dominate. This is obvious from the curve with parameters 
$\{R, b_c\} = \{2~\mbox{Mpc}, 1.5\}$
in Figure \ref{fig:pk}. 
On the other hand, when the bias
parameter is small, say $b_c \le 1$, 
the dark matter term dominates
over the Poisson fluctuation term at large length-scales ($k <
0.25~\mbox{Mpc}^{-1}$), 
which can be seen from the curve
with $\{R, b_c\} = \{2~\mbox{Mpc}, 1\}$. There is some  difference
between this curve and the DM model because of the 
$\R b_c W(kR)$ factor in equation (\ref{eq:P_HI_patchy}). In fact, the
the PR and DM models  exactly coincide at large scales for 
$\{R, b_c\} = \{2~\mbox{Mpc}, 0\}$ which we have not shown separately
in Figure \ref{fig:pk}.
At small length-scales ($k \gtrsim 0.25~\mbox{Mpc}^{-1}$) the amplitude
of the  dark matter power
spectrum becomes less than that of the Poisson fluctuation term, and
hence $P_{\rm HI}({\bf k})$ is independent of $b_c$ 
(one can see that the curve having $b_c = 1$ 
overlaps with the one having higher bias factor $b_c = 1.5$).
As mentioned earlier, at small length-scales ($k R \gg 1$) we expect
the $W^2(k R)$ to decay rapidly as $(kR)^{-4}$, and as a consequence 
$P_{\rm HI}({\bf k})$ will basically trace the dark matter. 
It may be  noted that  for $R=2 {\rm Mpc}$ we do not notice this
behaviour all the way  till $k = 10~\mbox{Mpc}^{-1}$ which is shown in the
Figure \ref{fig:pk}.

In addition to the effects considered above, 
the random motions within clusters
could significantly modify the signal by
 elongating  the HI clustering pattern along the line of sight   
[the Finger of God (FoG) effect].   We have incorporated this effect 
by multiplying the power spectrum $P_{\rm HI}({\bf k})$ with  an  extra 
Lorentzian term $(1+k^2_{\parallel} 
 \sigma_P^2/a^2 H^2)^{-1}$ (\citealt{sheth,ballinger}) where $\sigma_P$ is the one
 dimensional pair-velocity dispersion in relative galaxy velocities.

\section{Results}

\begin{figure}
\includegraphics[width=150mm]{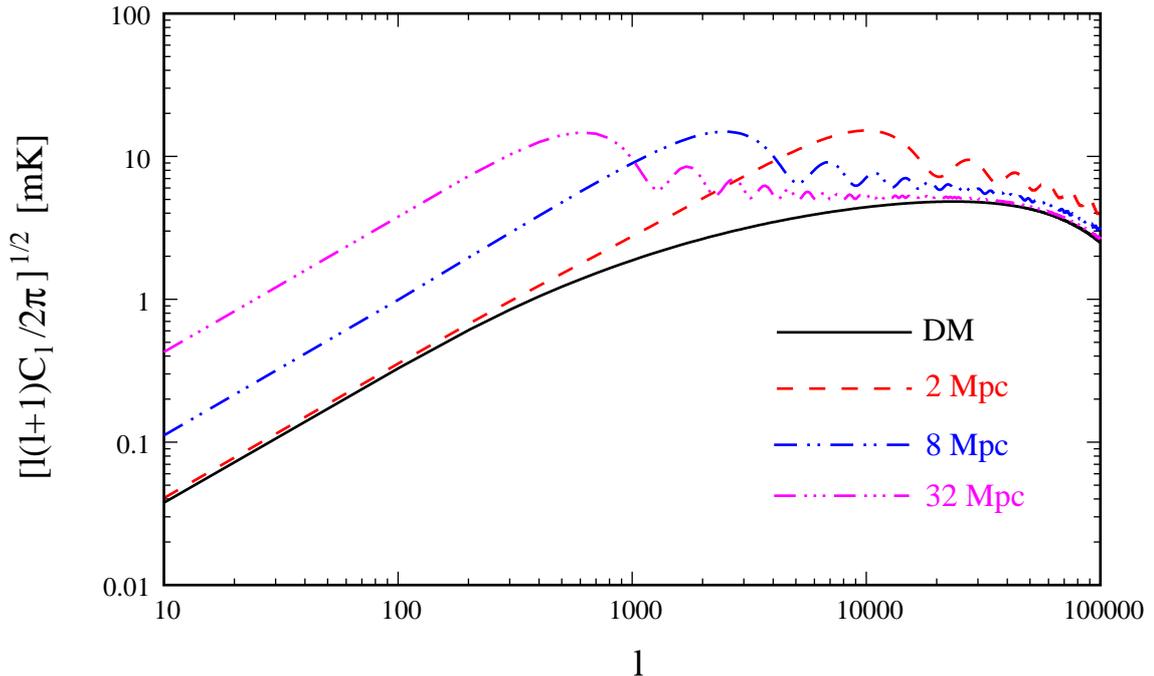}
\caption{The angular power spectrum of   HI brightness temperature
  fluctuations for different models of the HI distribution assuming
  $b_c=1$ and $\sigma_p = 0$. } 
\label{fig:cl}
\end{figure}

We first consider the angular power spectrum $C_l(\Delta \nu)$ at
$\Delta \nu=0$ for which  the results are shown in Figure  \ref{fig:cl}.
As discussed earlier,  $C_l(0)$ is essentially the 2D power spectrum
of HI 
fluctuations evaluated at the 2D Fourier mode $l/r_{\nu} \approx l \times
10^{-4} {\rm Mpc}^{-1}$.  
The results for the DM model serve as the fiducial case against which
we compare different possibilities for patchy reionization.

 For large bubble size ($R \ge 8~\mbox{Mpc}$)  the HI 
signal is  dominated by  Poisson fluctuations and it is well described
through 
\begin{equation}
\sqrt{l (l+1) C_l} \propto \sqrt{x_{\rm HII}} \, \bar{T} \,
 \frac{R}{r_{\nu}} \,  l
\label{eq:scl}
\end{equation}
 on scales larger  than the bubble.  At these angular scales  
the HI signal is substantially enhanced compared to  the DM model. 
 For smaller bubble size, the large
angle signal is sensitive to the bias $b_c$. The signal is very
similar to the DM model for $b_c=0$ and it is suppressed for higher
bias. In all cases (large or small bubble size), the signal is Poisson
fluctuation dominated on scales comparable to the bubble size and it 
 peaks at $l \approx \pi r_{\nu}/R$, with no dependence on $b_c$. The
 HI signal traces the dark matter on scales which are much smaller
 than the bubble size.    

\begin{figure}
\rotatebox{270}{\includegraphics[width=150mm]{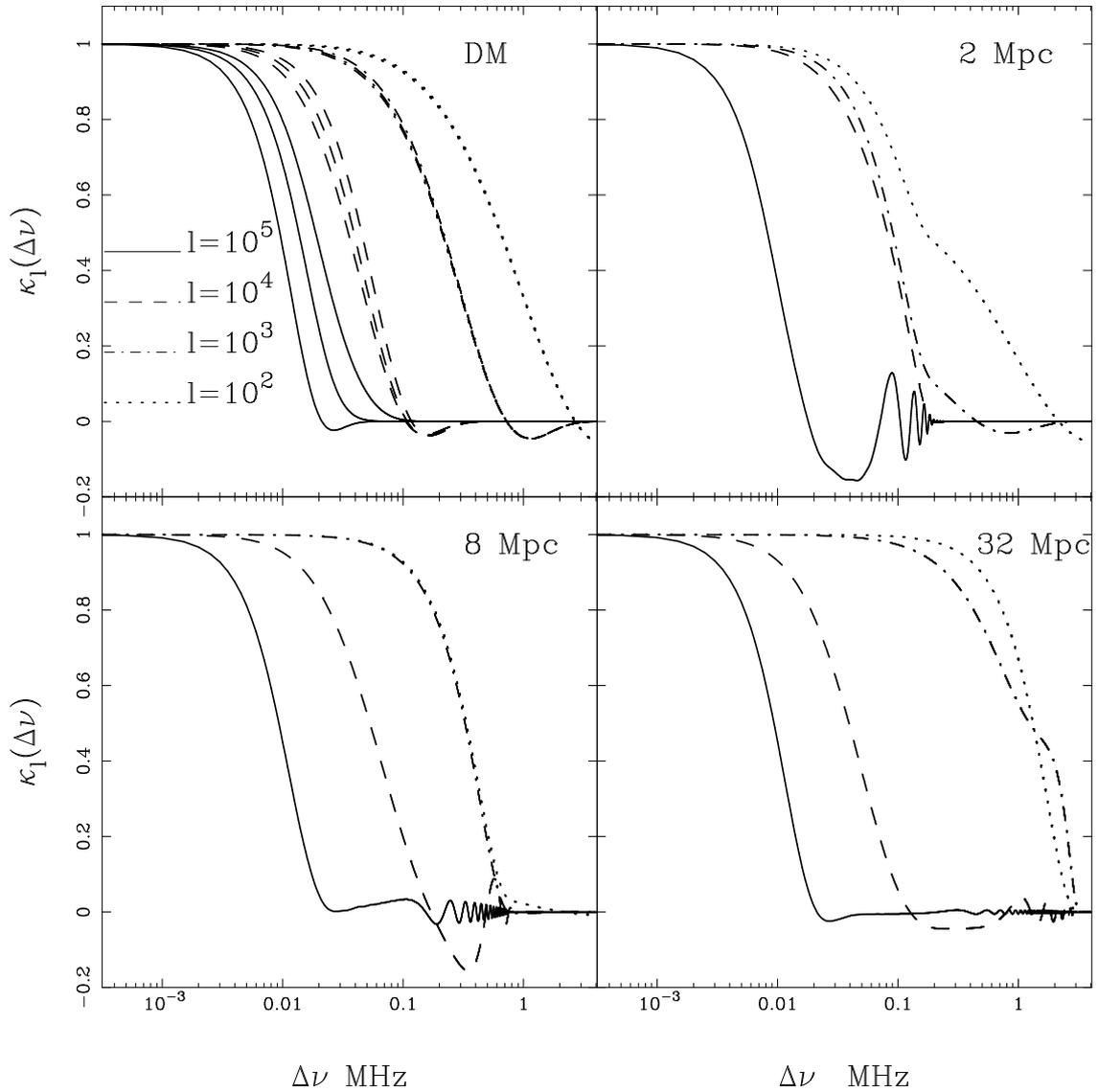}}
\caption{The frequency decorrelation function
$\kappa_l(\Delta \nu)$ [defined in equation (\ref{eq:kappa_l})] 
at $l=10^2,10^3,10^4,10^5$. Results are shown for the DM model and the
PR model with $b_c=1$ and the $R$ values shown in the figure. For the
DM model, we show results incorporating the 
FoG effect using $\sigma_P= 20 \, {\rm and } \, 40 \, {\rm km/s}$. 
For each $l$ value $\kappa_l(\Delta \nu)$ decreases  faster for
$\sigma_P=0$ and slowest for $\sigma_P=40\, {\rm km/s}$. There is a
significant  change due to the 
FoG effect only at $l \ge 10^{4}$. 
}
\label{fig:kappa_l}
\end{figure}

We next consider the behavior of $\kappa_l(\Delta \nu)$, the
frequency decorrelation function shown in  Figure \ref{fig:kappa_l}. 
For the DM model (upper left panel) where the HI fluctuations trace
the dark matter we 
find that the frequency difference $\Delta \nu$ over which the HI
signal remains correlated reduces monotonically  with increasing $l$.  
For example, while for $l=100$ $\kappa_l(\Delta \nu)$ falls to
$\sim 0.5$  at $\Delta \nu \sim 500 \, {\rm KHz}$, it
occurs much faster ($\Delta \nu \sim 10 \, {\rm KHz}$) for $l=10^5$. 
Beyond the first zero crossing $\kappa_l(\Delta \nu)$ becomes
negative (anti-correlation) and exhibits  a few highly damped
oscillations very close to zero. These oscillations arise from the
$\cos$ term in equation (\ref{eq:fsa}). 
The change in the
behavior of $\kappa_l(\Delta \nu)$ for the DM model arising from 
the FoG  effect is also shown in the same panel.
\citet{wh05} have proposed that $\sigma_p$ is
expected to have a value $\sim 30 \, {\rm km/s}$ at $z \sim 8$;  
in view of this, 
we show results for  $\sigma_p=20 \, {\rm and} \, 40 \,   {\rm km/s}$.  
We find that there is a discernible change at $l \ge 10^{4}$, and the
FoG effect causes the signal to remain correlated for a larger value
of $\Delta \nu$.  For $\sigma_p=20 \, {\rm km/s}$, 
the change is at most $15 \%$ for $l=10^{4}$ and
around $100 \%$ at $l=10^5$. Though we have not shown it explicitly, 
we expect similar changes due to FoG effect in the PR model also. 

\begin{figure}
\includegraphics[width=150mm]{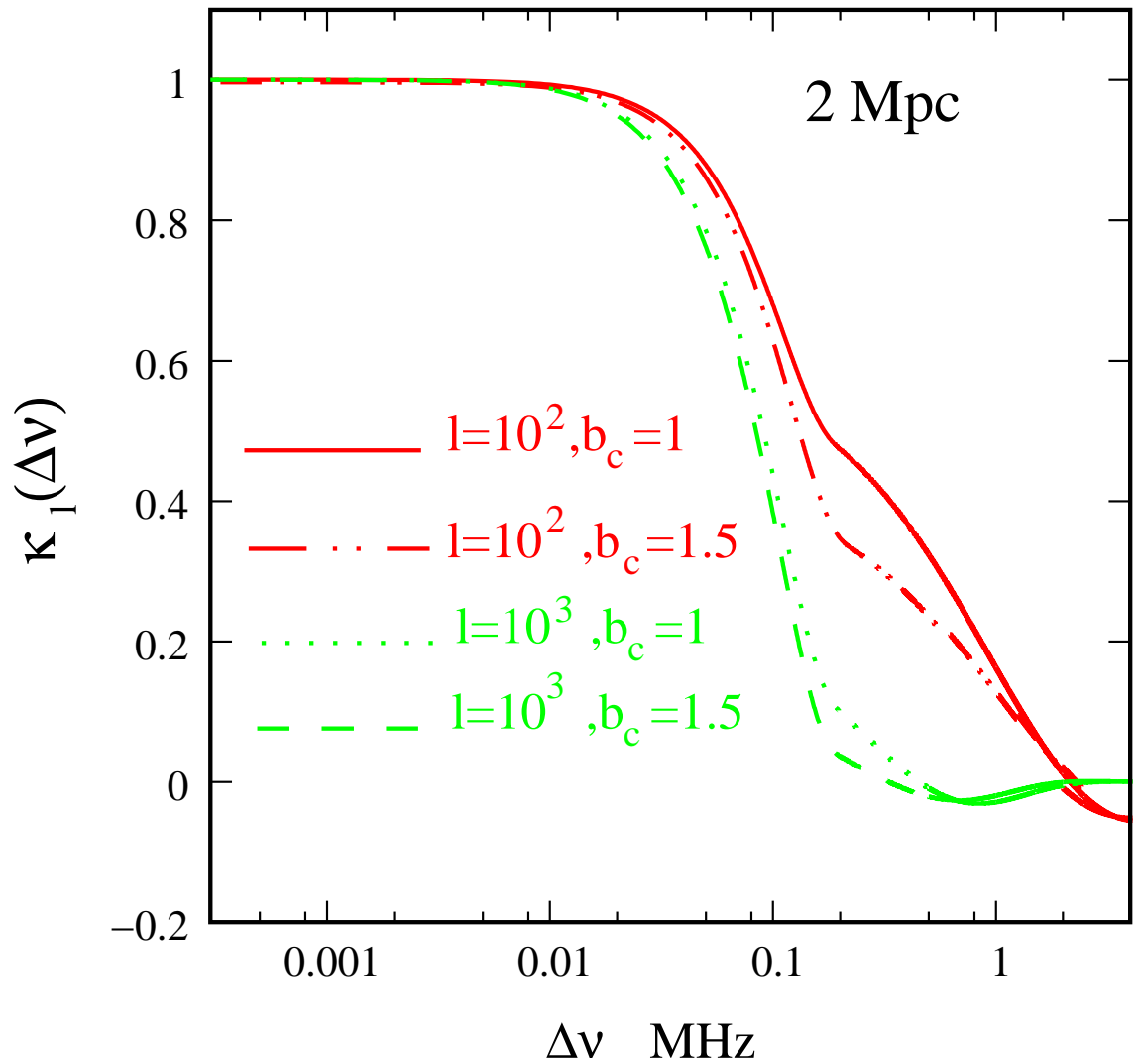}
\caption{This shows the $b_c$ dependence of the frequency decorrelation
  function $\kappa_l(\Delta \nu)$ for the PR model with $R=2 \, {\rm Mpc}$.}
\label{fig:kl2_bias}
\end{figure}

The patchy reionization model shows distinct departures from the DM
model in the behavior of $\kappa_l(\Delta \nu)$. This reflects the
imprint of the bubble size and the bias on the $\Delta \nu$
dependence. For $R=2 \, {\rm Mpc}$ and $b_c=1$ (upper right panel) the
large $l$ ($l > 1000$, comparable 
to bubble size) behavior is 
dominated by the Poisson fluctuation of the individual bubbles which
makes $\kappa_l(\Delta \nu)$ quite distinct from the DM model. Notice
that for $l=10^3$, $\kappa_l(\Delta \nu)$ falls faster than the DM
model whereas for $l=10^4$ it falls slower than the DM model causing
the $l=10^3$ and $10^4$  curves to nearly overlap.  
 The oscillations seen in $C_l$ as a function 
of $l$ in Figure \ref{fig:cl} are also seen in the $\Delta \nu$
dependence of $\kappa_l(\Delta \nu)$ at large $l$ ($10^5$).  
The behavior at $l=10^2$ is a combination of the dark matter and the
ionized bubbles, and is sensitive to $b_c$. For $b_c=1$, the initial
decrease in  $\kappa_l(\Delta \nu)$ is much steeper  than the DM model
with a sudden break after which the curve flattens. Figure
\ref{fig:kl2_bias} shows the $b_c$ dependence for $l=10^2$ and $10^3$. 
The bias dependence is weak for $l=10^3$ where the Poisson
fluctuations begin to dominate. For $l=10^2$, changing $b_c$ has a
significant affect only near the break in $\kappa_l(\Delta \nu)$
leaving much of the curve unaffected. For a smaller bubble size we
expect a behavior similar to $R=2 {\rm Mpc}$, with the $b_c$
dependence being somewhat more pronounced and the Poisson dominated
regime starting from a larger value of $l$.

For large bubble size $(R \geq 8 {\rm Mpc})$ the large angle
HI signal ($l < \pi r_{\nu}/R$) is entirely determined by the Poisson
fluctuations where the  
signal is independent of $l$. This is most clearly seen for $R=8 {\rm
  Mpc}$ where the $\kappa_l(\Delta \nu)$ curves for
$l=10^2$ and $l=10^3$  are identical.  For both $R=8 {\rm Mpc}$ and
$32 \, {\rm Mpc}$ the large $l$ behavior of $\kappa_l(\Delta \nu)$
approaches that of the DM model. 

\begin{figure}
\includegraphics[width=150mm]{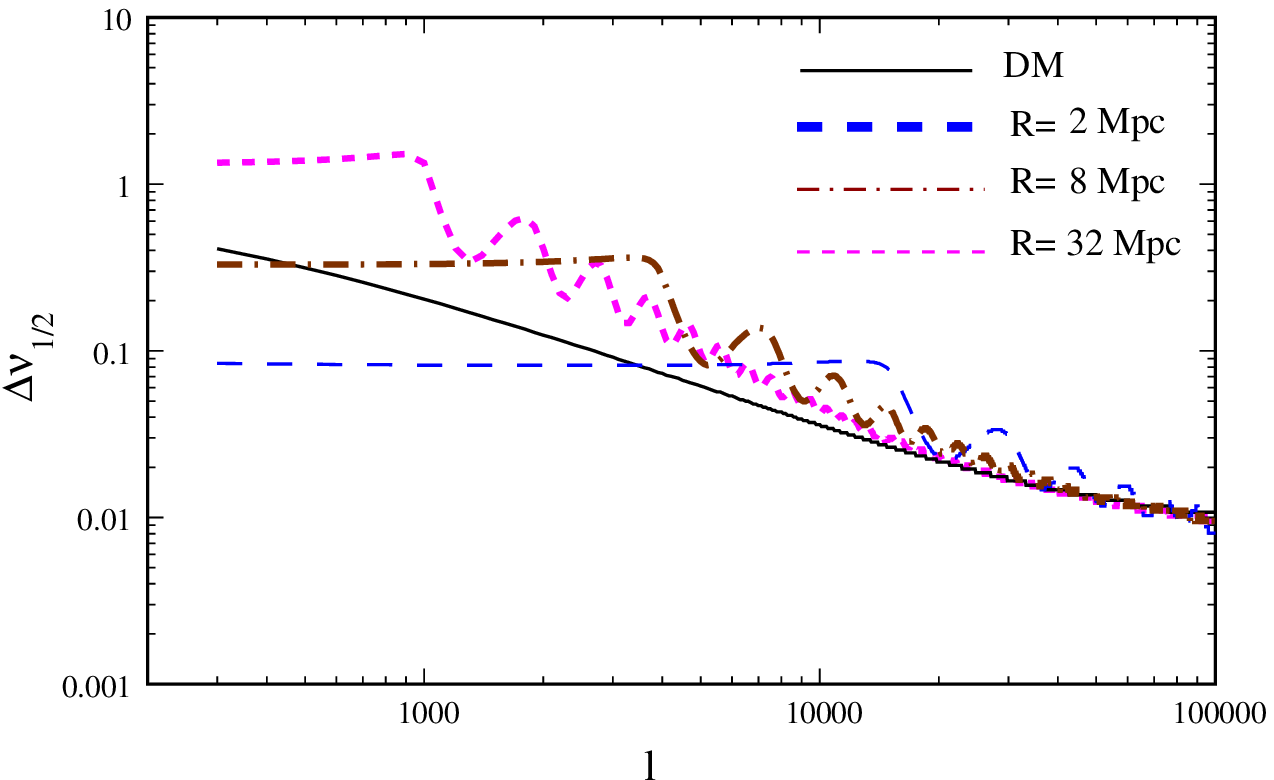}
\caption{This shows $\Delta \nu_{1/2}$ vs. $l$ for the DM model and
  the PR  model with $b_c=1$ for the $R$ values shown in the figure.  }
\label{fig:ldel}
\end{figure}

In the final part we quantify the frequency difference $\Delta \nu$
across  which the HI signal at two different frequencies remain
correlated.   To be more precise, we study  the behavior  of   
$\Delta \nu_{1/2}$  which is defined such that $\kappa_l(\Delta
\nu_{1/2})=1/2$ {\it ie.} the correlation falls to  50\% of  
its peak value at $\Delta \nu=0$. We study this for different angular
scales (different $l$) for the various models of HI distribution
considered here. 
The main aim of this exercise is to determine the  frequency
resolution that would be required to   study the HI fluctuations on a
given angular scale $l$. Optimally one would like to use a frequency
resolution smaller than $\Delta \nu_{1/2}$. A wider frequency channel
would combine different uncorrelated signals whereby the signal would
cancel out. Further, combining such signals would not lead to an
improvement in the signal to noise ratio. Thus it would be fruitful to
combine the signal at two different frequencies only as long as they
are correlated and not beyond, and we use $\Delta \nu_{1/2}$ to
estimate this.   
The plot of $\Delta \nu_{1/2}$ vs $l$ for the different HI 
models is shown in Figure \ref{fig:ldel}.

We find that for the DM model $\Delta \nu_{1/2}$ falls monotonically
with $l$ and the relation  is  well approximated by a power law
\begin{equation}
\Delta \nu_{1/2}= \, 0.2 {\, \rm MHz} \times
\left( \frac{l}{10^3} \right)^{-0.7} 
\end{equation}
which essentially says that $\Delta \nu_{1/2}\sim 0.66 \, {\rm MHz}$ on
$1^{\circ}$ angular scales,  $\Delta \nu_{1/2}\sim 0.04 \, {\rm MHz}$
on $1'$ angular scales and $\Delta \nu_{1/2}\sim 2 \, {\rm KHz}$
on $1''$ angular scales.

For the PR model, as discussed earlier, $\kappa_l(\Delta \nu)$ is $l$
independent on  angular scales larger than the bubble size $l < \pi
r_{\nu}/R$. As a consequence $\Delta \nu_{1/2}$ also is independent of
$l$ and it depends only on the bubble size $R$. This can be well
approximated by   
\be
\Delta \nu_{1/2} \approx 0.04~\mbox{MHz}~ \left(\f{R}{\mbox{Mpc}}\right)
\e
which given a large value $1.3 \, {\rm MHz}$ for $R=32 \, {\rm Mpc}$
while it falls  below the DM model for  $0.08 \, {\rm MHz}$ for $R=2
\, {\rm   Mpc}$. The large $l$ behavior of $\Delta \nu_{1/2}$
approaches the DM model though there are oscillations which persist
even at large $l$. 

We note that our findings are consistent with the earlier findings of
\citet{bharad05a} whereas they significantly different from the 
results  of \citet{santos} who assume frequency channels of  $1
\, {\rm MHz}$ which is too large.

\section{Implications for Separating Signal from Foregrounds}

Astrophysical foregrounds are expected to be several order of
magnitude stronger than the 21 cm signal. The MAPS foreground
contribution at a frequency $\nu$ can be parametrized as \citep{santos} 
\be
C_l(\Delta \nu)=A \left(\frac{\nu_f}{\nu} \right)^{ \bar{\alpha}}
\left(\frac{\nu_f}{\nu+\Delta \nu} \right)^{ \bar{\alpha}}
\left(\frac{1000}{l}\right)^{\beta}  I_l(\Delta \nu)
\label{eq:fg_1}
\e
where  $\nu_f=130$ MHz and $\bar{\alpha}$ is the mean spectral
index. The actual spectral 
index varies with line of sight across the sky and this causes the
foreground contribution to decorrelate with increasing frequency
separation $\Delta \nu$ which is quantified through the foreground
frequency 
decorrelation function $I_l(\Delta \nu)$  \citep{zal}
 which has been  modeled as 

\be
I_l(\Delta \nu)=\exp\left[ - \log_{10}^2 \left(1 +\frac{\Delta \nu}{\nu}
 \right)/2   \xi^2 \right] \,.
\e 
\begin{figure}
\includegraphics[width=150mm]{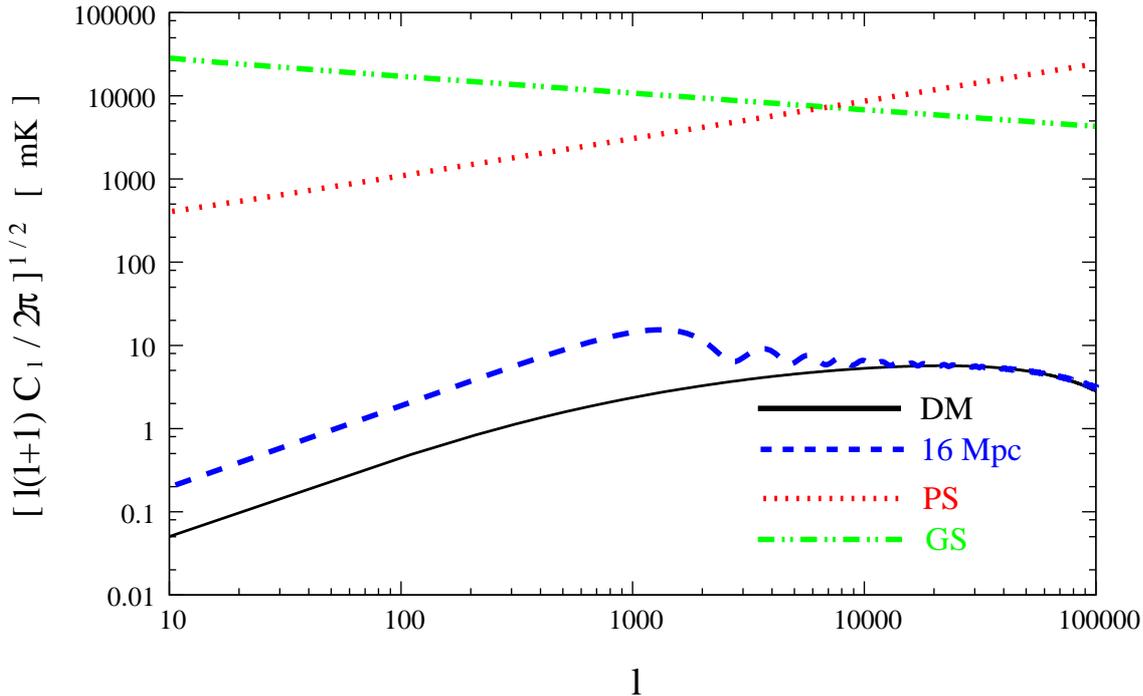}
\caption{Angular power spectrum $C_l(0)$ at $\nu=129 \,  {\rm  MHz}$
  for the two most dominant foreground components, the   diffuse
  galactic synchrotron radiation (GS) and the extragalactic point  
  sources (PS) assuming $S_{cut}=0.1 \, {\rm mJy}$. The expected
  signal is also shown  for the DM model and the PR model with $R= 16
  \,   {\rm Mpc}$. }

\label{fig:cl_fg}
\end{figure}

We consider the two most dominant  foreground
components namely extragalactic point sources and the 
diffuse synchrotron radiation from our own galaxy.  Point sources above
a flux level $S_{cut}$ can be identified in high-resolution images and
removed. We assume $S_{cut}=0.1 {\rm mJy}$ and adopt the parameter
values from  Table~1  of \citet{santos} for $A$, $\bar{\alpha}$,
$\beta$ and $\xi$. Figure \ref{fig:cl_fg} shows   the expected
$C_l(0)$ for the  signal and 
foregrounds. The galactic synchrotron radiation dominates at large
angular scales $l<10,000$ while the extragalactic point sources
dominate at small angular scales. For all values of $l$, the 
foregrounds are at least two orders of magnitude larger than the
signal.

The foregrounds have a continuum spectra, and the contributions 
at a frequency separation  $\Delta \nu$  are expected to be
highly correlated. For $\Delta \nu=1 \, {\rm MHz}$, the foreground
decorrelation function $I_l(\Delta \nu)$ falls by only $2 \times
10^{-6}$ for the galactic synchrotron radiation and by $3 \times
10^{-5}$ for the point sources. In contrast, the HI decorrelation
function $\kappa_l(\Delta \nu)$ is nearly constant at very small
$\Delta \nu$ and then has a sharp drop well within  $1 \, {\rm MHz}$, and 
is largely uncorrelated  beyond. 
This holds the promise of allowing  the signal to be separated from
the foregrounds.  A possible  strategy 
is to cross-correlate different frequency channels of    the full data
which has both  signal and foregrounds, and to  
use the distinctly different  $\Delta \nu$ dependence to separate the
signal from the foregrounds  \citep{zal}. An alternate approach 
is to subtract  a best fit continuum spectra along each line of sight
\citep{wt06} and then determine the  power spectrum. 
 This is expected to be an  effective foreground subtraction method
 in data  with very  low noise levels. 
We consider the former 
approach here, and discuss the implications of our results.

MAPS characterizes the joint $l$ and $\Delta \nu$ dependence 
which is expected to be different for the signal and the 
foregrounds. For a fixed $l$,  it will be possible to separate the
two with relative ease  at a frequency separation $\Delta \nu$ if  the
decrement in  the signal $C_l(0)[1-\kappa_l(\Delta \nu)]$  is more
than that of  the foregrounds $C_l(0)[1- I_l(\Delta \nu)]$.  
Note that because the foregrounds are much stronger  than the
HI signal, a very small decorrelation of the foreground
contribution  may cause a decrement in $C_l(\Delta \nu)$  which is
larger than 
that due to the  signal. 
We use  $F_l(\Delta \nu)$ defined as the ratio of the two
decrements 
\be 
F_l(\Delta \nu)=\frac{\{C_l(0) [1-\kappa_l(\Delta \nu)]\}_{\rm Signal}}
{\{C_l(0) [1-I_l(\Delta \nu)]\}_{\rm Foregrounds}}
\label{eq:Fl}
\e
to asses the feasibility of separating the HI signal from the
foregrounds. This gives an estimate of the
accuracy at which the $\Delta \nu$  dependence of the foreground 
 $C_l(\Delta \nu)$ has to be  characterized  for the
signal to be detected. 
\begin{figure}
\includegraphics[width=160mm]{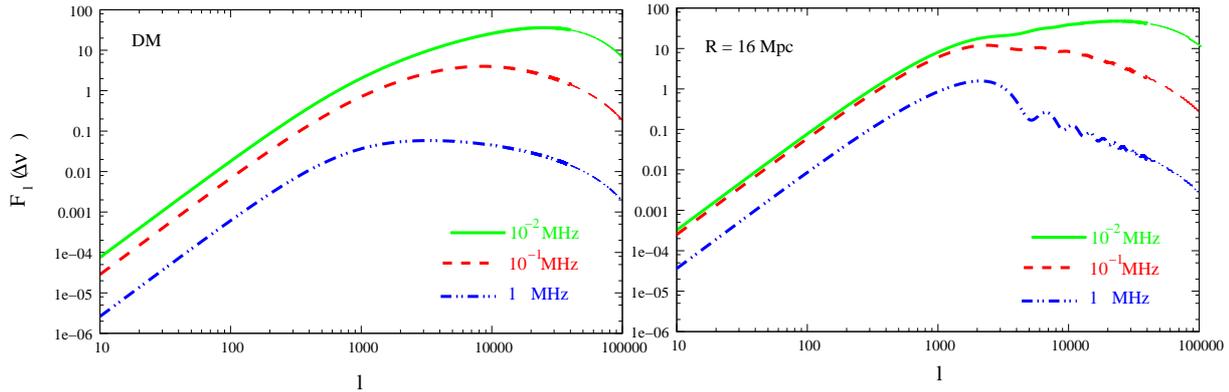} 
\caption{This shows $F_l(\Delta \nu)$ (defined in equation \ref{eq:Fl})
for the $\Delta \nu$ values shown in the figure. We consider both 
the DM model(left panel) and the PR model(right panel).}
\label{fig:Fl}
\end{figure}
 Note  we assume  that the  
$\left(\frac{\nu_f}{\nu+\Delta \nu} \right)^{ \bar{\alpha}}$ 
term in eq. (\ref{eq:fg_1}) can be  factored out before considering the
decrement in the  
foreground. Figure \ref{fig:Fl} shows the results for the DM model and
the PR model with $R=16 \, {\rm Mpc}$. 
First we note that $F_l(\Delta \nu)$ peaks at the angular scales
corresponding to $l \sim 10,000$  $(ie. \, 2^{'})$ and the prospects
of separating the signal from the foregrounds are most favorable  
at these scales. A detection will be possible in the range 
$l>1000$, $\Delta \nu \le 10 {\rm KHz}$ and $l>400$,$\Delta \nu
\le 100 {\rm KHz}$ for the DM and PR models respectively
provided  the  $\Delta \nu$ dependence of the
foregrounds $C_l(\Delta \nu)$ can be characterized with an
uncertainty less than order unity. The $l$ and $\Delta \nu$ range
would  increase if the $\Delta \nu$ dependence of the foreground
$C_l(\Delta \nu)$ were characterized to $10 \%$ accuracy. The largest
angular scales ( $l<100$ ) would require an accuracy better than  $1
\%$ which would possibly set the limit for forthcoming observations. 

The angular modes  $l=1,000$ and $l=10,000$ correspond to baselines
with antenna separations of  $\sim 300 \, {\rm  m}$  and 
$\sim 3 \, {\rm km}$ respectively. This baseline range is quite well
covered by the GMRT, and also the forthcoming interferometric
arrays.  This is possibly the optimal range for a detection. 
A possible detection  strategy would be to use  the $\Delta
\nu$ behavior of $C_l(\Delta \nu)$ in the range where $F_l(\Delta
\nu) \ll 1$ to characterize the foreground contribution. This can be
extrapolated to predict the foreground contribution at small $\Delta
\nu$ and any excess relative to this prediction can be interpreted as
the HI signal. A very precise determination of the
$\Delta \nu$ dependence of the foreground contribution would require a
very large $\Delta \nu$ range in the region where $F_l(\Delta \nu) \ll
1$, and a bandwidth of $\sim 10 {\rm MHz}$ would be appropriate. On the
other hand, at $l \sim 10,000$ the HI  $C_l(\Delta \nu)$ 
decorrelates  within $\sim 50 {\rm KHz}$ [or equivalently $F_l(\Delta
\nu)$ shows a considerable drop between $10 \, {\rm KHz} $ and $100 \,
   {\rm KHz} $, see Figure \ref{fig:Fl}], and it would be desirable to
   have   a frequency resolution better than  $\sim 10 {\rm KHz}$ 
 to optimally differentiate  between the signal and   the
 foregrounds. A lower resolution of $\sim 20 \, {\rm KHz}$ would
 possibly suffice at $l \sim 1,000$, particularly if the PR model
 holds.

\newpage

%\clearpage{\pagestyle{empty}\cleardoublepage} %%%%%%%%%%%%%%%%%%%%

 %\newpage
 \setcounter{section}{0}
 \setcounter{subsection}{0}
 \setcounter{subsubsection}{2}
 \setcounter{equation}{0}
 %\pagenumbering{arabic}

%-------------------------------------------
\chapter[Detecting  Ionized Bubbles  in Redshifted  21 cm Maps]
{\bf \textbf {Detecting  Ionized Bubbles  in Redshifted  21 cm Maps
\footnote{{\bf \em { This chapter is adapted from the paper  ``Detecting  ionized bubbles  in redshifted  21 cm maps''
 by \citet{kkd2}.}}}}}

\section{Introduction}
 In this Chapter we consider the possibility
of detecting ionized bubbles in redshifted $21 \, {\rm cm}$ HI maps. 
An ionized bubble embedded in  HI will appear as a
decrement in the background  redshifted 21 cm radiation. This decrement
will typically span across several pixels and frequency channels in 
redshifted 21 cm maps. Detecting this is a big challenge
because the HI signal ($\sim 1 \,{\rm mJy}$ or lower ) will be
buried in foregrounds  which are expected to be at least $2-3$ orders of 
magnitude larger. 
An objective detection criteria which optimally combines the  
entire signal in the  bubble while minimizing contributions from
foregrounds, system noise and other such sources 
is needed to search for ionized bubbles.   The noise in different
pixels of maps obtained from radio-interferometric observations is
correlated  (eg. \cite{thomp}), and  it is most convenient  to
deal with   visibilities instead.  These are the primary quantities
that are measured in radio-interferometry.

In this Chapter we develop a
visibility based formalism to detect an ionized bubble
or conclusively rule it out in radio-interferometric observations of
HI at  high redshifts. We apply our formalism for detecting ionized bubbles 
to make predictions for the GMRT and for one of the forthcoming instruments,
namely the MWA. For both telescopes we investigate the feasibility of
detecting the bubbles, and in situations where a detection is
feasible we predict the required observation time. For both telescopes 
we make predictions for observations only at a single frequency ($150
{\rm MHz}$), the aim here being  to demonstrate the utility of our
formalism and not  present an exhaustive analysis of the feasibility of
detecting ionized bubbles in different scenarios and circumstances. 
For the GMRT we have used the telescope parameters from their
website, while for the MWA  we use the telescope parameters from 
\citet{bowman06}.

The outline of the Chapter is as follows:
In Section 3.2 we discuss various sources which are expected to
contribute in low frequency radio-interferometric observation, this
includes the signal expected from an ionized bubble. In Section 3.3 we
present the formalism for detecting an ionized bubble, and in Section
3.4 we present the results and discuss its implications. 
The cosmological parameters used
throughout this Chapter are  those determined as the best-fit mean
values for a flat $\Lambda$CDM model by
WMAP 3-year data release,  i.e.,  
$\Omega_m=0.23, \Omega_b h^2 = 0.022, n_s = 0.96, h = 0.74, \sigma_8
= 0.76$ (\citealt{spergel}).

\section{Different Sources that Contribute to Low Frequency Radio
  Observations} 

The quantity measured in radio-interferometric observations is the
visibility $V(\u,\nu)$ which is measured in a number of frequency
channels $\nu$ across a frequency bandwidth $B$ for every pair of
antennas 
in the array. For an antenna pair, it is convenient to use $\u={\vec
  d}/\lambda$ to quantify the antenna separation ${\vec d}$  projected
in the plane perpendicular to the line of sight    in units of the
observing wavelength $\lambda$. We refer to $\u$ as a baseline. 
The visibility 
is related to the specific intensity pattern on the sky $I_{\nu}(\th)$
as 
\be
V(\u,\nu)=\int d^2 \theta A(\th) I_{\nu}(\th)
e^{ 2\pi \imath \th \cdot \u}
\label{eq:1}
\e
where $\th$ is a two dimensional vector in the plane of the sky with
origin at the center of the field of view, and $A(\th)$ is the 
beam  pattern of the individual antenna. For the GMRT this can be well 
approximated by Gaussian $A(\th)=e^{-{\theta}^2/{\theta_0}^2}$ where 
$\theta_0 \approx 0.6 ~\theta_{\rm FWHM}$ and we use the values
$2.28 ^{\circ} $ for $\theta_0$ at $150 \, {\rm MHz}$ for the GMRT. Each
MWA antenna element 
consists of $16$  crossed dipoles distributed uniformly in a square
shaped tile, and this is stationary with respect to the earth.
The MWA beam pattern is quite complicated, and it depends on the
pointing angle  relative to the zenith \citep{bowman07}. Our analysis
largely deals with the beam pattern within $1^{\circ}$ of the pointing
angle where it is reasonable to approximate the beam as being
circularly symmetric (Figures  3 and 5 of \citealt{bowman07} ). 
We  approximate the MWA antenna beam pattern as a Gaussian with
$\theta_0= 18^{\circ} $  at $153 \, {\rm MHz}$.
Note that the MWA primary beam pattern is better
modeled as $A(\th)\propto \cos^2(K \theta)$, but  a Gaussian gives a
reasonable approximation in the center of the beam which is the region
of interest here. 
Equation (\ref{eq:1}) is valid
only under the assumption that the field of  view is small so that it
can be well approximated by a plane,  or under the unlikely
circumstances that all the  antennas are coplanar.

The visibility  recorded in $150 \, {\rm MHz}$ radio-interferometric
observations is a combination of three separate contributions 
\be
V(\vec{U},\nu)=S(\vec{U},\nu)+N(\vec{U},\nu)+F(\vec{U},\nu)
\label{eq:2_2}
\e
where $S(\vec{U},\nu)$ is the HI signal that we are interested in,
$N(\vec{U},\nu) $ is the system noise which is inherent to the
measurement and $F(\vec{U},\nu)$ is the contribution from other
astrophysical sources  referred to as the foregrounds. Man-made 
radio frequency interference (RFI) from cell phones and other
communication  devices are also expected to contribute to the measured
visibilities.
 Given the lack of a detailed model for the RFI
contribution, and anticipating that it may be possible to remove it
before the analysis, we do not take it into account here.    

\subsection{The HI signal from ionized bubbles}

According to models of reionization by UV sources, the early stages
of reionization are characterized by ionized HII regions around
individual source (QSOs or galaxies). As a first approximation, 
we consider these regions as ionized spherical bubbles characterized 
by three parameters, namely, its comoving radius $R_b$, the redshift
of its center $z_c$ and the position of the center determined by the
two-dimensional vector in the sky-plane $\th_c$.
The bubble is assumed to be embedded in an uniform intergalactic
medium (IGM)
with a neutral hydrogen fraction $x_{\rm HI}$. 
We use $\rn$ to denote the  comoving distance to the
redshift where the HI
emission, received at a frequency $\nu=1420 \, {\rm MHz}/(1+z)$,
originated, and define $\rnp=d \, \rn/d \, \nu$. The planar section
through the bubble at a comoving distance $\rn$ is a disk of comoving
radius $R_{\nu}=R_b \sqrt{1- (\Delta \nu/\Delta \nu_b)^2}$ where 
$\Delta \nu=\nu_c-\nu$ is the distance from the the bubble center $\nu_c$
in frequency space with $\nu_c=1420 \, {\rm MHz}/(1+z_c)$ and 
$\Delta \nu_b=R_b/r'_{\nu_c}$ is the bubble size in the 
frequency space.
The bubble, obviously, extends from 
$\nu_c-\Delta \nu_b$ to  $\nu_c+\Delta \nu_b$
in frequency and  in each frequency channel 
within this frequency range the image of the ionized bubble
is a circular disk of angular radius $\theta_{\nu}=R_{\nu}/r_{\nu}$;
the bubble  is  not seen in  HI  beyond this frequency range. 
Under such assumptions, the  specific intensity of the
redshifted HI emission is  
\be 
 I_{\nu}(\th)=\bar{I}_{\nu} x_{\HI}
\left[1-\Theta\left(1- \frac{\mid \th-\th_c\mid}{\theta_\nu} \right) \right] 
\Theta \left(1- \frac{\mid \nu -\nu_c \mid}{\Delta \nu_b} \right)
\label{eq:3}
\e
where $\bar{I_{\nu}}=2.5\times10^2\frac{Jy}{sr} \left (\frac{\Omega_b
  h^2}{0.02}\right )\left( \frac{0.7}{h} \right ) \left
(\frac{H_0}{H(z)} \right ) $  is the radiation background from the 
uniform HI distribution 
and $\Theta(x)$ is the Heaviside
step function. 

The soft X-ray emission from the quasar responsible for the ionized
region is expected to heat the neutral IGM in a shell around the
ionized bubble. The HI emission from this shell is expected to be
somewhat  higher than $\bar{I}_{\nu}$ \citep{wyithe04a}. We do not
expect this to make a very big contribution, and we do not consider
this here. 

If we assume that the angular extent of
the ionized  bubble is  small compared to the angular scale of primary
beam {\it ie.} $\theta_\nu \ll \theta_0$,  we can take
$A(\th)$ outside the integral in 
eq. (\ref{eq:1}) and write the signal  as 
$A(\th_c) \int d^2 \theta I_{\nu}(\th)
e^{ 2\pi \imath \th \cdot \u}$, which essentially involves a
Fourier 
transform of the circular aperture $\Theta\left(1- \mid \th-\th_c\mid
r_{\nu}/R_{\nu}\right)$. For example, a bubble
of radius as large as 40 {\rm Mpc} at $z = 8.5$ would have an
angular size of only $\theta_\nu \approx 0.25^{\circ}$  which satisfies the
condition  $\theta_\nu  \ll  \theta_0$. 
In a situation where the bubble is at the center of the field of view, 
the visibility is found to be 
\be
S_{\rm center}(\u,\nu)=-\pi \bar{I_{\nu}} x_{\HI} \theta^2_\nu 
\left [ \frac{2 J_1(2 \pi U \theta_\nu
  )}{2 \pi U \theta_\nu}\right ] 
\Theta \left(1- \frac{\mid \nu -\nu_c \mid}{\Delta \nu_b} \right)
\label{eq:s_c}
\e
where $J_1(x)$ is the first order Bessel function. 
Note that $S_{\rm center}(\u,\nu)$  is real and it is the Fourier
transform of a circular 
aperture.  
The uniform HI background also contributes  
 $\bar{I_{\nu}} \pi \theta_0^2 {\rm e}^{-\pi^2 \theta_0^2 U^2}$ to the
visibility, but  this has been  dropped as it is quite insignificant
at the baselines 
of interest.  Note that the approximations used in eqs. (\ref{eq:s_c}) 
have been tested extensively by comparing the values with 
the numerical evaluation of the integral in eq. (\ref{eq:1}). 
We find that the two 
match to a high level of accuracy for the situations of interest
here. 
In the general situation where the bubble is shifted 
by $\th_c$ from the center of the field of
view, the visibility is given by
\be
S(\u,\nu) = {\rm e}^{-\theta_c^2/\theta_0^2} e^{2\pi i 
  \u \cdot \th_c}   S_{\rm center}(\u,\nu)
\label{eq:5}
\e
i.e., there is a phase shift of ${\rm e}^{2\pi i   \u \cdot \th_c}$ and 
a ${\rm e}^{-\theta_c^2/\theta_0^2}$  drop in 
the overall amplitude.

\begin{figure}
\includegraphics[width=150mm]{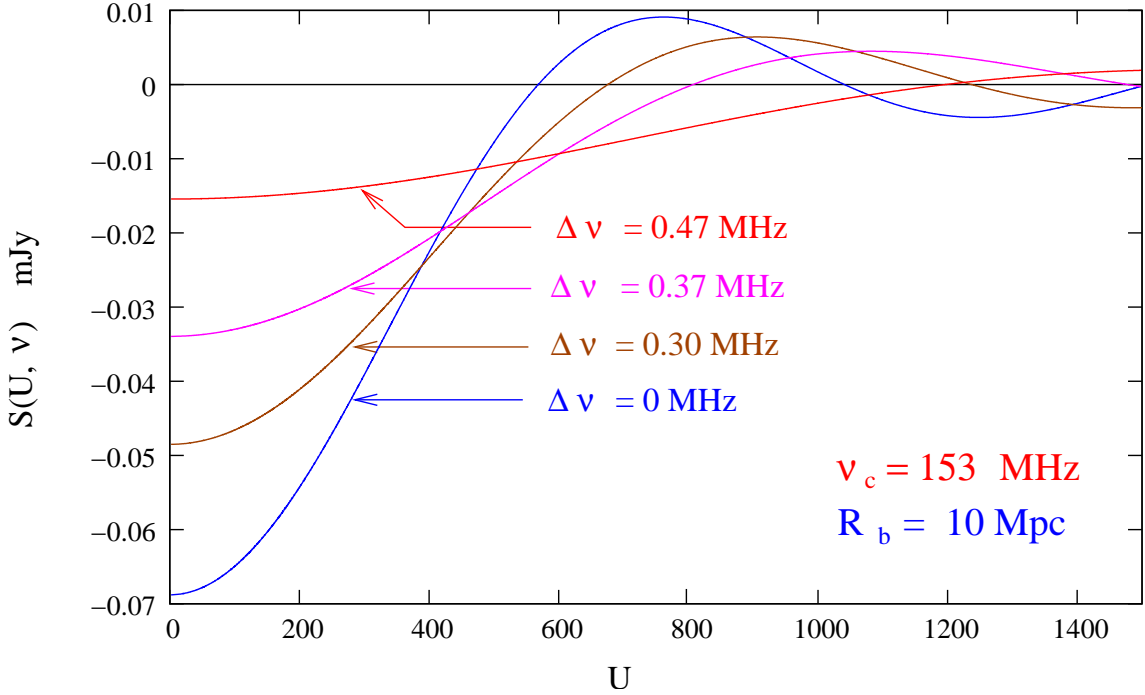}
\caption{Signal from a spherical ionized bubble of comoving
  radius $10 \, {\rm Mpc}$ as a function of baseline $U$ for
  different frequency channels.}  
\label{fig:bubu}
\end{figure}
\begin{figure}
\includegraphics[width=150mm]{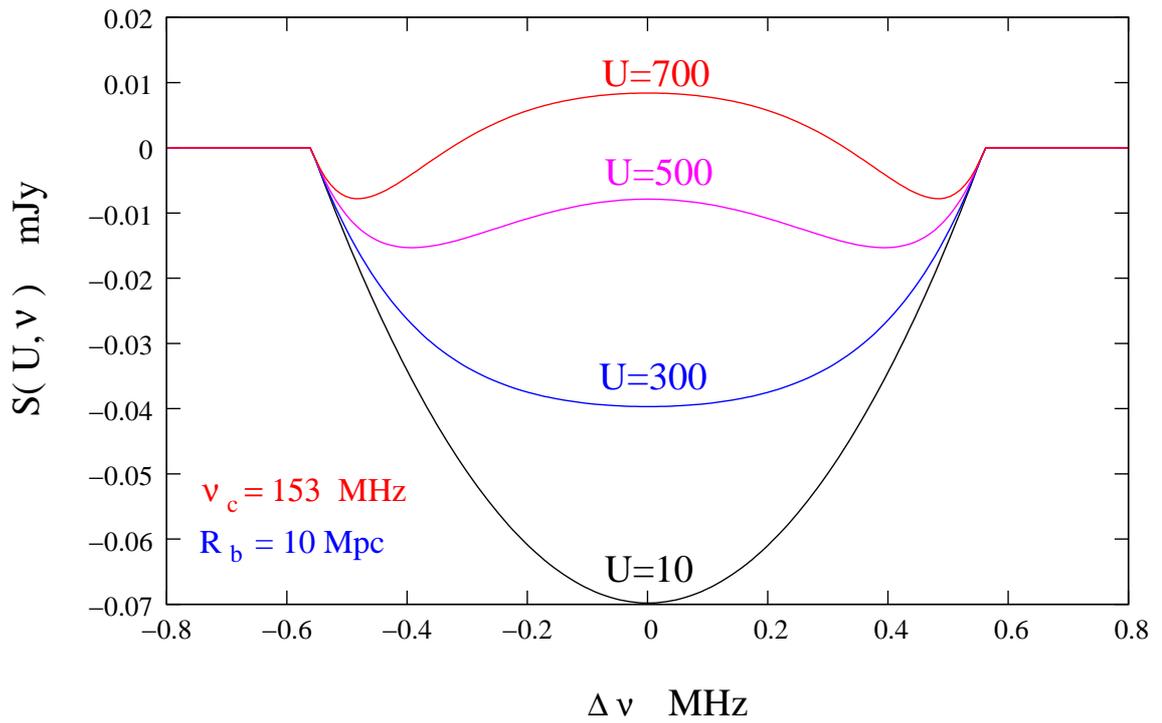}
\caption {Signal from a spherical  ionized bubble of comoving
  radius $10 \, {\rm Mpc}$ as a function of $\Delta \nu=\nu - \nu_c$   
 for different baselines.}
\label{fig:bubnu}
\end{figure}

Figures \ref{fig:bubu} and \ref{fig:bubnu} show the $U$ and $\Delta
\nu$ dependence of the visibility signal from an ionized
bubble with $R_b=10 \, {\rm Mpc}$ located at the center of the field
of view  at  $\nu_c= 153 \, {\rm MHz}$ ($z_c=8.3$),
assuming $x_{\rm HI}=1$. The signal extends over $\Delta \nu = \pm
\Delta \nu_b$  where $\Delta \nu_b=0.56 \, {\rm MHz}$. The extent in
frequency  $\Delta \nu_b = R_b/r'_{\nu_c}$ scales $\propto R_b$ when the
bubble size is varied. The Bessel function $J_1(x)$ has the first
zero crossing at $x=3.83$. As a result, 
the signal $S(\u,\nu)$ extends to $U_0=0.61
\rn [R_b \sqrt{1- (\Delta \nu/\Delta \nu_b)^2}]^{-1}$ where it has the
first zero crossing, and $U_0$  scales with the bubble size 
as $U_0 \propto 1/R_b$.  
The peak value of the signal is $S(0,\nu)=\pi
\bar{I}_{\nu} (R_b/\rn)^2  \sqrt{1- (\Delta \nu/\Delta \nu_b)^2}$ and
scales as $S(0,\nu) \propto R_b^2$ if the bubble size is
varied. We see that the peak value of the signal is $S(0,\nu_c)=70
\,{\rm   \mu Jy}$ for bubble size $R_b=10 \,{\rm Mpc}$ and would
increase  to  $1.75 \,{\rm  mJy}$ if $R_b=50 \,{\rm Mpc}$. 
Detecting these ionized bubbles will  be a big challenge because  the
signal is  buried in noise and foregrounds which are both
considerably larger in amplitude. 
Whether we are able to detect the ionized bubbles
or not  depends critically on our ability to construct optimal filters
which  discriminate  the signal from other contributions.    

\subsection{HI fluctuations}
In the previous sub-section, we assumed the ionized bubble
to be embedded in a perfectly uniform IGM. In reality, however, 
there would be fluctuations in the HI distribution in the IGM which,
in turn,
would contribute to the visibilities.
This contribution to the HI signal can be treated as a random variable
$\hat{S}(\u,\nu)$  with zero mean $\langle \hat{S}(\u,\nu) \rangle
=0$, whose  statistical properties are characterized by the
two-visibility correlation $\langle \hat{S}(\u_1,\nu_1)
\hat{S}(\u_2,\nu_2) \rangle$. This is related to $P_{\rm HI}({\bf k})$
the power spectrum of the 21 cm radiation efficiency in redshift space  
\citep{bharad04} through 
\begin{eqnarray}
\langle \hat{S}(\u_1,\nu) \hat{S}^{*}(\u_2,\nu+\Delta \nu)\rangle 
\!\!\!\!\!&=&\!\!\!\!\!
\delta_{\u_1,\u_2} 
\frac{\bar{I}^2_{\nu} \theta_0^2}{2 \rn^2} \nonumber 
 \\ \!\!\!\!\!&\times  & \!\!\!\!\!\int_0^{\infty}
\de \kp \, P_{\rm HI}(\k)   
\cos(\kp \rnp \Delta \nu)
 \label{eq:a17} 
\end{eqnarray}
where $\delta_{\u_1,\u_2}$ is the Kronecker delta {\it ie.} different
baselines are uncorrelated,  To estimate the contribution from the HI
fluctuations we make the simplifying assumption 
that the HI  traces the dark matter,  which  gives 
$P_{\rm HI}(\k)=\bar{x}^2_{\HI}\left( 1+  \mu^2 \right)^2 
P(k)$ where $P(k)$ is the dark matter power spectrum and $\mu$ is the
cosine of the angle between $\k$ and the line of sight.  
This assumption is reasonable because the scales of interest are much
larger  than the Jeans length $\lambda_J\sim 10-100 \, \rm{kpc}$, and we
expect the HI to cluster in the same way as the dark matter.  

In addition to the above, there could be other contributions
to the HI signal too. For example, there would be 
several other ionized regions in the field of view
other than  the bubble under consideration.
The Poisson noise from these ionized patches  will increase the HI 
fluctuations and  there will also be an overall
drop in the contribution because of the reduced neutral fraction. 
These effects will  depend on the reionization model, and the
simple assumptions made  would only provide a representative
estimate of the actual contribution.  Figure \ref{fig:fg} shows the
expected contribution from the HI fluctuations (HF) to the individual 
visibilities for GMRT and MWA.  
Note that while  this can be considerably larger than the signal that
we are trying to detect (particularly when the bubble size is  small), 
there is a big difference between the two.
The signal from the bubble is correlated across different
baselines and frequency channels whereas the contribution from random
HI fluctuations is uncorrelated at different baselines and it become
uncorrelated beyond a certain frequency  
separation $\Delta \nu$ \citep{bharad05a,kkd1}. 

\subsection{Noise and foregrounds}

\begin{figure}
\includegraphics[width=150mm]{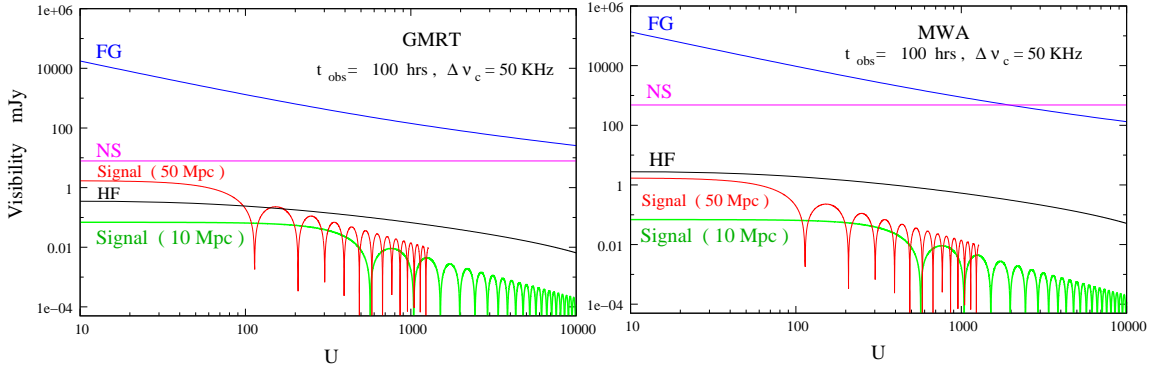}
\caption{The magnitude of the different contributions to the  visibility
 $V(\u,\nu)$   at $\nu=153 \,  {\rm  MHz}$ as a function of $U$. The 
signal,  foregrounds (FG), noise (NS) and HI fluctuations
(HF) contributions are shown for the GMRT (left) and MWA (right). The expected
  signal is shown  for  bubbles with radius $R=10 \, \rm{Mpc}$ and
 $R=50 \, \rm{Mpc}$.
The noise is estimated for a single baseline assuming an observation 
time $t_{obs}=100 \,
 {\rm  hrs}$ and channel width $\Delta \nu_c= 50 \, {\rm KHz}$.
 }
\label{fig:fg}
\end{figure}

The system noise contribution $N(\u,\nu)$ in each baseline and
frequency channel is expected to be an independent Gaussian random
variable with zero mean ($\langle \hat{N} \rangle =0$) and  whose
variance is independent of   $\u$  and $\nu_c$. The predicted
rms. noise contribution  is
(\citet{thomp})  
\be
\sqrt{\langle \hat{N}^2
  \rangle}=\frac{\sqrt2k_BT_{sys}}{A_{eff}\sqrt{\Delta \nu_c  
      \Delta t}}
\label{eq:rms}
\e
where $T_{sys}$ is the total system temperature, $k_B$ is the Boltzmann
constant, $A_{eff}$ is the effective collecting  area of each
antenna, $\Delta \nu_c$ is the channel width and $\Delta t$ is the correlator
integration time. Equation (\ref{eq:rms}) can be rewritten as
\be
\sqrt{\langle \hat{N}^2  \rangle}
=C^x \left (\frac{\Delta \nu_c}{1
   \rm{MHz}} \right )^{-1/2}\left ( \frac{\Delta
    t}{1 \rm{sec}}\right )^{-1/2}
\label{eq:noise_2}
\e
where $C^x$ varies for different interferometric arrays. Using the
GMRT parameters 
$T_{sys}=482\rm{K}$ and $A_{eff}/2 k_B=0.33 \, {\rm{K/Jy}}$ at $153 {\rm
      MHz}$ gives $C^x=1.03 {\rm Jy}$ for the GMRT where as for MWA
  $T_{sys}=470\rm{K}$ and $A_{eff}/2 k_B=5\times10^{-3} \, {\rm{K/Jy}}$
  \citep{bowman06} gives $C^x=65.52{\rm Jy}$.
The rms noise  is reduced by a factor $\sqrt{\Delta t/t_{obs}}$ if we
average over $t_{obs}/\Delta t$ independent observations where
$t_{obs}$ is the total observation  time. 
Figure \ref{fig:fg} shows   the expected noise
for a single baseline at $153 {\rm MHz}$ for $\Delta \nu_c= 50 \, {\rm
  KHz}$ and  an observation  time of $100\,  {\rm hrs}$ for both the
 GMRT and MWA. Though $T_{sys}$ is nearly equal for the GMRT and the
 MWA, the noise in a single  baseline is   expected to be
$~60$ times larger for MWA than that for the GMRT. This is a
because the individual antennas have a much
 larger collecting area at the GMRT as compared to the MWA. 
 The fact that the MWA has
many more antennas $(N=500)$ as compared to the GMRT $(N=30)$ 
compensates for this. Note that nearly half (16) of the GMRT
antennas are at very large baselines which are not particularly
sensitive to the signal on the angular scales  the ionized bubble, and
only the other 14 antennas in the $1 \, {\rm km} \times 1 \, {\rm km}$
central square will contribute towards detecting the signal.  
For both the GMRT and the MWA, $T_{sys}$ is dominated by the sky
contribution $T_{sky}$ with the major contribution coming from our
Galaxy. We expect $T_{sys}$ to vary depending on whether the source is
in the Galactic plane or away from it. The value which we have used is
typical for directions off the Galactic plane. Further, the noise
contribution will also be baseline dependent which is not included
in our analysis.

Contributions from astrophysical foregrounds are expected 
to be several order of
magnitude stronger than the HI  signal. 
Extragalactic point sources and synchrotron radiation from our Galaxy 
are predicted to be the most dominant foreground components. Assuming
that the foregrounds    
are randomly distributed, with possible clustering, we have $\langle
\hat{F}(U,\nu) \rangle =0 $ for all the baselines other than 
the one at zero spacing ($U=0$), which is not considered in this work.
The  statistical properties are characterized by the
two-visibility correlation $\langle \hat{F}(\u_1,\nu_1)
\hat{F}(\u_2,\nu_2) \rangle$. We express this (details in
Appendix \ref{sec:vis})  in terms of  the multi-frequency angular
power spectrum 
(hereafter MAPS) $C_{l}(\nu_1\, 
\nu_2)$   of the brightness temperature fluctuations at the  
frequencies $\nu_1$ and $\nu_2$ as \citep{santos,kkd1} 
\begin{eqnarray}
\langle \hat{F}(\u_1,\nu_1) \hat{F}(\u_2,\nu_2) \rangle
 =
\delta_{\u_1,-\u_2}  
\pi \left(\frac{   \theta_1^2 \theta_2^2}{\theta_1^2+
  \theta_2^2} \right )
\nline
\hspace{2cm} \left(\frac{\del B}{\del
  T}\right)_{\nu_1} \left(\frac{\del B}{\del T}\right)_{\nu_2} 
 C_{2  \pi   U_1}(\nu_1\, \nu_2).
\label{eq:11}
\end{eqnarray}
where $(\del B/\del  T)_{\nu}=2 k_B \nu^2/c^2$ is the
conversion factor to specific intensity, and we have assumed that the
primary beam pattern $A(\theta)=e^{-\theta^2/\theta_0^2}$ is frequency
dependent through $\theta_0 \propto \nu^{-1}$ and use $\theta_1$ and
$\theta_2$ to denote the value of $\theta_0$ at $\nu_1$ and $\nu_2$
respectively. Note that the foreground contribution to  different 
baselines are expected to be uncorrelated.

For each component of the foreground the MAPS is modeled as 
\be
C_{l}(\nu_1\, \nu_2)=A \left(\frac{\nu_f}{\nu_1} \right)^{ \bar{\alpha}}
\left(\frac{\nu_f}{\nu_2} \right)^{ \bar{\alpha}}
\left(\frac{1000}{l}\right)^{\beta}  I_l(\nu_1\, \nu_2).
\label{eq:fg}
\e
where  $\nu_f=130 \,{\rm  MHz}$,  and for each foreground
 component 
 $A$, $\beta$ and $\bar{\alpha}$
are the amplitude, the power law index of the angular power spectrum 
 and the mean spectral  index respectively.  The actual spectral  
index varies with line of sight across the sky and this causes the
foreground contribution to decorrelate with increasing frequency
separation $\Delta \nu=|\nu_1-\nu_2|$ which is quantified through the
foreground frequency  decorrelation function $I_l(\nu_1\, \nu_2)$
\citep{zal}  which has been  modeled as 
\be
I_l(\nu_1\, \nu_2)=\exp\left[ - \log_{10}^2 \left(\frac{\nu_2}{\nu_1}
 \right)/2   \xi^2 \right] \,.
\e
We consider the two most dominant  foreground
components namely extragalactic point sources and the 
diffuse synchrotron radiation from our own galaxy.  Point sources above
a flux level $S_{cut}$ can be identified in high-resolution continuum
images and removed. We note that absence of large baselines at the MWA
restricts the angular resolution, but it may be
possible to use the large frequency bandwidth $32 \, {\rm MHz}$ to
identify continuum point sources in the frequency domain. 
$S_{cut}$ depends on $\sigma$ the rms. noise  in the
image. We use   $S_{cut}=5 \sigma$ where $\sigma$ is the   
rms noise in the image given by (assuming 2 polarizations)
\be
\sigma=\frac{C^x}{\sqrt{2 N_b}}\left (\frac{B}{1
   \rm{MHz}} \right )^{-1/2}\left ( \frac{t_{obs}}{1 \rm{sec}}\right
)^{-1/2} 
\label{eq:12}
\e
where  $N_b=N(N-1)/2$ is the  number of  independent baselines, 
$N$ is the number of antennas in the array,
$B$ is the total frequency bandwidth and 
 $t_{obs}$ the  total observation time. For $t_{obs}=100 \, {\rm hrs}$
and $B= 6 \, {\rm MHz}$ we have $S_{cut}=0.1 {\rm mJy}$ for the GMRT and
using $B= 32 \,{\rm MHz}$ it gives $S_{cut}=0.2 \, {\rm  mJy}$  for
the MWA. 
 The value of $S_{cut}$ will be smaller for longer observations, but
reducing $S_{cut}$ any further does not make any difference to our
results so we hold $S_{cut}$ fixed at these values for the rest of our
analysis. The confusion noise from the unresolved
point sources is a combination of two parts, the Poisson contribution
due to the discrete nature of these sources and the clustering
contribution. The amplitude of  these two contributions have 
different $S_{cut}$ dependence. The parameter values that we have used
are listed in Table~\ref{tab:parm}. We have adopted the parameter
values from \citet{santos} and incorporated the $S_{cut}$ dependence
from \citet{dimat1}.

\vspace{.2in}
\begin{center}
\begin{table}
\caption{Parameters values used for characterizing different
  foreground contributions}
\label{tab:parm}
\begin{tabular}{|c|c|c|c|c|}
\hline
Foregrounds & $A ({\rm mK^2})$ & $\bar{\alpha}$ & $\beta$ & $\xi$\\ 
\hline
Galactic synchrotron & $700$ & $2.80$ &$2.4$ & $4$ \\
\hline
Point source & $61\left (\frac{S_{cut}}{0.1 {\rm mJy}}\right )^{0.5}$ 
& $2.07$ & $1.1$ &$2$ \\
(clustered part) & & & & \\
\hline
Point source & $0.16\left (\frac{S_{cut}}{0.1 {\rm
    mJy}}\right )^{1.25}$ & $2.07$ & $0$ & $1$ \\
(Poisson part) & & & & \\
\hline
\end{tabular}
\end{table}
\end{center}
\vspace{.2in}

Figure \ref{fig:fg} shows   the expected  foreground contributions 
 for the GMRT and MWA. 
The galactic synchrotron radiation is the most dominant foreground
 component  at large  angular scales ($U<1000$ for GMRT and $U<2000 $
 for MWA), while the clustering of the unresolved extragalactic point sources
 dominates at small angular scales. For all 
values of $U$, the foregrounds are at least four 
orders of magnitude larger than the signal, and also considerably
larger than the noise.

 The MWA has been designed with the detection of
 the statistical  HI fluctuation signal in mind, and hence  it
 is planned to have a very large field of view. 
 The foreground  contribution to a
 single baseline 
is expected to be $~10$ times stronger for the MWA than for the  GMRT 
because of a larger field of view.  As we shall show later, the
 increased foreground contribution is not a limitation for detecting
 HII bubbles. 
The foregrounds have a continuum spectra, and the contribution 
at two different frequencies at a  separation  $\Delta \nu$  are
expected to be 
highly correlated. For $\Delta \nu=1 \, {\rm MHz}$, the foreground
decorrelation function $I_l(\Delta \nu)$ falls by only $2 \times
10^{-6}$ for the galactic synchrotron radiation and by $3 \times
10^{-5}$ for the point sources. In contrast, the signal from an ionized
bubble peaks at a frequency corresponding to the bubble center and falls
rapidly with $\Delta \nu$ (Figure \ref{fig:bubnu}).
This holds the promise of allowing  the signal to be separated from
the foregrounds.

\section{Formalism for Detecting the Ionized Bubble}

We consider a radio-interferometric observation of 
duration $t_{obs}$,  carried out over the frequency range $\nu_1$ to
$\nu_2$.  
The HI signal from an ionized bubble, if it is present in the
data,  will be buried in foregrounds and
noise both of which are expected to be much larger. In this Section we
present a filtering technique aimed at detecting the signal from an
ionized bubble if it is present in our observations.
To detect the signal from an ionized bubble of radius $R_b$ with
center at redshift $z_c$ (or frequency $\nu_c$ )  and at an angle
$\th_c$ from the center of the field of view, 
we introduce an estimator $\E[R_b,z_c,\th_c]$ defined as 
\be
\hat{E}=  \left[ \sum_{a,b} S_{f}^{\ast}(\u_a,\nu_b)
\hat{V}(\u_a,\nu_b) \right]/\left[   \sum_{a,b} 1 \right]
\label{eq:13}
\e
where $S_f(\u,\nu)$ is a filter which has been constructed to detect
the particular ionized bubble. 
Here $\u_a$ and $\nu_b$ refer to the
different baselines and frequency channels in our observations, and in
eq. (\ref{eq:13}) we are to sum over all independent data points
(visibilities).  
Note that the estimator $\E$ and the filter $S_f(\u,\nu)$ both depend
on $[R_b,z_c,\th_c]$, the parameters of the bubble we wish to
detect, but we do not show this explicitly. The values of these
parameters will be clear from the context.  

We shall be working in the continuum
limit where the two sums in eq.~(\ref{eq:13}) can be replaced by
integrals and we have 
\be
\E =\int d^2U \int d\nu  \, \rho_N(\u,\nu) \, \, 
{S_f}^{\ast}(\u,\nu) \hat{V}(\u,\nu)  
\label{eq:14}
\e
$d^2 Ud\nu \, \rho_N(\u,\nu)$ is the fraction  of data points {\it
  ie.} baselines and  frequency channels in the
interval  $d^2 U \, d \nu$. Note that $\rho_N(\u,\nu)$ is   usually
frequency   dependent,  and it is normalized so that $\int d^2 U \,\int
d\nu \, \rho_N(\u,\nu)=1$. We refer to $\rho_N(\u,\nu)$ as the
normalized baseline distribution function. 

We now calculate $\langle \E \rangle$ the expectation value of the
estimator. Here the angular brackets  denote an average with respect
different realizations of the HI fluctuations, noise and foregrounds,
all  of which have 
been   assumed to be random variables with zero mean. This gives
$\langle \hat{V}(\u,\nu) \rangle =S(\u,\nu)$ 
and 
\be
\langle \E \rangle  =\int d^2U \, \int d\nu  \, \rho_N(\u,\nu) \, \, 
{S_f}^{\ast}(\u,\nu) S(\u,\nu)  
\label{eq:15}
\e

We next calculate the variance of the estimator which is the sum of
the contributions  from the noise (NS), the foregrounds(FG)  and the
HI fluctuations (HF)
\begin{eqnarray}
\langle (\Delta \E)^2 \rangle &\equiv& \langle (\E- \langle \E \rangle
)^2\rangle 
\nonumber \\ 
&=&\left <(\Delta \hat
E)^2 \right >_{{\rm NS}}+\left<(\Delta \hat E)^2 \right >_{{\rm FG
}} \,
 +\left<(\Delta \hat E)^2 \right >_{{\rm HF}} \,.\nonumber \\ 
\label{eq:16_3}
\end{eqnarray}
To calculate the noise contribution we go back to eq. (\ref{eq:13})
and use the fact that the noise  in different baselines and
frequency channels are  uncorrelated.  We have 
\be
\langle (\Delta \hat E)^2 \rangle_{\rm NS}
= \langle \hat{N}^2  \rangle \left[
 \sum_{a,b} \mid S_{f}(\u_a,\nu_b)\mid^2 \right] / \left[ \sum_{a,b} 1
  \right]^2 
\label{eq:17}
\e
which in the continuum limit is 
\bear
\langle (\Delta \hat E)^2 \rangle_{\rm NS}
&=& \left[ \langle \hat{N}^2 \rangle /\sum_{a,b} 1 \right] 
\nline
&\times& \int d^2U \, \int d\nu  \, \rho_N(\u,\nu) \, \, 
\mid S_{f}(\u,\nu)\mid^2
\label{eq:18}
\ear
The term $ \sqrt{\left[ \langle \hat{N}^2 \rangle /\sum_{a,b} 1
  \right]}$  is the same as $\sigma$, the rms. noise in the image
  (eq. \ref{eq:12}). We then have  
\be
\langle (\Delta \hat E)^2 \rangle_{\rm NS}
= \sigma^2 
\int d^2U \, \int d\nu  \, \rho_N(\u,\nu) \, \, 
\mid S_{f}(\u,\nu)\mid^2\,.
\label{eq:20.a}
\e

For  the foreground contribution we have 
\bear
\left <(\Delta \hat E)^2 \right >_{\rm{FG}}\!\!\!\!\!&=&\!\!\!\!\!
\int\de^2 U_1 \int \de^2 U_2 \int \de \nu_1 \int \de \nu_2
\nline 
\!\!\!\!\!&\times&\!\!\!\!\!
\rho_N(\u_1,\nu_1) \rho_N(\u_2,\nu_2)
{S_f}^{\ast}(\u_1,\nu_1) {S_f}^{\ast}(\u_2,\nu_2)
\nline 
\!\!\!\!\!&\times&\!\!\!\!\!
\langle \hat{F}(\u_1,\nu_1) \hat{F}(\u_2,\nu_2) \rangle 
\ear
In the continuum limit we have  (details given in 
Appendix \ref{sec:vis})
\bear
\langle \hat{F}(\u_1,\nu_1) \hat{F}(\u_2,\nu_2) \rangle &=&
\delta_D^{(2)}(\u_1+\u_2)  \left(\frac{\del B}{\del  T}\right)_{\nu_1}
\left(\frac{\del B}{\del  T}\right)_{\nu_2} 
\nline
&\times&
C_{2 \pi U_1}(\nu_1,\nu_2)
\label{eq:21}
\ear
which gives the variance of the foreground contribution to be 
\bear
\left <(\Delta \hat E)^2 \right >_{\rm{FG}}\!\!\!\!\!&=&\!\!\!\!\!
\int\de^2 U  \int \de \nu_1 \int \de \nu_2 \left(\frac{\del B}{\del  T}\right)_{\nu_1} \left(\frac{\del B}{\del
  T}\right)_{\nu_2} 
 \nline 
\!\!\!\!\!&\times&\!\!\!\!\!
\rho_N(\u,\nu_1) \rho_N(\u,\nu_2)
{S_f}^{\ast}(\u,\nu_1) {S_f}(\u,\nu_2)
 \nline 
\!\!\!\!\!&\times&\!\!\!\!\!
 C_{2 \pi U}(\nu_1,\nu_2)
\label{eq:fg1_3} 
\ear

We use eq. (\ref{eq:fg1_3}) to calculate  
$\left <(\Delta\hat E)^2 \right >_{\rm{HF}}$  too, 
 with the difference that we use 
the power spectrum 
$C_{2 \pi   U}(\nu,\nu + \Delta \nu)$  for the HI fluctuation  from
\citet{kkd1} instead of the foreground contribution. 

In an observation it will be possible to detect 
the presence of an ionized bubble having parameters $[R_b,z_c,\th_c]$ 
at, say 3-sigma confidence level,
if $\langle \hat{E} \rangle \ge 3 \sqrt{\langle (\Delta \hat{E})^2
  \rangle}$. In such a situation, an observed value $E_o$ can be
interpreted as a detection with  $99.7 \%$ (i.e., 3-sigma)
confidence  if $E_o > 3 \sqrt{\langle (\Delta \hat{E})^2
  \rangle}$. The presence of the ionized bubble can be ruled out at
the same level  of confidence if $\langle \hat{E} \rangle - E_o > ~3~
\sqrt{\langle (\Delta  \hat{E})^2   \rangle} $. 

\subsection{Baseline distribution}
\begin{figure}
\includegraphics[width=150mm]{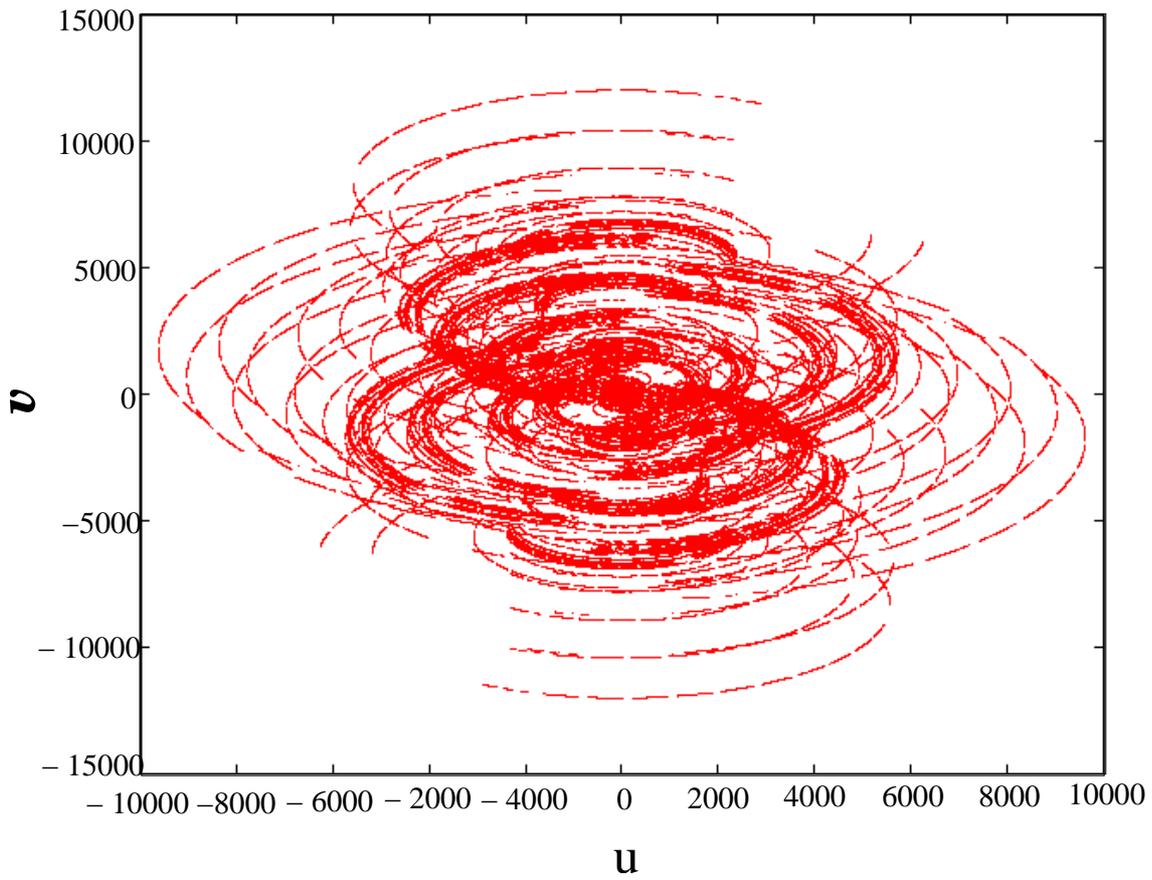}
\caption{This shows the baseline coverage for  $14 \, {\rm hrs}$ of 
 GMRT  $153 \, {\rm MHz}$ observation  at $ 45^{\circ}$ declination.}
\label{fig:4}
\end{figure}
\begin{figure}
\includegraphics[width=150mm]{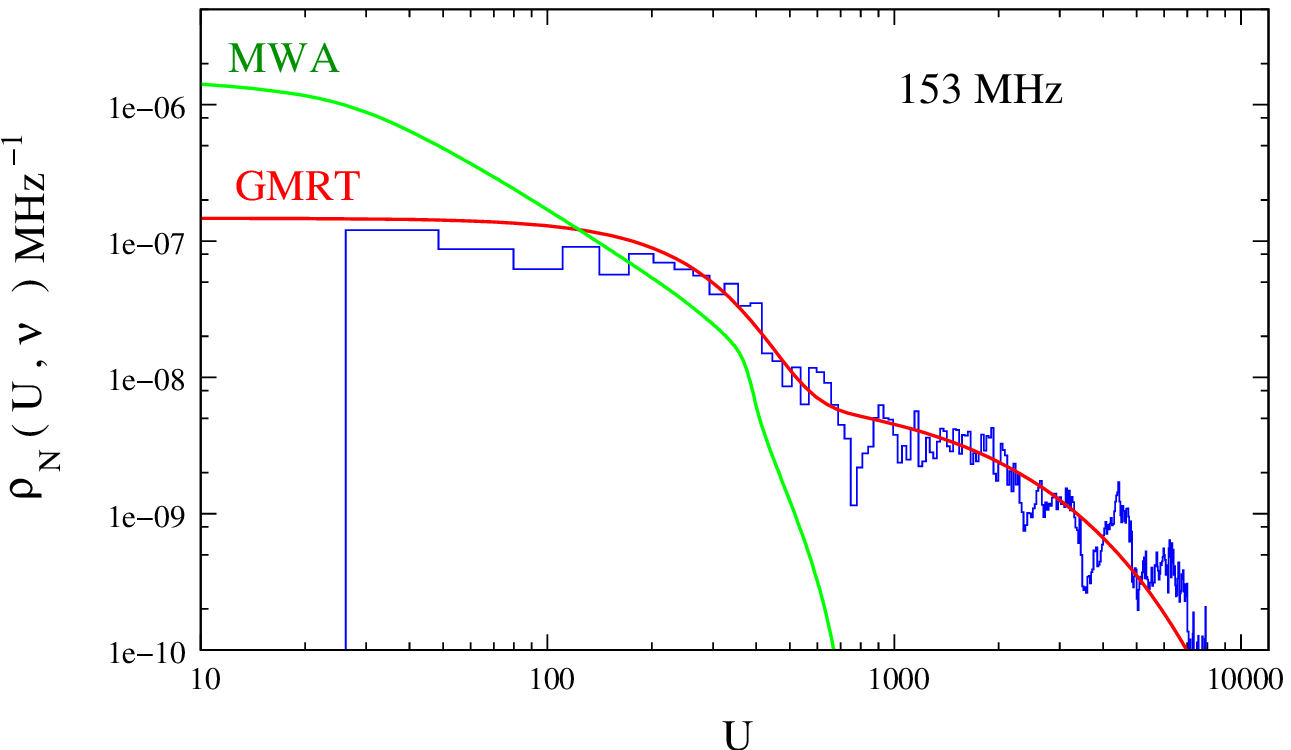}
\caption{This shows the normalized baseline distribution
  $\rho_N(U,\nu)$ for the GMRT and the MWA at $153 \, {\rm MHz}$. The
  wiggly curve  shows the actual values for the GMRT observation shown
  in Figure \ref{fig:4} and  the smooth curve is the analytic fit.} 
\label{fig:5}
\end{figure}

In this subsection we discuss the normalized baseline distribution
function $\rho_N(\u,\nu)$ which has been introduced earlier. 
Figure \ref{fig:4} shows the baseline 
coverage for $14 \, {\rm hrs}$ of observation towards a
 region at  declination $\delta=45^{\circ}$  with the GMRT  at
$153 {\rm MHz}$. In this figure  $u$ and $v$ refer to the Cartesian
 components of the baselines $\u$. Note that the baseline distribution
 is not exactly circularly symmetric. This asymmetry depends on the
source  declination which would be different for every observation. 
 We make the simplifying assumption that the baseline  distribution is
 circularly symmetric whereby  $\rho_N(\u,\nu)$ is a function of $U$. 
This considerably simplifies our analysis and gives reasonable 
estimates of what we would expect over a range of declinations.  
Figure \ref{fig:5} shows $\rho_N(\u,\nu)$ for the GMRT determined from
the baseline coverage shown in Figure  \ref{fig:4}. We find that this
is well described by the sum of a Gaussian and an exponential
distribution. The GMRT has a hybrid antenna distribution \citep{chengalur}
with $14$ antennas being randomly distributed in a central square
approximately $1 \, {\rm km} \times 1 \, {\rm km}$ and $16$ antennas being
distributed along a Y each of whose arms is $14 \, {\rm km}$ long. 
The Gaussian gives a good fit at small baselines in 
the central  square and the exponential fits the large baselines.  
Determining the best fit parameters using a least square gives 
\bear
\rho_N(\u,\nu)&=&\frac{1}{B}\left(\frac{\lambda}{{1 \, \rm km}}\right)^2 \left [ 0.21
 \exp \left ( 
  -\frac{{U^2 \lambda^2}}{2 a^2}\right ) \right.
\nline
&+& \left. 9.70 \times 10^{-3} \exp
  \left ( -\frac{U \lambda-b}{d}\right )\right ]
\label{eq:22}
\ear
where  $a=0.382 \, {\rm km}$, $b=0.986 \, {\rm km}$, $d=3.07 \, {\rm
  km}$ and $B$ is the frequency bandwidth which has a maximum value of
$6 \, {\rm MHz}$. 

Following \citet{bowman06} we assume that the MWA  antennas are
distributed within a radius of $0.750 \, {\rm km}$  with  the density of
antennas decreasing with radius $r$ as $\rho_{ant}(r)\propto
  r^{-2}$ and  with a  maximum density of one antenna per $18 {\rm
    m^2}$. The
normalized baseline distribution is  estimated in terms of 
$\rho_{ant}(r)$ and we have 
\bear
\rho_N(\u,\nu)&=&\frac{1}{4.4\times10^2 }\frac{1}{B}\left(\frac{ \lambda }{1 \,{\rm
    km}}\right)^2 \int_{
    r=0}^{\infty}\de^2r\rho_{ant}(
  r)
\nline
&\times&
\int_{\phi=0}^{2\pi}\rho_{ant}(|{\vec{r}} -\lambda \, \u \,|)\de \phi
\label{eq:23}
\ear
where the bandwidth $B$ is $32\, {\rm MHz}$, $|{\vec{r} }-\lambda \,
\u | = \left (  r^2 + U ^2\,\lambda ^2 - 
2  r\, \lambda \,
  U \, cos\phi \right )^{1/2}$. Note that $\rho_N(\u,\nu) $ depends on
the observed frequency.
Figure \ref{fig:5} shows the normalized baseline distribution function
$\rho_N(\u,\nu)$ for both  the GMRT and the MWA. We see that maximum
baseline for the GMRT is $U_{max}\sim  10,000$ whereas $U_{max}\sim
750$ for the MWA. However, the smaller baselines will be sampled more densely 
in the MWA as compared to the GMRT. 

\subsection{Filter}

It is a major challenge  to detect the signal which is expected to be
buried in noise and foregrounds both of which are much stronger
(Figure \ref{fig:fg}). It would be relatively simple to detect the
signal in a situation where there is only noise and no
foregrounds. The signal to noise ratio (SNR) is maximum  if we use
the  signal that we wish to detect as the filter ($ie. \, 
S_f(\u,\nu)=S(\u,\nu)$) and the SNR  has a value  
\bear
\frac{\langle ( \hat E) \rangle }{\sqrt{\langle (\Delta \hat
E)^2 \rangle_{\rm NS}}}\!\!\!\!\!&=&\!\!\!\!\!
\frac{1}{\sigma}
\left[ \int \, \de^2 U \, \int \de \nu \, \rho_{N}(\u,\nu) \mid
  S(\u,\nu) \mid ^2  \right]^{0.5}
\nline
\!\!\!\!\!&\propto&\!\!\!\!\! \sqrt{t_{obs}}\,.
\label{eq:s1}
\ear
The observing time necessary for a $3$-$\sigma$
detection (i.e., SNR $=3$) would be the least for
this filter. Note that the factor $-\pi x_{\rm HI}{\bar I_{\nu}}$
outside the signal (eq. \ref{eq:s_c}) is almost constant along the line of
sight of the ionized bubble. This factor does not affect  the value of the
quantity SNR$=\frac{\langle ( \hat E) \rangle }{\sqrt{\langle (\Delta \hat
E)^2 \rangle_{\rm NS}}}$. One can drop this term from the filter
$S_f(\u,\nu)$ without losing the effectiveness of the method. The
difficulty with  
using this filter is that  the foreground contribution to
$\sqrt{\langle   (\Delta \hat  E)^2 \rangle}$ is orders of magnitude
more than   $\langle ( \hat E) \rangle$.  The foregrounds, unlike the
HI signal,  are all expected to have a smooth frequency dependence and
one requires filters which incorporate this fact so as to reduce
the foreground contribution. We
consider two different filters which  reduce
the foreground contribution, but it occurs at the expense of
reducing the SNR, and  $t_{obs}$ would be more than that
predicted by eq. (\ref{eq:s1}).

The first filter (Filter I) subtracts out any frequency independent
component from 
the frequency range $\nu_c-B^{'}/2$ to $\nu_c+B^{'}/2$ with $B{'} \le B$
{\it ie.} 
\bear
S_f(\u, \,\nu)\!\!\!\!\!&=&\!\!\!\!\!\left(\frac{\lambda_c}{\lambda}\right)^2\left[ S(\u, \,\nu)\right.
\nline
\!\!\!\!\!&-&\!\!\!\!\! \left. \frac{\Theta(1-2 \mid \nu -\nu_c \mid/B^{'}) }{ B^{'}}
\int_{\nu_c-B^{'}/2}^{\nu_c + B^{'}/2}S(\u,\nu') \, \de \nu' \right ].
\nonumber\\ 
\label{eq:25}
\ear
This filter has the advantage that it does not require  any prior
knowledge about the foregrounds except that they have a continuous
spectrum. It has the drawback that there will be contributions from the
residual foregrounds as all the foregrounds are expected to have a
power law spectral dependence and not a constant. A larger value of
$B^{'}$  causes the SNR to 
increases,  and in the limit $B^{'} \rightarrow {\infty}$
the SNR  approaches the value given in
eq. (\ref{eq:s1}). Unfortunately the residues in the foregrounds also
increase with $B^{'}$. We use $B^{'}=4 \Delta \nu_b$ provided it is
less than $B $, and $B^{'}=B $ otherwise.  

The  frequency dependence of the total foreground contribution
can be expanded in Taylor series. Retaining terms only up to the first order
we have 
\be
C_l(\nu_1,\,\nu_2)=  C_l(\nu_c,\,\nu_c)
\left [1- \left (\Delta \nu_1+\Delta
  \nu_2 \right )\alpha_{eff} /\nu_c \right]
\label{eq:24}
\e
where  $\Delta \nu=\nu -\nu_c$ and $\alpha_{eff}=\frac{\sum_i \alpha^i
  \, A^i \left (1000/l\right)^{\beta_i}}{\sum_iA^i \left ( 
1000/l\right )^{\beta_i}}$ is the   effective spectral index,
here $i$ refers to the different foreground components. Note that
$\alpha_{eff}$ is $l$ dependent.  The second filter that we consider
(Filter II) allows for a linear frequency dependence of the
foregrounds and we have 
\bear
S_f(\u,\nu)\!\!\!\!\!&=&\!\!\!\!\!(1+\alpha_{eff}\Delta \nu /\nu_c)\left(\frac{\lambda_c}{\lambda}\right)^2 \left[
S(\u,
\,\nu) \right.
\nline
\!\!\!\!\!&-&\!\!\!\!\!\left.
\frac{\Theta(1-2 \mid \nu -\nu_c \mid/B^{'}) }{ B^{'}}
\int_{\nu_c-B^{'}/2}^{\nu_c + B^{'}/2}
S(\u,\nu') \, \de \nu'  \right].
\nonumber\\ 
\label{eq:26}
\ear
Note that for both the filters we   include an extra factor 
$\left(\lambda_c/\lambda\right)^2$. 
%which has not been shown in
%equations (\ref{eq:25}) and (\ref{eq:26}).  
This is introduced with
the purpose of canceling out the $\lambda^2$ dependence of
of the normalized baseline distribution function  $\rho_N(\u,\nu)$ and 
this  substantially reduces the foreground contribution.  

\section{Results and Discussions}

\begin{figure}
\includegraphics[width=150mm]{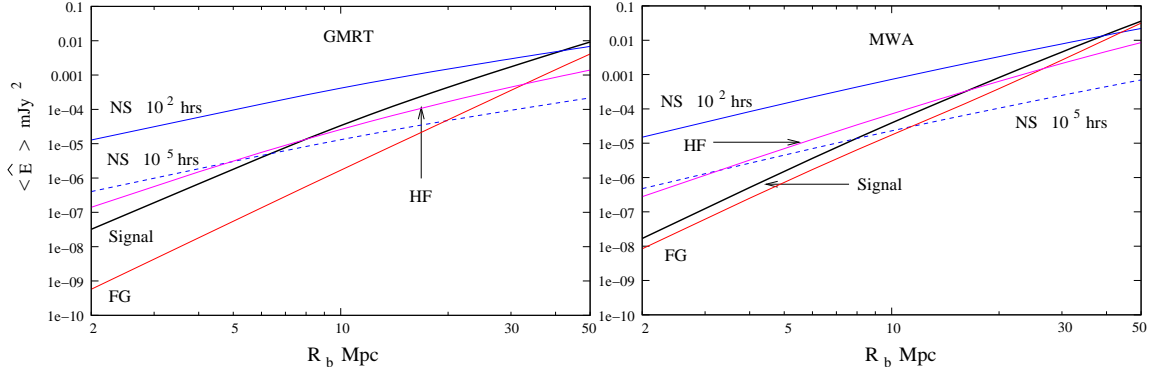}
\caption{The signal  quantified through the  expectation value of the 
  estimator $\langle \hat{E} \rangle$ for Filter I. The 
other components  (NS - Noise, FG - Foregrounds, HF - HI Fluctuations)
  are quantified through their   contribution to the 3-sigma fluctuation  
  $3 \times \sqrt{\langle  ( \Delta \hat{E}  )^2 \rangle}$.  
}
\label{fig:int_fil0}
\end{figure}

\begin{figure}
\includegraphics[width=150mm]{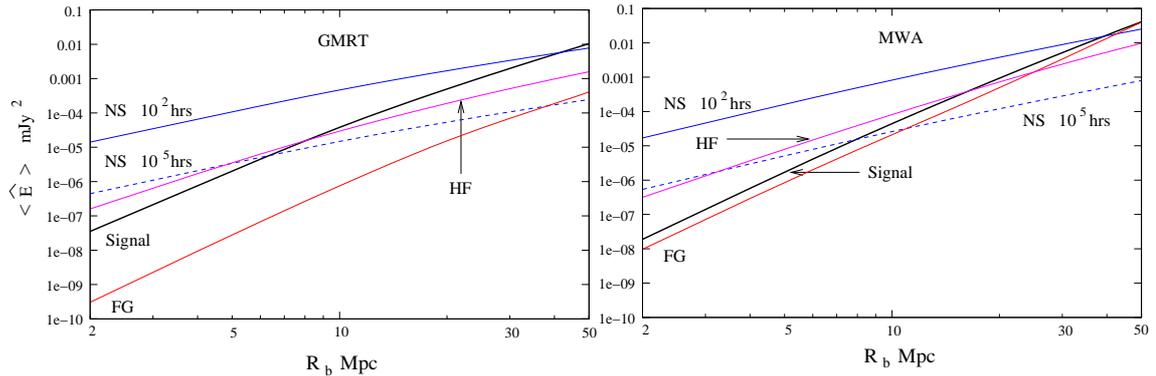}
\caption{Same as Figure \ref{fig:int_fil0} except that 
  Filter II is used instead of Filter I.}
\label{fig:alpha_eff}
\end{figure}

We first consider the most optimistic  situation where the bubble is
at the center of the field of view and the filter center is exactly
matched with the bubble center. 
The size distribution of HII regions are
quite uncertain, and  would depend on the reionization history and the
distribution of ionizing sources. However, there are some indications in
the literature on what could be the typical size of HII regions. For
example, \citet{wyithe05} deduce from proximity zone
effects that $R_b \approx 35$\,\rm{Mpc} at $z\approx 6$, which should be
considered as a lower limit. On the other hand,
\citet{furlanetto3} (Figure 1(a)) infer that the characteristic bubble size
$R_b>10 \,\rm{Mpc}$ at $z=8$ if the ionized fraction $x_i>0.75$ ($R_b \sim 50
\rm{Mpc}$ if $x_i \sim 0.9$). Theoretical models which match a variety of 
observations \citep{choudhury07} imply that $x_i$ could be as high
as $90 \, \%$ at $z \sim 8$, which would mean bubble sizes of
$\,\sim40-50$\,\rm{Mpc}. To allow for the large  variety of possibilities,   
 we have presented  results for a wide range of $R_b$ values from $2
\, {\rm Mpc}$ to $50 \, {\rm Mpc}$. We
restrict our analysis to a situation where the IGM  outside the bubble 
is completely neutral ($x_{\rm HI}=1$).  The signal would fall
proportional to 
$ x_{\rm HI}$ if the IGM outside the bubble were partially
ionized  ($x_{\rm HI}<1$).  
The expected signal  
$\langle \hat{E} \rangle$ and 3-sigma fluctuation $3 \times
\sqrt{\langle  ( \Delta \hat{E}  )^2 \rangle}$ from each of the
  different components discussed in Sections 2 and 3 as a function 
of bubble size $R_b$ are shown in Figures \ref{fig:int_fil0} 
and \ref{fig:alpha_eff}.
Both the figures show exactly the
same quantities, the only difference being that  they refer to Filter
I and Filter II respectively. 
A  detection is possible only in situations where $\langle \hat{E} 
  \rangle > 3 \times \sqrt{\langle  ( \Delta \hat{E}  )^2 \rangle}$, 
 the {\it rhs. } now refers to the total contribution to the
estimator variance 
from all the components. 

The signal is expected to scale as $R_b^3$ and the noise as  $R_b^{3/2}$ 
in a situation where the baseline distribution is uniform {\it ie.}
$\rho_N(\u,\nu)$ is independent of $U$.  This holds at $U <300$ for
the GMRT (Figure \ref{fig:5}), and the expected scaling is seen for
$R_b \ge 20 \, {\rm 
  Mpc}$. For smaller bubbles the signal extends to larger baselines
where $\rho_N(U,\nu)$ falls sharply, and the signal and the noise
both have a steeper $R_b$ dependence. The MWA baseline distribution is
flat for only a small $U$ range   (Figure \ref{fig:5}) beyond which it
drops. In this case the signal and noise are found to scale as $R_b^4$
and $R_b^2$ 
respectively. Note that the maximum baseline at  MWA is $U=750$, and 
hence a considerable amount of the signal is lost for  $R_b < 10 \, {\rm
  Mpc}$.

At  both the GMRT and the MWA, for $100 \, {\rm hrs}$ of observation,
the noise is larger than the signal for bubble  size $R_b \le 40 
\,{\rm   Mpc}$.  At the other extreme, for an  integration time of
$10^5 \, {\rm hrs}$   the noise is  below the signal for $R_b > 6
\,{\rm  Mpc}$ for the GMRT and $R_b > 8 \,{\rm   Mpc}$ for the MWA. 
The  foreground contribution  turns out to be smaller  than the
signal for the entire range of bubble sizes that we have considered, 
thus justifying our choice of filters. 
Note that 
Filter II is more efficient in foreground subtraction, but it requires
prior knowledge about the frequency dependence. 
For both the  filters the foreground  removal is more effective at the
GMRT 
than the MWA  because of the frequency dependence of
$\rho_N(\u,\nu)$. The  assumption  that this  is proportional to
$\lambda^2$ is valid only when $\rho(\u,\nu)$ is independent of $U$,
which, as we have discussed, is true  for a large $U$ range at the
GMRT. The $\lambda$ dependence is much more complicated at the MWA,
but we have not considered such details here as the foreground
contribution is 
anyway smaller than the signal. It should also be noted that the foreground
contribution increases at small baselines (eq. \ref{eq:fg}),  
and  is very sensitive to
the  smallest value of $U$ which we set at $U=20$ for our
calculations. Here it must be noted that our results  are valid only under the
assumption that the
foregrounds have a smooth frequency dependence. A slight deviation from this
and 
the signal will be swamped by the foregrounds. Also note that this filtering
method is effective only for the detection of the bubbles and not for
the statistical HI fluctuations signal. 

The contribution from the HI fluctuations impose a lower limit on the
size of the bubble which can be detected. However long be the
observing time,  it will not be
possible to detect bubbles of size  $R_b<8 \, {\rm Mpc}$ using  the
GMRT and  size $R_b<16 \, {\rm Mpc}$ using the MWA. The HI fluctuation
contribution increases at small baselines. The problem is particularly
severe at MWA because of  the  dense sampling of the small baselines
and the  very large field of view. We note that the MWA is being
designed with the detection  of the statistical HI
fluctuation  signal in mind, and hence it is not surprising  that this
contribution is quite large. 
For both  telescopes it 
 may be possible to reduce this  component by cutting off the filter
 at small baselines. We have not explored this possibility in this work
because  the enormous
 observing times required to detect such small bubbles makes
 it unfeasible with the GMRT or MWA.

\begin{figure}
\includegraphics[width=150mm]{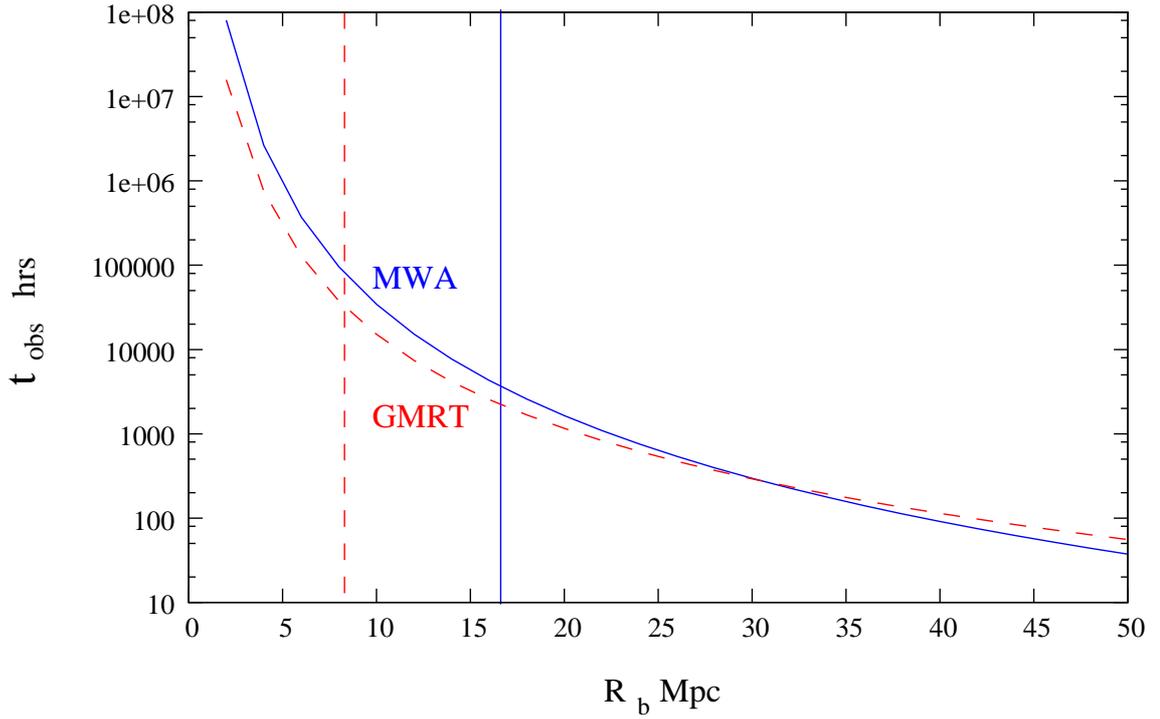}
\caption{The observing time $t_{obs}$  that would be required for a $3
  \, \sigma$ detection of  a bubble of radius $R_b$ provided it is at
  the center of the 
field of view. The vertical lines shows the lower limit (due to HI
fluctuations) where a detection will be possible ($R_b=8 \, {\rm Mpc}$
for GMRT and $R_b=16 \, {\rm Mpc}$ for MWA).}
\label{fig:R_Tobs}
\end{figure}

Figure \ref{fig:R_Tobs} shows the observation time that would be
required to detect bubbles of different sizes using Filter I for 
GMRT and the MWA. Note that the observing time shown here refers to a
$3 \, \sigma$ detection which is possibly adequate for targeted
searches centered on observed quasar position. A more stringent
detection criteria at the $5 \, \sigma$ level would be apropriate for
a  blind search. The observing time would go up by a factor of $3$ for
a $5 \, \sigma$ detection.  
The
observing time is similar for Filter II and hence we do not show this 
separately. 
In calculating the observing time we have only taken into
account the noise contribution as the other contributions do not change
with time. The  value of $R_b$ below which a detection  is not
possible due to the HI fluctuations is shown by vertical lines 
for both telescopes. 
We see that with $100 \,{\rm hrs}$ of observation both the telescopes
will be able to detect bubbles with $R_b>40\, {\rm   Mpc}$ while 
bubbles with  $R_b>22 {\rm  Mpc}$ can be detected with $1000 \,{\rm
  hrs}$ of observation. 

The possibility of detecting a bubble is less when the bubble
centre does not coincide with the centre of the field of view.
In fact, the SNR  falls as $e^{-\theta_c^2/\theta_0^2}$ if the
bubble center is shifted  away by $\theta_c$ from the center of the
field of view and the filter is  also shifted so that its center
coincides with 
that of the bubble.  There will be a corresponding increase 
$t_{obs} \propto e^{2 \theta_c^2/\theta_0^2}$ in the observing time
required to detect the bubble. It will be possible to detect bubbles
only if they are located near the center of the field of view
($\theta_c \ll \theta_0$), and the required observing time increases
rapidly with $\theta_c$ for off-centered bubbles.

\begin{figure}
\includegraphics[width=150mm]{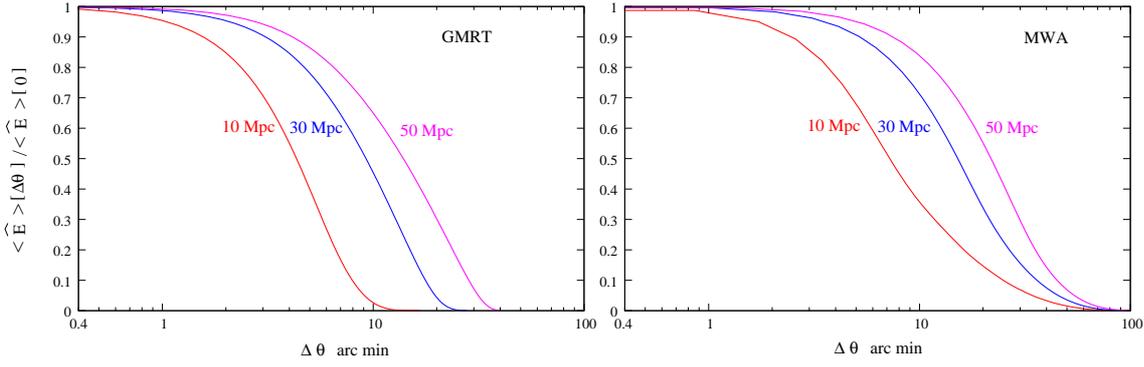}
\caption{The Overlap between the signal and the filter when there is a
  mismatch $\Delta \theta$ between the centers of the bubble and the
  filter for GMRT (left) and MWA (right). The results are shown for 
different bubble sizes.}
\label{fig:dtheta}
\end{figure}

When searching for bubbles in a particular observation it will be
necessary to consider filters corresponding to all possible  value of 
$R_b$, $\nu_c$ and $\th_c$. A possible strategy would be to
search at  a discrete set of values in the range of $R_b$, $\nu_c$ and
$\th_c$ values where a detection is feasible. 
The crucial issue here would be the choice of the sampling density
 so that we do not miss out an ionized bubble whose
parameters  do not exactly  coincide with any of the values in  
the discrete set  and  lie somewhere in between.  To illustrate this
we discuss the considerations for choosing and optimal value of 
$\Delta \theta_c$ the sampling interval for $\th_c$. We use 
$\langle \hat{E} \rangle [\Delta  \theta]$   to denote the expectation
value of the estimator when there is a  mismatch $\Delta \theta$
between the  centers   of the bubble and the filter. The ratio
${\rm   Overlap}=\langle \hat{E} \rangle[\Delta \theta]/\langle 
\hat{E}\rangle[0]$, shown in   Figure \ref{fig:dtheta} for 
GMRT (left panel) and MWA (right panel), 
 quantifies the overlap between the signal and the
filter   as $\Delta \theta$ is varied. We see that the choice of
$\Delta \theta$ would depend on the size of the bubble we are trying to
detect and  it would be smaller for the GMRT as compared to the
MWA. Permitting the Overlap to drop to $0.9$ at the middle of the sampling
interval, we find that it is $8^{'}$ at 
the GMRT and $20^{'}$ at  the MWA for  $R_b=50 \, {\rm Mpc}$.  

\begin{figure}
\includegraphics[width=150mm]{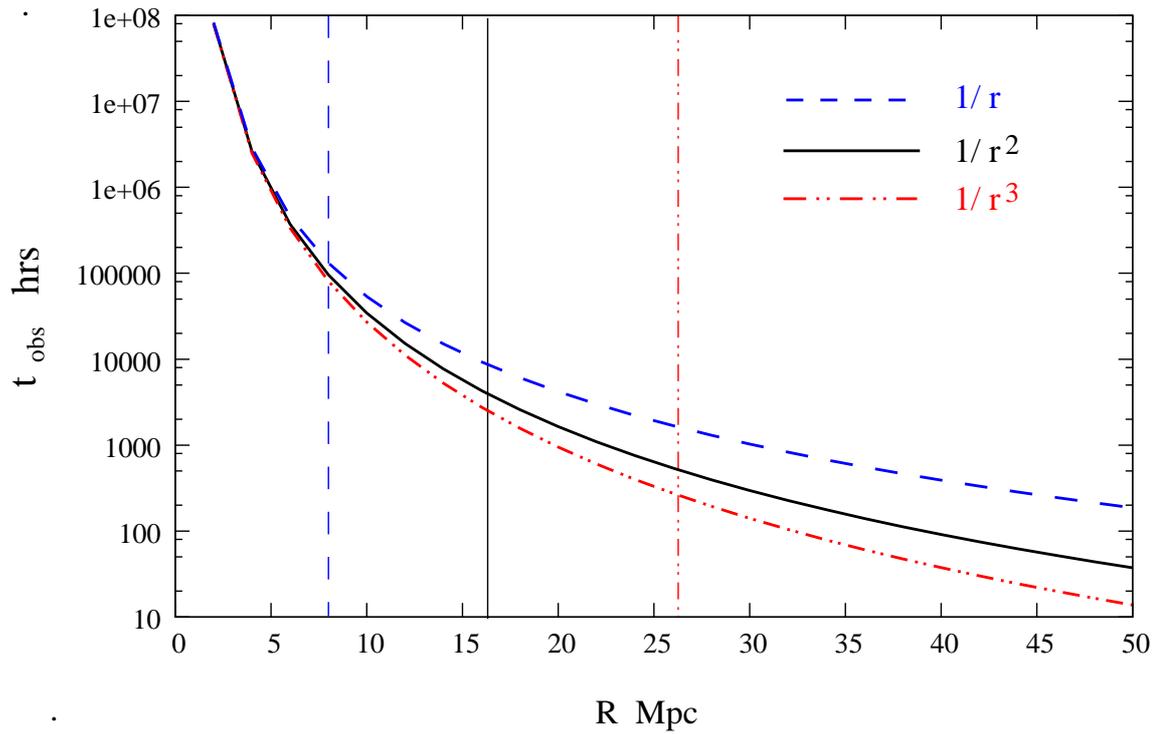}
\caption{Same as the Figure \ref{fig:R_Tobs} considering three
  different antenna distributions 
$\rho_{ant}(r)\propto 1/r$,\, $1/r^2$,\, $1/r^3$ \, for the MWA.}
\label{fig:R_Tobs_rho}
\end{figure}

The MWA is  yet to be  constructed, and it may be possible
that  an  antennae distribution different from
$\rho_{ant}(r)\propto 1/r^2,\,$ may improve the prospects of detecting
HII bubbles.  We have tried out $\rho_{ant}(r)\propto  1/r \, {\rm
  and} \, 1/r^3$ for which the results are shown in 
Figure~\ref{fig:R_Tobs_rho}.  We find that the required  
integration time falls considerably for the $1/r^3$ distribution
whereas the opposite occurs for $1/r$. 
For example, for $R_b=50 \,{\rm Mpc}$ the integration time increases
by  $5$ times for $1/r$ and decreases by $3$ times for  $1/r^3$ as
compared to $1/r^2$. Based on this we expect the integration time to
come down if the antenna distribution is made steeper, but this occurs
at the expense of increasing the HI fluctuations and the
foregrounds. We note that for the $1/r^3$ distribution the foreground
contribution is more than the signal, but it may be possible to
overcome this by modifying the filter. The increase in the HI
fluctuations is inevitable, and it restricts the smallest bubble that
can be detected to $R=26 \, {\rm MPc}$ for $1/r^3$. In summary, the
$1/r^2$ distribution appears to be a good compromise between reducing
the integration time and increasing the HI fluctuations and
foregrounds.

Finally we examine some of the assumptions made in this work.
First, the Fourier
relation between the specific intensity and the visibilities
(eq. \ref{eq:1}) will be valid only near the center of the field of
view and  full three dimensional wide-field imaging is needed away
from the center. As the feasibility  of detecting a bubble  away from the 
center falls rapidly, we do not expect the wide-field effects to be very
important. Further, these effects are most significant at large
baselines whereas  most of the signal from ionized bubbles is in the
small  baselines.

Inhomogeneities in the IGM will affect the propagation of ionization
fronts, and the ionized bubbles are not expected to be exactly
spherical \citep{wyithe05}. This will cause a mismatch between the
signal and the filter which in turn will degrade the SNR. In addition
to this, in future we plan to address a variety of other issues like
considering different observing frequencies and  making predictions
for the other upcoming  telescopes. 

Terrestrial signals  from television, FM radio,
satellites, mobile communication etc., collectively referred to as
RFI,  fall in the same frequency band as the redshifted $21 \rm cm$ signal
from the reionization epoch.  These are
expected to be much stronger than the expected $21 {\rm cm}$ signal, and
it is necessary to quantify and characterize the RFI. 
 Recently \citet{bowman07} have  characterized   the RFI for the 
 MWA  site on the frequency range $80$ to $300 \rm MHz$. They find an
 excellent  RFI environment except for a few channels which are
 dominated by satellite  communication signal.  The impact  of RFI on
 detecting ionized bubbles is an important issue which we plan to
 address in future. 

The effect of polarization leakage is another issue we postpone for
future work. This could cause polarization structures on the sky to
appear as frequency dependent ripples in the foregrounds intensity
. This could be particularly severe for the MWA.

\newpage

%\clearpage{\pagestyle{empty}\cleardoublepage} %%%%%%%%%%%%%%%%%%%%

 %\newpage
 \setcounter{section}{0}
 \setcounter{subsection}{0}
 \setcounter{subsubsection}{2}
 \setcounter{equation}{0}
 %\pagenumbering{arabic}

%-------------------------------------------
\chapter[Simulating Matched Filter Search for Ionized
    Bubbles]
{\bf \textbf {Simulating Matched Filter Search for Ionized Bubbles\footnote{{\bf \em { This Chapter is adapted from the paper  ``Simulating the impact of HI fluctuations on  matched filter search for ionized bubbles in redshifted 21 cm maps''
 by \cite{kkd3}}}.}}}

\vspace{3.5cm}
\section{Introduction}
 
In the previous Chapter we present an analytic framework for predicting 
the expected value and the standard deviation $\sigma$ 
of the matched filter estimator for the detection of a 
 spherical ionized  bubble of comoving radius $R_b$. We identify three
 different contributions  to $\sigma$, namely
 foregrounds, system  noise and the fluctuations in the HI outside  
the bubble that we are trying to detect.  Our analysis shows 
that the matched filter  effectively removes the foreground  
contribution so  that it falls  below the signal. Considering the 
system noise for the GMRT and  the MWA  we find that a $3 \, \sigma$ detection 
will be possible  for a bubble  of comoving
radius $R_b \ge 40 \, {\rm Mpc}$  in $100 \, {\rm hrs}$ of observation
and $R_b \ge 22 \, {\rm Mpc}$ in $1000 \, {\rm hrs}$ of observation
for both the instruments. The HI fluctuations, we find, impose a
fundamental restriction on bubble detection. Under the assumption that
the HI outside the ionized bubble traces the dark matter we find that 
 it is not possible to detect bubble of
size $R_b \le 8 \, {\rm Mpc}$  and $R_b \le 16 \, {\rm Mpc}$ at the
GMRT and MWA respectively. Note that the matched filter technique 
is valid for both,  a targeted search around  QSOs as well as  for a blind
search in a random direction.

Here we validate the visibility based matched filter
technique introduced in the previous Chapter through simulations of bubble
detection. Our simulations are   capable of 
handling interferometric arrays with widely different 
configurations like the GMRT and the MWA , the two instruments that we
consider here.  As mentioned earlier, the  fluctuations in the HI
outside the target bubble impose a fundamental  restriction for bubble
detection. The analytic approach of the Chapter 3 assumes that the HI
outside the bubble traces the dark matter.  In this Chapter we
carry out simulations that incorporate this assumption and use these
to assess the impact of HI fluctuations for bubble detection. We also
use the simulations to determine the accuracy to which the GMRT and
the MWA will be able to determine the size  and the position  of an 
ionized bubble, and test if this is limited due to the presence of HI
fluctuations. 
In a real situation  a typical FoV is expected to  
contain  several ionized patches besides the one that we are trying to
detect. We  use simulations to assess the impact of HI fluctuations for
bubble detection  in patchy reionization scenarios.

The outline of the Chapter is as follows. Section  4.2 presents a 
brief description of how we simulate  21-cm maps for three different
scenarios of the HI distribution, one where the HI traces the dark
matter and two with patchy reionization.  Subsections 4.2.1 and 4.2.2
respectively discuss how the simulated maps are converted into
visibilities and  how the matched filter analysis is simulated.
We present our results in Section 4.3. Subsections 4.3.1, 4.3.2 and 4.3.3
present results for bubble detectability, size determination and
position determination under the assumption that the HI outside the
bubble traces the dark matter. Section 4.3.4 presents results for bubble
detectability in patchy reionization scenarios. We discuss redshift dependence 
of bubble detection in Section 4.4 and present our summary
in Section 4.5.

For the GMRT we have used the telescope parameters from their 
website, while for the MWA  we use the telescope parameters from
\citet{bowman06}. The cosmological parameters for a flat ($k=0$)
$\Lambda$CDM model used
throughout this paper  are   
$\Omega_m=0.3, \Omega_b h^2 = 0.022, n_s = 1., h = 0.74, \sigma_8
= 1$.

\section{Method of Simulation}
We have simulated the detection of the HI signal of an ionized bubble
whose center is at
redshift $z_c=6$ which corresponds to $\nu_c=203 \, {\rm MHz}$. 
 The choice of $z$ value  is guided by the fact that we expect large 
ionized regions towards the end of reionization $z \gtrsim 6$
\citep{wyithe,furlanetto3}.  
Our aim here is to validate the analytic calculations of
the Chapter 3 and hence the exact value of $z$ is not very important.

We consider  four  scenarios of reionization for bubble detection. 
In the first three scenarios  there is  a spherical ionized bubble,
the one that we are trying to detect, at the center of the FoV. This 
bubble has comoving radius
$R_b$ and  is embedded in HI that traces the dark matter. 
In the first scenario  there is  a single bubble  in the field 
of view. We refer to this  as the SB scenario. In this scenario the
HI fraction $x_{\rm HI}$ is assumed to be 
uniform outside the bubble. The uncertainty due to the HI
fluctuations is expected to be lowest  
in this scenario because of the absence of patchiness. This is the
most optimistic scenario  for bubble detection.

In the next  two scenarios, we  attempt to quantify the effect of
patchy reionization (PR)  outside the bubble that we are trying to detect 
by introducing many other, possibly overlapping, bubbles in the FoV.  
Unfortunately, there is no obvious way to fix the sizes of these bubbles
from any theoretical models as they depend crucially on the nature of
reionization sources and other physical factors. 
In scenario PR1, we assume that the large HII
regions which we are trying to detect are surrounded by many
small ionized regions whose sizes are fixed by
the following procedure: we assume the globally averaged
neutral fraction $x_{\rm HI}$ to be $\sim 0.5$; the reason for this
choice is that the effects of patchiness
would be most prominent when typically half of the IGM is ionized. 
Given the value of $x_{\rm HI}$, we try to obtain a reasonable 
estimate of the size of the background bubbles from available
models. For example, semi-numeric simulations of patchy reionization  
\citep{mesinger07} predict that  the bubble size distribution  peaks around
$5 \,{\rm Mpc}$ when $x_{\rm HI} = 0.61$ (see their Fig 6).
We thus choose the spherical background bubbles to have radii $6
\,{\rm Mpc}$ and compute 
the number of 
background bubbles by demanding that the resulting neutral fraction
is 0.5. The bubble centres are chosen such that they trace the 
underlying dark matter distribution. At the end, the value of 
$x_{\rm HI}$ turns out to be slightly higher 0.62 because of overlap of the 
bubbles. Note that because of these overlaps, 
the shapes of the resulting ionized regions would not always be 
perfectly spherical.
In this scenario, we have essentially attempted
to capture a situation where there are many small, possibly
overlapping  ionized regions produced by galaxies and a few large
ionized regions (like the one that we are trying to detect) produced
by QSOs. 

Since the choice of the background bubble size is not robust 
by any means,
we consider a different scenario PR2  where these bubbles have the same 
comoving radius $R_b$ as the bubble that we are trying to detect. 
The centers of these extra
bubbles trace the dark matter distribution as in PR1.
The number of bubbles is fixed by the globally averaged 
$x_{\rm HI}$ which we take to be 0.62 same as in PR1.
The PR2 scenario represents a situation where we
predominantly  have  large  ionized regions produced 
either by rare luminous sources or
through the overlap  of several small ionized regions in the later
stages of reionization.

A  particle-mesh (PM) N-body code   was used to simulate the dark
matter distribution.  In the Chapter 3 we show that the
HI signal of the ionized bubble is largely concentrated at small
baselines or large angular scales,  thus a very high
spatial resolution is not required. 
We have used a  grid spacing of $2 \,{\rm Mpc}$  for the simulations. 
This is adequate for bubbles in the range 
 $4 \le R_b \le 50 \, {\rm Mpc}$
that we consider. The simulations use 
 $256^3$ particles on a $256^3$ mesh. For the 
GMRT  a single N-body simulation was cut into $8$ equal cubes  of size
$256 \, {\rm 
  Mpc}$ on each side. Considering that each cube may be viewed 
along three different directions, we have a total of $24$ different
realization of the dark matter distribution. Each cube  corresponds to
$18 \, {\rm MHz}$ in frequency and 
$\sim 2\degr$ in angle which is comparable to the GMRT FoV
 which has FWHM=$1.7\degr$ at $203 {\rm MHz}$.  The MWA FoV is
much larger (FWHM=$13\degr$). Here eight independent N-body 
simulations were used.  Viewing these along three different directions
gives twenty four different realizations of the dark matter
distribution. Limited computer memory restricts the simulation size 
and   the angular extent  $(\sim 4^{\degr})$ is considerably smaller
than the MWA FoV. We do not expect this to affect the signal but the
contribution from the HI fluctuations  outside the bubble is
possibly underestimated for the MWA.

\begin{figure}
\includegraphics[width=40mm, angle=270]{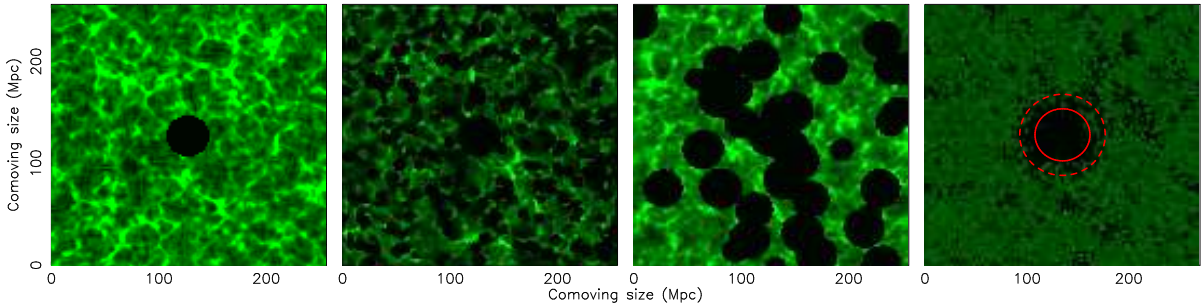}
\caption{This shows HI images on slices through the center of the
  bubble for the  four different
  scenarios SB, PR1, PR2 and SM (from left to right). In first three
  panels  the 
  central,  circular dark region of radius $R_b=20 \, {\rm Mpc}$ shows
  the 
  HII bubble  that we are trying to detect. The  HI outside this bubble 
  traces the dark  matter distribution. In the SB scenario(left)
  the hydrogen neutral 
  fraction is $x_{\rm HI}=1$  outside the bubble. In the 
  PR1  scenario (2nd from left) the extra    bubbles are all of a fixed
  comoving  radius $6 \, {\rm Mpc}$. In the  PR2  scenario (3rd from
  left) the   extra   bubbles have the same   comoving radius as the
  bubble that   we are trying to detect.   In both the   PR1 and
  PR2 reionization scenarios 
  the centers of the extra  bubbles trace the dark matter distribution
  and  $x_{\rm    HI}=0.62$. In the SM scenario (right) the central
  region up to radius $27\, {\rm Mpc} $ is fully ionized (marked with
  solid circle) and beyond that region up to radius $42\, {\rm Mpc} $
  the region is partially filled with HI patches (dashed circle). The
  mean neutral fraction is $x_{\rm HI}=0.5$. These simulations are
  all   for the   GMRT.} 
\label{fig:image}
\end{figure}

The dark matter density contrast $\delta$ was used to calculate the 
redshifted $21-{\rm cm}$ specific intensity $I_{\nu}=\bar{I}_{\nu}
x_{\rm HI} (1+\delta)$   for each grid point of our simulation.
Here $\bar{I}_{\nu}=2.5\times10^2\frac{Jy}{sr} \left (\frac{\Omega_b 
h^2}{0.02}\right )\left( \frac{0.7}{h} \right ) \left
(\frac{H_0}{H(z)} \right ) $ and $x_{\rm HI}$ the hydrogen neutral
fraction is $0$ inside the ionized bubbles and $1$ outside. 
%{\bf to be deleted We do
%not consider the effect of peculiar velocity  \citep{bharad04}
%because we do not 
%expect this to make a very big contribution for bubble detection.}
The simulated boxes are transformed to frequency and sky coordinate. 
Figure \ref{fig:image} shows  the HI image on a slice through the
center of the bubble of radius $R_b=20 \, {\rm Mpc}$.  The mean 
neutral fraction $\bar{x}_{\rm HI}$
is $\sim 1$ in the SB scenario, while it is $\sim
0.62$ for the two PR simulations shown here.

The three scenarios discussed above consider only spherical bubbles,
and the  only departures from sphericity arise from bubble
overlap. It is important to  assess how well  
 our bubble detection technique works   for non-spherical bubbles,   
which we do using  ionization maps produced by the semi-numeric (SM) 
approach.  In particular, we use maps obtained by the method of
\cite{choudhury08}. 
Essentially, these maps are produced by incorporating an excursion-set
based technique for identifying ionized regions given the density
distribution and the ionizing sources \citep{zahn07,mesinger07,geil07}.
In addition, the method of \cite{choudhury08} incorporate
inhomogeneous recombination and self-shielding of high-density regions
so that it is consistent 
with the ``photons-starved'' reionization scenario implied
by the Ly$\alpha$ forest data \citep{bolton07,choudhury08a}.
%To illustrate on how our matched-filtering technique works for such 
%maps, 
We use a simulation box of size 270 Mpc with $2000^3$ particles
which can resolve collapsed halos as small as $\approx 10^9 M_{\odot}$.
The ionization maps are generated at a much lower resolution
with a
grid size of $2.7$ Mpc. The box 
corresponds to $19\, {\bf MHz}$ in frequency and $\sim2\degr$ in angle 
comparable to the GMRT FoV. We have assigned luminosities to the collapsed
halos such that the mean neutral fraction $x_{\bf HI}=0.5$. The most massive
halo (mass $\sim 10^{13} M_{\odot}$) identified  in the box 
is made to coincide with the 
box centre and we assume that
it hosts a luminous QSO; its luminosity and age are chosen 
such that it would produce a 
spherical HII region of comoving size $\approx$ 27 Mpc in a completely homogeneous neutral medium [see, e.g., equation (8) of \cite{geil07}]. However, the actual ionized region is far from spherical both
because of the surrounding bubbles from other halos and also
because of inhomogeneous recombination.
We find visually from the maps (see the rightmost panel of Figure
\ref{fig:image}) that the HII region is fully ionized up to radius
$\approx 27$ Mpc. Beyond that the region is partially filled with
neutral patches. This patchy ionized region extends up to radius $\sim
42$ Mpc and then merges with the average IGM.  The fully ionized
region and the region with HI patches are marked with two circles. We
use this box for GMRT as three independent realizations viewing the
box along three different directions.  For the MWA we need a much
larger simulation box which requires substantially more 
computing power, beyond the resources available to us at present. 
Hence we do not consider the MWA for this scenario. 

\subsection{Simulating visibilities}
The quantity measured in radio-interferometric 
observations is the visibility $V(\u,\nu)$ which is related to the specific
intensity pattern on the sky $I_{\nu}(\th)$ as 
\be
V(\u,\nu)=\int d^2 \theta A(\th) I_{\nu}(\th)
e^{ 2\pi \imath \th \cdot \u}
\label{eq:vis}
\e
Here the baseline $\u={\vec d}/\lambda$ denotes 
the antenna separation ${\vec d}$  projected in the plane 
perpendicular to the line of sight  in units of the observing
wavelength $\lambda$, $\th$ is a two dimensional vector in the plane
of the sky with
origin at the center of the FoV, and $A(\th)$ is the 
beam  pattern of the individual antenna. For the GMRT this can be well 
approximated by Gaussian $A(\th)=e^{-{\theta}^2/{\theta_0}^2}$ where 
$\theta_0 \approx 0.6 ~\theta_{\rm FWHM}$ and we use the values
$1.7\degr$  for $\theta_0$  at $203\, {\rm MHz}$ corresponding to
the redshift $z=6$ for the GMRT. The MWA beam pattern is expected
to be quite complicated, and depends on 
the pointing angle relative to the zenith \citep{bowman07}. Our analysis
largely deals with the beam pattern within $2^{\circ}$ of the pointing
angle where it is reasonable to approximate the beam as being
circularly symmetric (Figures  3 and 5 of \citealt{bowman07} ). We  
approximate the MWA antenna beam pattern as a Gaussian. 

We consider  $128$ frequency channels across  $18 {\rm MHz}$
bandwidth. The image $I_{\nu}(\theta)$ at each channel is multiplied
with  the   telescope beam pattern $A(\vec {\theta},\nu)$.  The
discrete Fourier transform (DFT) of the product $I_{\nu}(\theta) \,
A(\vec {\theta},\nu)$  gives the 
complex visibilities $\hat V(\vec{U},\nu)$. The GMRT simulations 
have baselines in the range $30.5 \le U \le 3900$
which is adequate to capture the HI  signal from  ionized bubbles
which  is expected to be confined to  small baselines $U<1000$. 

\begin{figure}
\includegraphics[width=85mm, angle=270]{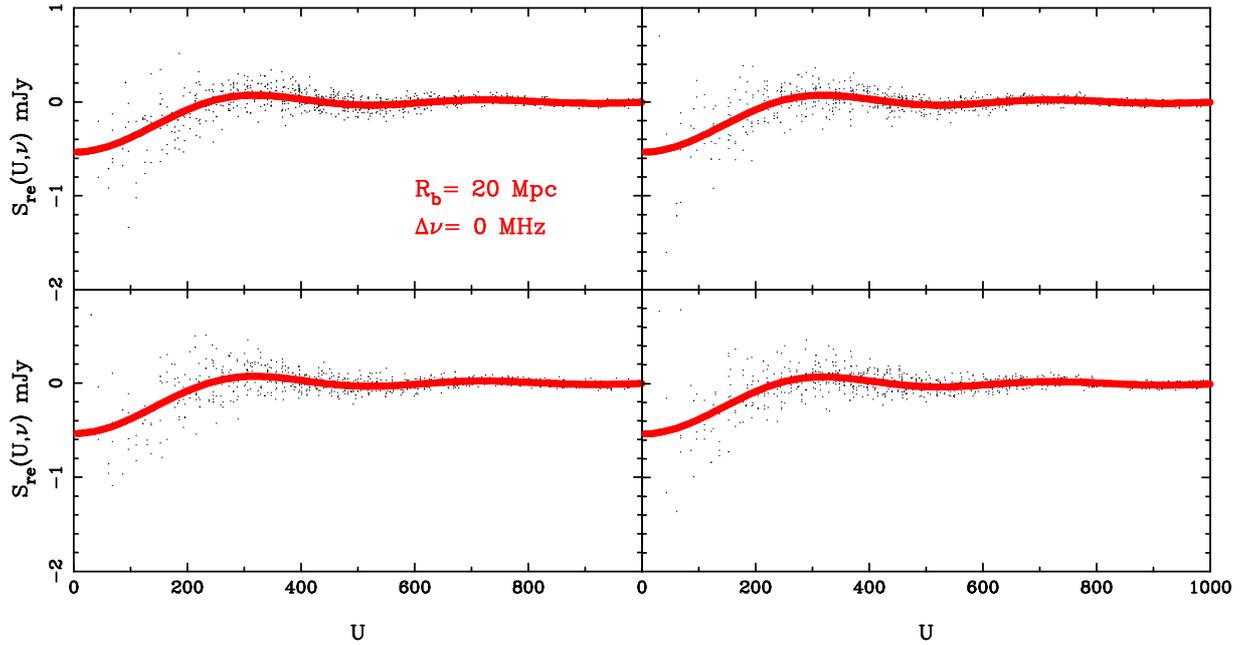}
\caption{This shows the visibility signal (real part) from a frequency
  slice   through the   center of a spherical ionized bubble 
  of comoving radius $20 \, {\rm Mpc}$ embedded in HI. 
  The solid  curves show the expected signal assuming that the bubble is
  embedded in uniformly   distributed HI. The data points show
  the visibilities for a few randomly chosen baselines from our
  simulation of the SB scenario.  The difference between the data
  points and  the solid curve
  is due to the fluctuations in the HI outside the bubble.  Each panel
  corresponds to a different   realization of the simulation. }  
\label{fig:vis_U}
\end{figure}
\begin{figure}
\includegraphics[width=85mm, angle=270]{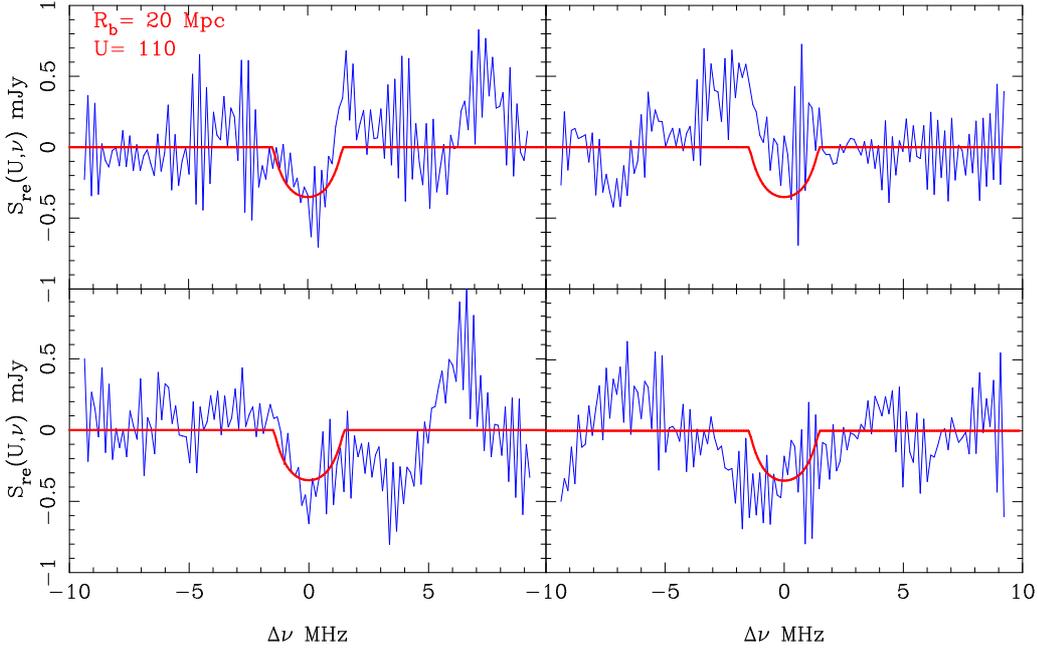}
\caption{Same as the previous figure except that $U$ is fixed at $110$
  while  the frequency varies , and  $\Delta \nu=\nu-\nu_c$.}
\label{fig:vis_nu}
\end{figure}

The visibility  recorded in radio-interferometric
observations is actually a combination of  several  contributions 
\be
V(\vec{U},\nu)=S(\vec{U},\nu)+HF(\vec{U},\nu)+N(\vec{U},\nu)+F(\vec{U},\nu)
\,. 
\label{eq:2}
\e
where $S(\u , \nu)$ is the HI signal that we are interested in,
$HF(\u,\nu)$ is contribution from the fluctuating HI outside the
bubble that we
are trying to detect, $N (\u,\nu)$ is the system noise which
is inherent to the measurement and $F (\u,\nu)$ is the contribution
from other astrophysical 
sources referred to as the foregrounds.
The signal  $S(\u,\nu)$ from an ionized bubble of comoving radius
$R_b$ embedded in an uniform HI distribution  can be analytically
calculated (Chapter 3). The solid curve in Figures \ref{fig:vis_U} and
\ref{fig:vis_nu}  show the expected signal for $R_b=20 \, {\rm
  Mpc}$.  
The $U$ extent,  frequency extent and peak value of the signal scale
as  $ R_b^{-1}$, $R_b$ and $R_b^2$ respectively for other values of 
$R_b$.    Note that   $S(\u,\nu)$ is real when the bubble is at the 
center of the FoV.

The data points shown  in Figures \ref{fig:vis_U} and \ref{fig:vis_nu}  
are the real part of a few randomly chosen visibilities determined
from the simulation of a  $R_b=20 \, {\rm Mpc}$ bubble in the
SB scenario. The deviations from the analytic predictions are due to
the HI fluctuations  $HF(\u,\nu)$  {\it ie.} in the SB scenario the HI
outside the bubble traces the dark matter fluctuations. Notice that
these fluctuations  are often so prominent that the signal cannot be
made out.   We expect even larger fluctuations  in the other three scenarios which incorporate patchiness of reionization.

The system noise contribution $N(\u,\nu)$ in each baseline and
frequency channel is expected to be an independent Gaussian random
variable with zero mean ($\langle \hat{N} \rangle =0$) and  
variance $\sqrt{\langle \hat{N}^2  \rangle}$is independent of   $\u$
and $\nu_c$. We use (Chapter 3)  
\be
\sqrt{\langle \hat{N}^2  \rangle}
=C^x \left (\frac{\Delta \nu_c}{1
   {\rm MHz}} \right )^{-1/2}\left ( \frac{\Delta
    t}{1 \rm{sec}}\right )^{-1/2}
\label{eq:noise}
\e
where $C^x$ has values $0.53{\rm Jy}$
and $54.21{\rm Jy}$ for the GMRT and the MWA respectively (Chapter 3).

The contribution from astrophysical foregrounds $F(\vec{U},\nu)$ is
expected  to be several order of magnitude stronger than the HI
signal.   The foregrounds are predicted to have a featureless,
continuum spectra 
whereas the signal is expected to have a dip 
 at $\nu_c$ (Figure \ref{fig:vis_nu}). This difference holds the
 promise of allowing us to separate the signal from the foregrounds.

\subsection{Simulating signal detection}

The signal component $S(\u,\nu)$  in the observed visibilities 
$V(\u,\nu)$ is expected to be  buried deep in other 
contributions many of which are orders of magnitude larger. Detecting
this is a big challenge. For optimal signal detection 
we consider the  estimator (Chapter 3) 
\be
\hat{E}= \sum_{a,b} S_{f}^{\ast}(\u_a,\nu_b)
\hat{V}(\u_a,\nu_b) 
\label{eq:estim0}
\e
where 
$S_f(\u,\nu)$ is a filter which has been constructed to detect
a particular ionized bubble, $\hat{V}(\u_a,\nu_b)$ refer to the
observed visibilities and  $\u_a$ and $\nu_b$ refer to the
different baselines and frequency channels in the observation. 
The filter $S_f(\u,\nu)$  depends on $[R_f,z_c,\th_c]$
the  comoving  radius, redshift and angular position of the bubble
that we are trying to detect.  We do not show this  
explicitly,  the values of these parameters will be clear
from the context.

The  baselines  obtained using DFT in our simulations are uniformly
distributed on a plane. In real observations, the baselines will have
a complicated distribution depending on the antenna layout and
direction of observation. We incorporate this through the normalized
baseline distribution function $\rho_N(U,\nu)$  which  is defined such that 
$d^2 Ud\nu \, \rho_N(\u,\nu)$ is the fraction  of data points {\it
  ie.} baselines in the
interval  $d^2 U \, d \nu$ and $\int d^2 U \,\int d\nu \,
\rho_N(\u,\nu)=1$.  
We use the functional forms of   $\rho_N$ determined in the Chapter 3  for
the GMRT and the MWA. 

Using  the  simulated visibilities, we evaluate the estimator 
as
\be
\hat{E}= \, (\Delta U)^2 \, \Delta \nu \, 
 \sum_{a,b} S_{f}^{\ast}(\u_a,\nu_b)
\hat{V}(\u_a,\nu_b) \rho_N(\u_a,\nu_b) 
\label{eq:estim1_4}
\e
where the sum is now over the baselines and frequency channels in the
simulation.

The filter $S_f(\u, \,\nu)$ (Filter I of  Chapter 3)  is defined as 
\bear
S_f(\u,
\,\nu)\!\!\!\!\!&=&\!\!\!\!\!\left(\frac{\nu}{\nu_c}\right)^2\left[
  S(\u, \,\nu)\right. 
\nline
\!\!\!\!\!&-&\!\!\!\!\! \left. \frac{\Theta(1-2 \mid \nu -\nu_c
  \mid/B^{'}) }{ B^{'}} 
\int_{\nu_c-B^{'}/2}^{\nu_c + B^{'}/2}S(\u,\nu') \, \de \nu' \right
]. \nonumber\\ 
\label{eq:rms_fg}
\ear
where the first term $ S(\u, \,\nu)$ is the expected signal of the
bubble that we are trying to detect. We note that this term is the
matched filter that gives the maximum signal to noise ratio (SNR).  The
second term involving the Heaviside function $\Theta(x)$ 
subtracts out any frequency independent  component from  the frequency
range $\nu_c-B^{'}/2$ to $\nu_c+B^{'}/2$. The latter term is
introduced to subtract out the foreground contributions. The 
$(\nu/\nu_c)^2$ term accounts for the fact that
$\rho_N(U,\nu)$ changes with frequency (equivalently wavelength).

We have used the $24$  independent realizations
of the simulation   for the first three scenarios to  determine the
mean  $\langle\hat{E}\rangle$ and 
the  variance $\langle(\Delta\hat{E})^2\rangle$ of the 
estimator. The high computational requirement restricts us to use 
just  $3$ realizations  for the SM scenario.
 Only the
  signal is correlated with the filter, and only this  is 
expected to contribute to the  mean  $\langle\hat{E}\rangle$. All the
other components are uncorrelated with the filter and they are
expected to contribute only to the variance
$\langle(\Delta\hat{E})^2\rangle$. 
 The variance is a sum of three contributions (Chapter 3)
\begin{equation}
 \langle(\Delta\hat{E})^2\rangle= \langle(\Delta\hat{E})^2\rangle_{HF} +
 \langle(\Delta\hat{E})^2\rangle_{N} +
 \langle(\Delta\hat{E})^2\rangle_{FG} \,.
\end{equation}
 The simulations give an estimate of
 $\langle(\Delta\hat{E})^2\rangle_{HF}$  the contribution from HI
 fluctuations.  We do not include system noise explicitly 
in our  simulations. The noise contribution from a single visibility 
 (eq. \ref{eq:noise})   is used  to estimate 
 $ \langle(\Delta\hat{E})^2\rangle_{N}$ (eq. 3.19 of Chapter 3).  Under the assumed foreground
 model, the foreground contribution $
 \langle(\Delta\hat{E})^2\rangle_{FG} $ is predicted to be smaller
 than the signal and we do not consider it here.

\section{Results}

We first  consider the detection  of an ionized bubbles and the
estimation  of its parameters in the SB scenario where there is only 
a single bubble in the FoV. We consider the most optimistic situation
where the bubble is located in the center. 
 In reality this can  only be achieved in targeted observations of 
 ionized bubbles around luminous QSOs. In a blind search, the bubble
 in general will be located at some arbitrary position in the FoV, and
 not the center.  It has already been mentioned that the  
foregrounds can be removed by a suitable choice of the filter. 
Further, the system noise can, in principle,   be  reduced by
increasing the observation time. The HI fluctuations outside the
bubble impose a fundamental restriction on bubble detection. 

\subsection{Restriction  on bubble detection}
We have carried out simulations for different values of the  bubble
radius $R_b$ chosen uniformly at an interval of $2 \, {\rm Mpc}$
in the range $4$ to $50 \, {\rm Mpc}$.  In each case 
we consider only the most  optimistic situation where 
the bubble radius $R_f$ used in the filter is precisely
matched to   $R_b$. In reality  it is necessary to try filters of
different radius  $R_f$ to determine which gives the best match.  

\begin{figure}
\includegraphics[width=57mm, angle=270]{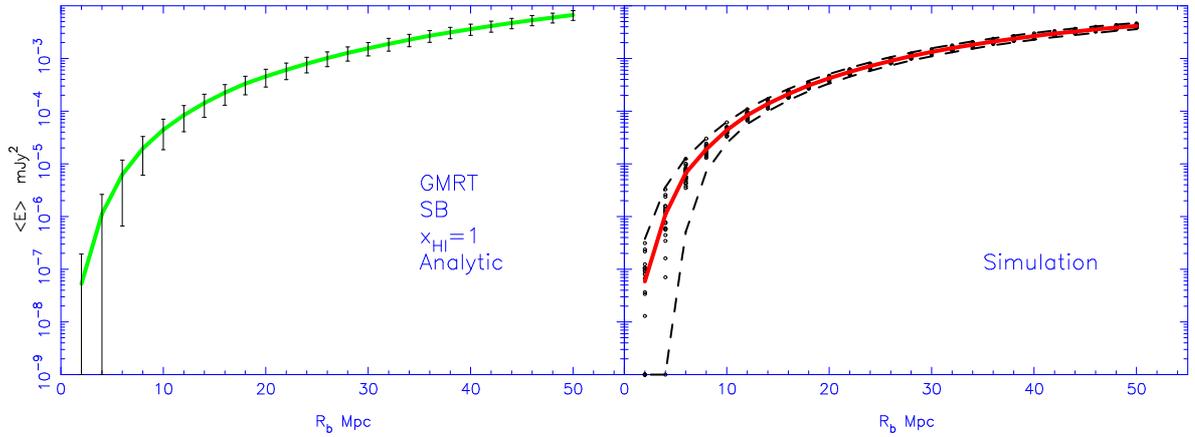}
\caption{The estimator $\hat{E}$ (defined in equation
  \ref{eq:estim0}) for bubble size
  $R_b$ ranging from $4\,{\rm Mpc}$ to $50\,{\rm Mpc}$ for the GMRT in
  the SB  scenario.  It is assumed that the filter is exactly matched
  to the  bubble. The left panel shows the analytic predictions for
  the  mean  estimator 
  $\langle\hat{E}\rangle$  and the $3-\sigma$ error-bars
  due to the HI fluctuations. The solid and  the dashed lines 
in the right panel respectively  show the   $\langle\hat{E}\rangle$
  and the   $3-\sigma$  envelope determined from  the simulations.  
The data points in the right panel show  $\hat{E}$ in the individual
  realizations.  
}
\label{fig:R_Es}
\end{figure}

\begin{figure}
\includegraphics[width=57mm, angle=270]{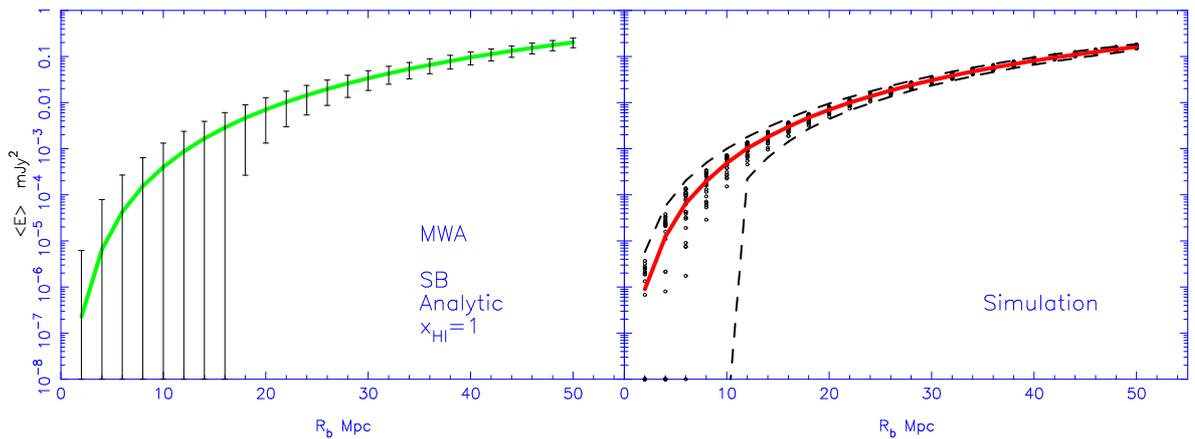}
\caption{Same as the Figure \ref{fig:R_Es} for the MWA.}
\label{fig:R_EsM}
\end{figure}

Figures \ref{fig:R_Es} and  \ref{fig:R_EsM} 
shows the results  for the GMRT and the MWA
respectively.   We  compare the analytic predictions of the Chapter 3 (left 
panel) with  the prediction of our simulations (right
panel). The  analytical predictions for  the mean value 
$\langle\hat{E}\rangle$ arising from the signal 
 and 
$  \sqrt{\langle(\Delta\hat{E})^2\rangle}_{HF}$  due to the HI 
fluctuations  are respectively calculated using equations (3.15) and
(3.22) of the Chapter 3. The  signal 
depends on the bubble radius $R_b$ and the mean neutral fraction
which is taken to be $x_{\rm HI}=1$.  The uncertainty due to the
HI fluctuations is calculated  using the  dark matter power spectrum
under the assumption that the HI traces the dark matter. 

We find that $\langle\hat{E}\rangle$ and
$\sqrt{\langle(\Delta\hat{E})^2\rangle}_{HF}$ determined from the
simulations is in rough agreement with the analytic predictions.  
The mean $\langle\hat{E}\rangle$ is in very good agreement 
for $R_b>6 \, {\rm Mpc}$, there is a slight  discrepancy for smaller
bubbles arising from  the finite grid size $(2 \, {\rm
  Mpc}$ in  the simulation). 
The HI fluctuations $\sqrt{\langle(\Delta\hat{E})^2\rangle}_{HF}$ are  
somewhat  underestimated by the simulations. This is  more
pronounced for the MWA  where the limited box size of our  simulations
results in a   FoV  which is considerably smaller than the actual
antennas.  We note that the  $24$ different
values of $\hat E$ determined from the different realizations of the
simulation  all lie within $\langle\hat{E}\rangle \pm 
3 \sqrt{\langle(\Delta\hat{E})^2\rangle}_{HF}$  determined from the
analytic predictions. 

The good agreement between the simulation
results and the analytical predictions  is particularly important
because each  is based on several approximations,
many of which  differ between  the two methods.  Our results
show that the effect of these approximations, though present, 
are well under control. The  analytical method has the advantage 
that it is very   easy to  calculate and  can be evaluated 
very quickly at an extremely low computational cost. 
 Unfortunately, its utility is mainly  limited to the
SB scenario and it cannot be easily applied to an arbitrary PR scenario
with a complicated  HI  distribution. Simulations, though
computationally more cumbersome and expensive, are useful  in such a
situation. It is thus important to test that the two methods 
agree for the SB scenario where both of them  can be
applied. Note that the HI fluctuation predicted by the  
SB scenario sets the lower limit for the HI fluctuation in any of the
PR models. It is expected that patchiness will increase the HI
fluctuations above the SB predictions.

It is meaningful to attempt bubble detection at, say $3\sigma$
confidence level, only if  $\langle \hat{E} \rangle \ge 3
\sqrt{\langle (\Delta \hat{E})^2   \rangle_{HF}}$. The HI fluctuations
overwhelm the signal in a  situation where this condition is not 
satisfied, and bubble detection is not possible. 
In a  situation where this condition is satisfied, an
observed value $E_o$ of the estimator can be interpreted as a
$3\sigma$ detection if $E_o > 3
\sqrt{\langle (\Delta   \hat{E})^2   \rangle}$.
The simulations show that a $3-\sigma$ detection is not possible 
for $R_b \leq 6 \,{\rm Mpc}$ and $R_b \leq 12 \,{\rm Mpc}$  
at  the GMRT and  MWA respectively. As noted earlier, the HI
fluctuations are somewhat under predicted in the simulations 
and the analytic predictions  $R_b \leq 8 \,{\rm Mpc}$  and $R_b \leq
16 \,{\rm Mpc}$ respectively, are somewhat larger. 

The limitation on the bubble size $R_b$ that can be detected 
is larger  for the MWA as compared to  the  GMRT. This is because of  
two reasons, the first being the fact that the MWA  has a very 
dense sampling of the small baselines where the HI fluctuation
are very large, and the second being the large FoV. 
In fact, the baseline distribution of the experiment has a significant
role  in determining the quantum of HI fluctuations and thereby
determining the lower cut-off for  bubble detection.
Looking for an optimum baseline distribution for bubble detection is
also an issue which we plan to address in future. In a situation where
the antenna layout is already in place, it may possible to tune the
filter to reduce the HI fluctuations. 

We have not considered the effect of peculiar velocities
\citep{bharad04} in our simulations.   From equation (3.22) of the Chapter 3 
we see that the HI fluctuations scale as 
$\sqrt{\langle (\Delta \hat{E})^2   \rangle_{HF}} \propto
\sqrt{C_{l}}$, where $C_{l}$ is the  HI multi-frequency angular power   
spectrum (MAPS). The $C_l$s increase  by a  factor $\sim 2$ due to 
peculiar velocities, whereby $\sqrt{\langle (\Delta \hat{E})^2
  \rangle_{HF}}$ goes up by  a factor $\sim 1.5$. This  increase does
not significantly change our results, and is small compared to the
other uncertainties in the PR models.

The  signal  $\langle \hat{E} \rangle$ and the HI
fluctuations  $\sqrt{\langle (\Delta \hat{E})^2   \rangle_{HF}}$  both
scale as $\propto \bar{x}_{\rm HI}$, and the lower limit for bubble
detection  is unchanged for smaller neutral fractions.

\subsection{Size determination}

\begin{figure}
\includegraphics[width=80mm, angle=270]{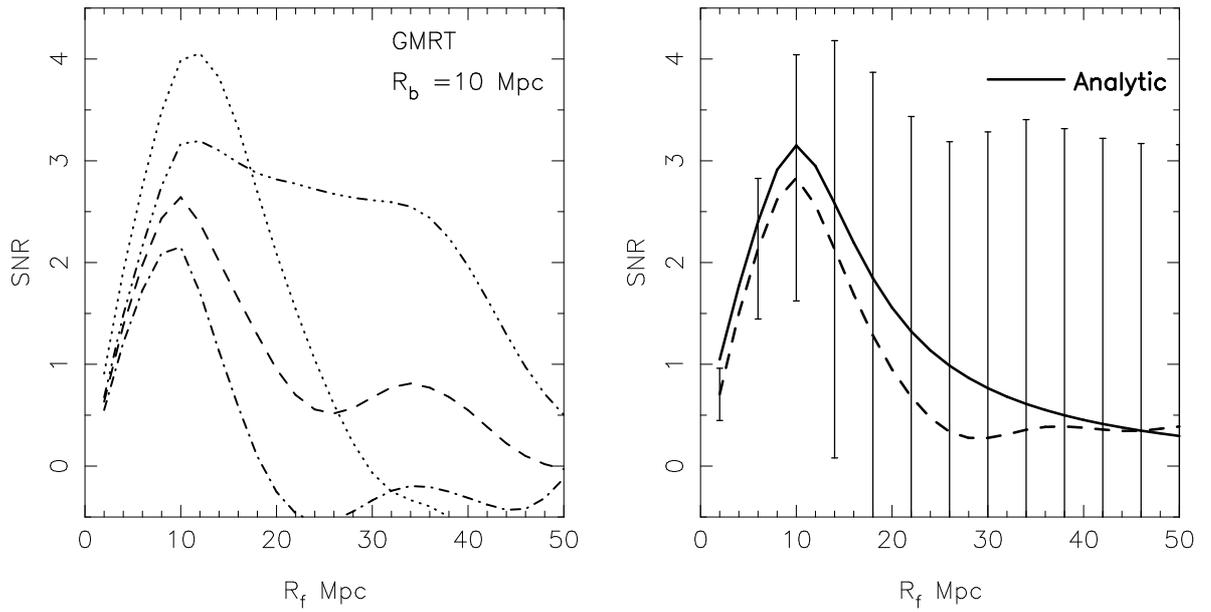}
\caption{The SNR $=\langle\hat{E}\rangle/\sqrt{\langle (\Delta \hat
    E)^2 \rangle_{\rm NS}}$ for  $1000\,{\rm
  hrs}$ observation with the GMRT 
as a function of the filter size $R_f$ for the
  case where the actual  bubble size is $R_b=10 \, {\rm Mpc}$ . The left panel shows $4$ different realizations 
    of the simulation.  The right panel shows the mean SNR and
    $3-\sigma$ error-bars
    calculated using 24 realizations. The solid line
    shows the analytical predictions. }
\label{fig:find_R10}
\end{figure}

In this subsection we estimate the accuracy to which it will be
possible to determine the bubble radius $R_b$. This, in general, is 
an unknown quantity that has to be determined from the
observation by trying  out filters with different values of 
 $R_f$. In the matched filter technique we  expect the predicted
SNR (only system noise) ratio   
\begin{equation}
{\rm SNR}=\frac{\langle\hat{E}\rangle}{\sqrt{\langle (\Delta \hat
    E)^2 \rangle_{\rm NS}}}
\label{eq:snr}
\end{equation}
to peak when the filter is exactly matched to the signal 
{\it ie.} $R_f=R_b$. The  solid line in the right panel of  Figure
\ref{fig:find_R10} shows this  for  $R_b=10\,{\rm
  Mpc}$. We find that the SNR peaks exactly when the  filter size 
$R_f=10\,{\rm   Mpc}$. We propose that this can be used to
observationally determine  $R_b$. 
For varying $R_f$, 
we consider  the ratio of the  observed value $E_o$ to the expected
system noise $\sqrt{\langle   (\Delta \hat     E)^2 \rangle_{\rm
    NS}}$, referring to this as  the  SNR. The $R_f$ value
where this SNR peaks gives  an estimate  of the actual bubble size $R_b$.  
The observed SNR will differ from the predictions of
eq. (\ref{eq:snr}) due to the  HI fluctuations outside the bubble.
These variations will differ from realization to realization and this 
can  introduce uncertainties in   size estimation.  We have used 
the simulations to estimate this.

The left panel of Figure \ref{fig:find_R10} shows the SNR as a
function of $R_f$ for $4$
different realizations of the simulation for the GMRT with bubble size 
$R_b=10 \, {\rm Mpc}$.  We see that for $R_f \leq R_b$ 
the SNR shows a very similar behavior in all the  realizations, and
it  always peaks  around $10 \,{\rm Mpc}$ as expected. For $R_f  > R_b$   
the  behavior of the SNR as a function of $R_f$ shows 
considerable variation  across the realizations. In some cases the
drop in SNR away from the peak is quite rapid whereas in others it is
very gradual (for example, the dashed-dot-dot curve). In many cases
there is an   spurious extra peak in the SNR at  an $R_f$ value that is
much larger than $R_b$. These spurious peaks do not pose a problem for
size determination as they are well separated from $R_b$ and can be
easily distinguished from the genuine peak. 

The error-bars in the right panel of   Figure \ref{fig:find_R10} show
the $3-\sigma$ fluctuation in the simulated  SNR determined from $24$
realizations of the simulation. Note that the fluctuations  at
different $R_f$  are correlated. Although the overall 
amplitude changes from one realization to another, 
the shape of the curve in the vicinity of $R_f=R_b$ is nearly
invariant across  all the realizations. In all of the $24$
realizations we can identify a well defined peak at the expected value 
$R_f=R_b$.  

\begin{figure}
\includegraphics[width=80mm, angle=270]{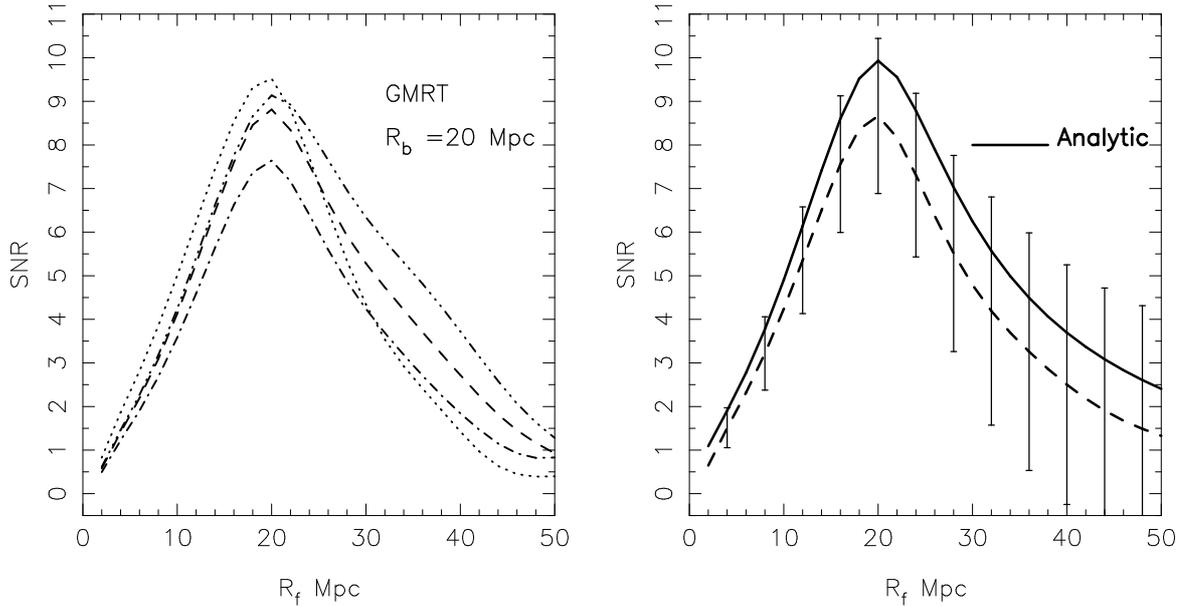}
\caption{Same as the Figure \ref{fig:find_R10} for
$R_b=20 \, {\rm Mpc}$  for the GMRT.}
\label{fig:find_R20}
\end{figure}

\begin{figure}[h]
\includegraphics[width=80mm, angle=270]{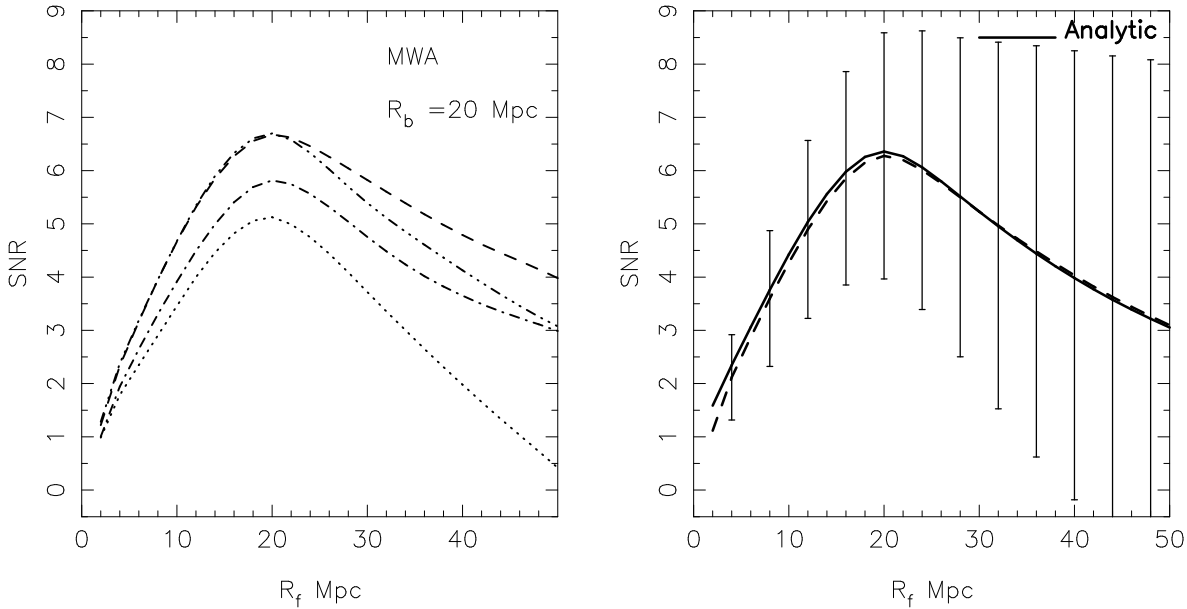}
\caption{Same as the Figure \ref{fig:find_R10}  for
$R_b=20 \, {\rm Mpc}$  for the MWA.}
\label{fig:find_R20M}
\end{figure}

Figures  \ref{fig:find_R20} and \ref{fig:find_R20M} show
the results for a similar analysis with $R_b=20 \, {\rm Mpc}$ for the
GMRT and the MWA respectively. It is not possible to detect a bubble
 of size $R_b=10 \, {\rm Mpc}$ with the MWA,  and hence we do not show
this. Here again,we find that  for all the realizations of the
simulations  the SNR peaks at $R_f=R_b$. The relative variations in
the SNR across the realizations is much less for $R_b=20 \, {\rm Mpc}$
as compared to $10 \, {\rm Mpc}$ and there are no spurious peaks. 
 Also, for the same bubble size  the
variations are smaller for the GMRT as compared to the MWA.  We do not
find any spurious peaks for $R_b=20 \, {\rm Mpc}$.

A point to note is that the mean SNR determined from the simulations
is somewhat smaller than the analytic predictions, both being shown in
the right panels of Figures \ref{fig:find_R10},  \ref{fig:find_R20} and
Figure \ref{fig:find_R20M}.  
 There are a couple of reasons that could account for this namely,
(i) the bubble in the simulation is  not exactly a sphere 
because of the finite grid size and  thus the match between the
filter and the signal is not perfect even when the sizes are same and
(ii)  the finite box-size imposes a minimum baseline beyond which the
signal is not represented in the simulation.

Based on our  results  we conclude that in the SB scenario for the
GMRT the accuracy to which the  bubble size can be determined  in our 
simulations is decided  by the resolution $2 \, {\rm Mpc}$ and not 
by the HI   fluctuations. In reality the limitation will come from the
angular resolution of the instrument which sets the limit at
$0.5 \, {\rm Mpc}$ for the GMRT and $8 \, {\rm Mpc}$ for the MWA.

The height of the SNR peak  depends on the neutral fraction and it 
can be used to observationally determine this. We find that the HI
fluctuations do not change the position of the peak but introduce
considerable variations in its height even if $x_{\rm HI}=1$. 
The HI fluctuations  restrict the accuracy to which the 
neutral fraction can be estimated, an issue that we propose to address
in future work.

\subsection{Determining the position}

In the previous two subsections, we have considered cases where the
bubble's position  is known. Here we assume that the bubble's size is
known and we estimate the accuracy to which its position can be
determined in the presence of HI fluctuations. The situation
considered here is  a blind search whereas the former is a
targeted search centered on a QSO. 

In a real situations it would be necessary to jointly determine four 
parameters the bubble radius $R_b$, two angular coordinates
($\theta_x, \theta_y$) and the central frequency $\nu_c$ from the
observation. However, to keep the computational requirement under
control,  in this analysis we assume that $R_b$  is known. The
bubble is placed  at the center of the FoV   and frequency band, and
we  estimate how well  the position can be recovered  from the
simulation. To determine the bubble's position we move  the center of
the filter to different positions and search for a peak in the SNR. 
To reduce the computational requirement,   this is done along one
direction at a time, 
keeping the other two directions fixed at the bubble's actual center.   
We have also carried out simulations where the bubble is located 
off-center. We do not explicitly show these results because they are
exactly the same as when the bubble is at the center except for the
fact  that the value of the peak SNR is lower because of the
primary beam pattern.   

\begin{figure}
\includegraphics[width=140mm, angle=270]{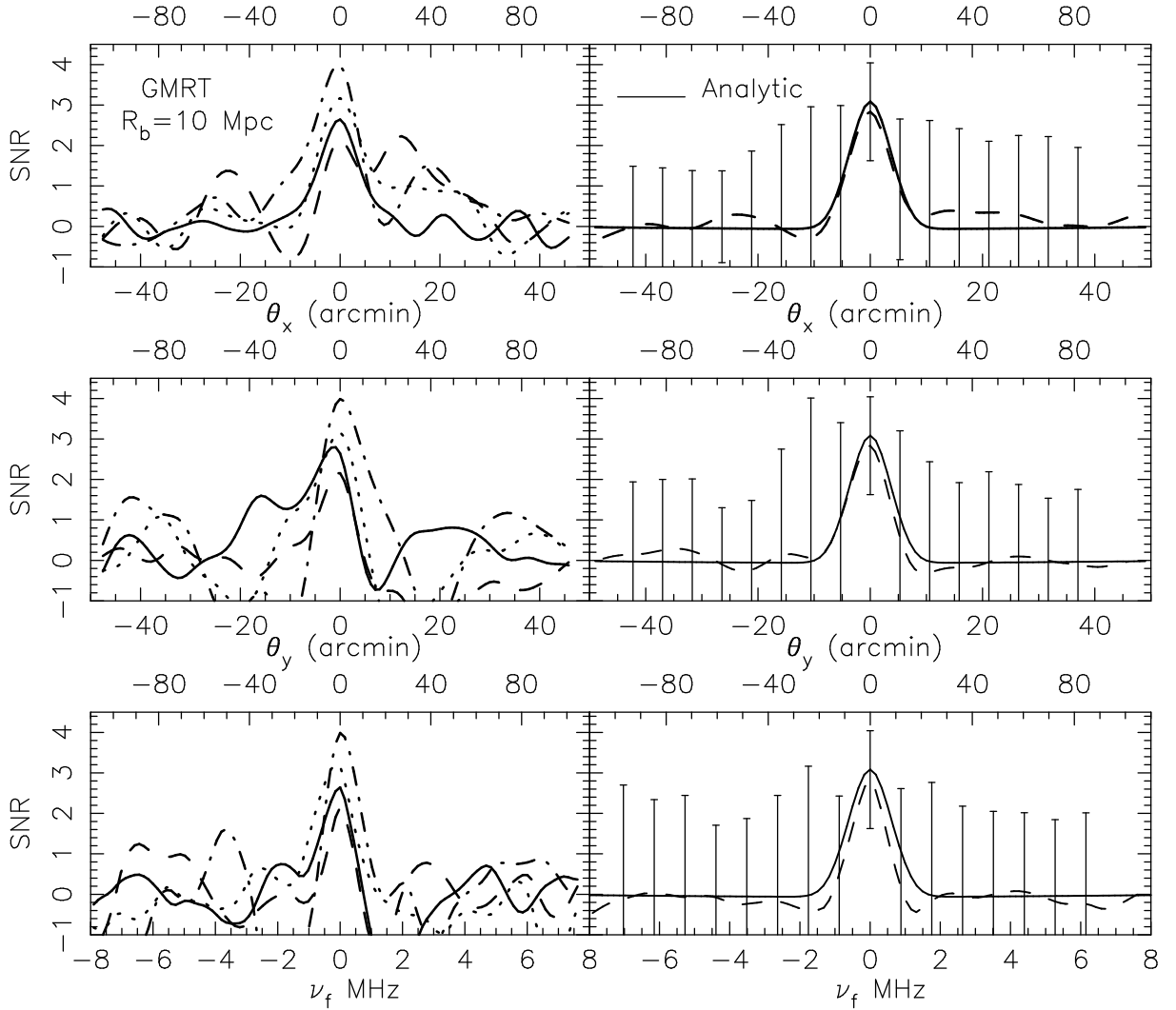}
\caption{The  SNR for $1000 \, {\rm hrs}$ GMRT observations 
for a   bubble of   size $R_b=10 \, {\rm Mpc}$
  located at the  center of the filed of view. The 
  filter  scans along  
  $\theta_x,\theta_Y,\nu_f$ (top, middle, bottom) to determine the
  bubble's position. 
 The left panel
  shows results for $4$  realizations of the SB simulation, the right
  panels show the mean (dashed curve) and $3-\sigma$   error-bars
  determined from $24$   realizations of the simulation and the
  analytic prediction  for the mean (solid curve).}
\label{fig:find_theta_nu10}
\end{figure}

Figure \ref{fig:find_theta_nu10} shows the results for $R_b=10\,
{\rm Mpc}$  for the GMRT. The left panel shows results for $4$  realizations of
the simulation, the right panels show the mean and $3-\sigma$  
determined from $24$ realizations of the simulations and the analytic
prediction for the mean value. In all cases a peak is seen at the
expected position matched with the bubble's actual center. 
The HI fluctuations pose a severe problem for determining the 
bubble's position as it introduces considerable fluctuations in the
SNR. In some cases these fluctuations are comparable to the peak at
the bubble's actual position     (see the dashed line in the upper
left panel). The possibility of  spurious peaks makes it difficult to
reliably determine the bubble's position.

\begin{figure}
\includegraphics[width=140mm, angle=270]{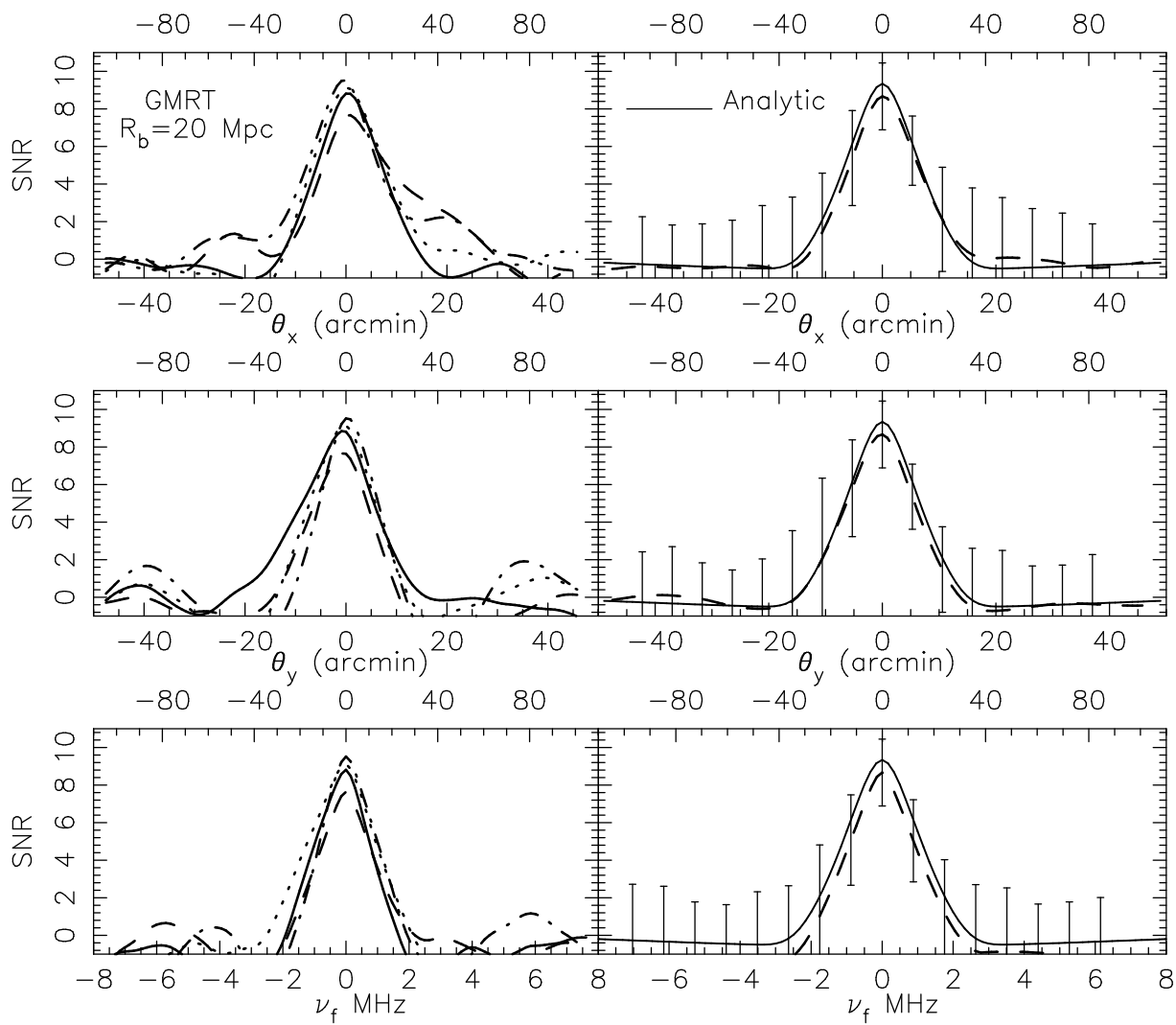}
\caption{Same as  the Figure \ref{fig:find_theta_nu10} for 
  $R_b=20 \, {\rm Mpc}$ for the GMRT.}
\label{fig:find_theta_nu20}
\end{figure}

\begin{figure}
\includegraphics[width=140mm, angle=270]{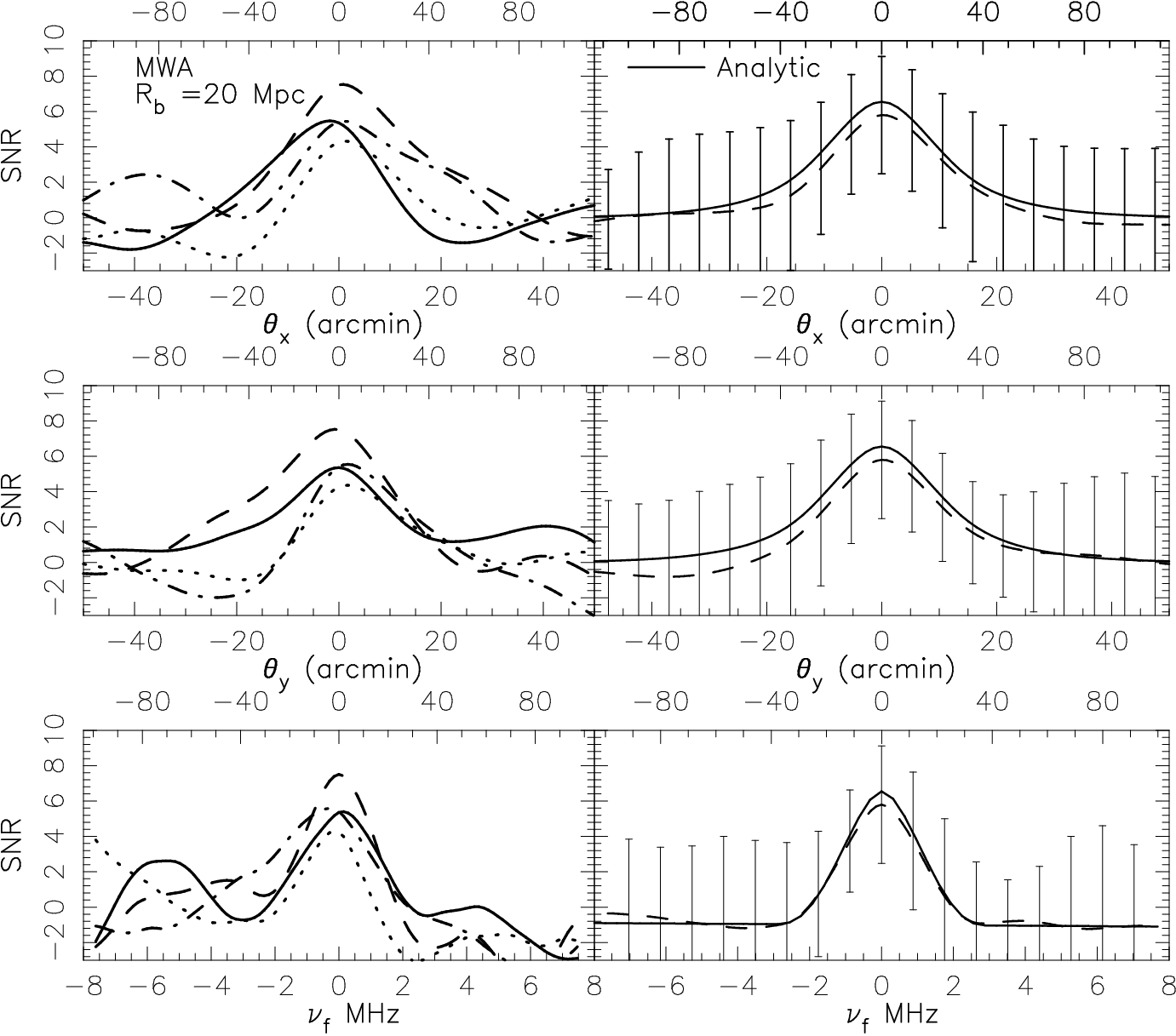}
\caption{Same as  the Figure \ref{fig:find_theta_nu10} for 
  $R_b=20 \, {\rm Mpc}$ for the MWA.}
\label{fig:find_theta_nu20M}
\end{figure}

We present the results for  $R_b=20 \,{\rm Mpc}$
in Figures \ref{fig:find_theta_nu20} and  \ref{fig:find_theta_nu20M} 
for the GMRT and MWA respectively. The HI fluctuations do not pose a
problem for determining the position of such bubbles using the
GMRT. In all the realizations of the GMRT simulations there is a  peak
at the expected position. The FWHM  $\sim 40 {\rm Mpc}$ 
is approximately the same along $\theta_x$,$\theta_y$ and $\nu_f$  
and is comparable
to the separation at which the overlap between the bubble and the
filter falls to half the maximum value. The HI fluctuations does
introduce spurious peaks, but these are quite separated from the actual
peak and have a smaller height. We do not expect these to be of
concern for position  estimation.

The MWA simulations all show a peak at the expected bubble
position. The FWHM along $\theta $  ($\sim 60 \, {\rm Mpc}$ )
is somewhat  broader than that along $\nu$ ($\sim 40 \, {\rm
  Mpc}$).    The low spatial resolution $\sim 8 \, {\rm Mpc}$ possibly  
contributes to increase the FWHM along $\theta$. The HI fluctuations
introduce spurious peaks whose heights are  $\sim 50 \, \%$ of the
height of the actual peak.

\subsection{Bubble detection in patchy  reionization }

\begin{figure}
\includegraphics[width=85mm, angle=270]{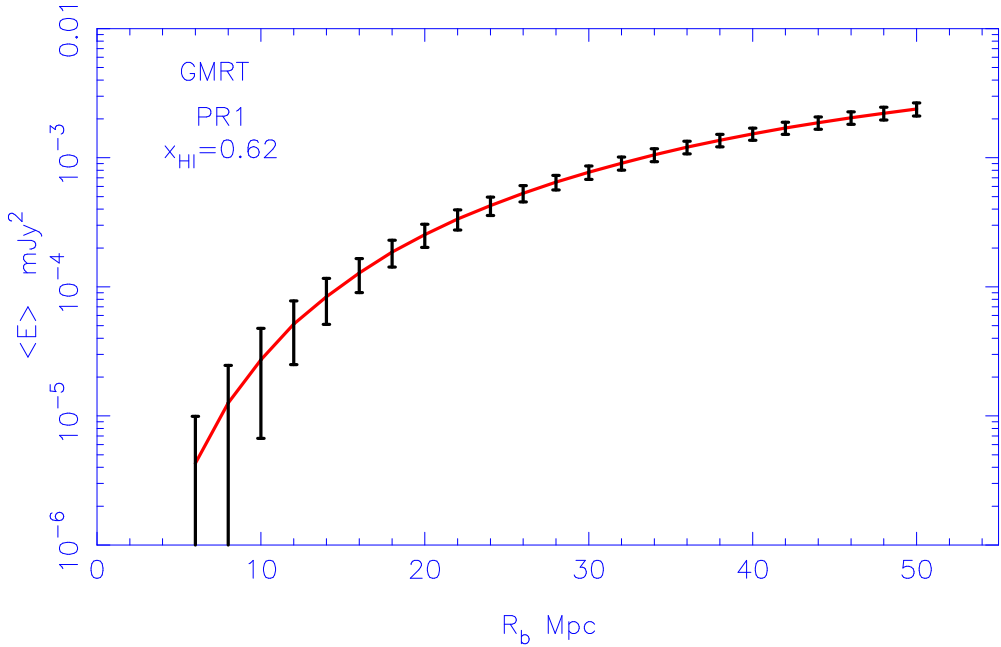}
\caption{The mean    $\langle\hat{E}\rangle$ and $3-\sigma$
  error-bars   of the estimator as a function of   $R_f$ for the GMRT
 estimated from    $24$ different realizations of the   PR1 scenario. 
In all cases the filter  is   exactly matched to the bubble.}
\label{fig:R_Es_pr1}
\end{figure}

\begin{figure}
\includegraphics[width=85mm, angle=270]{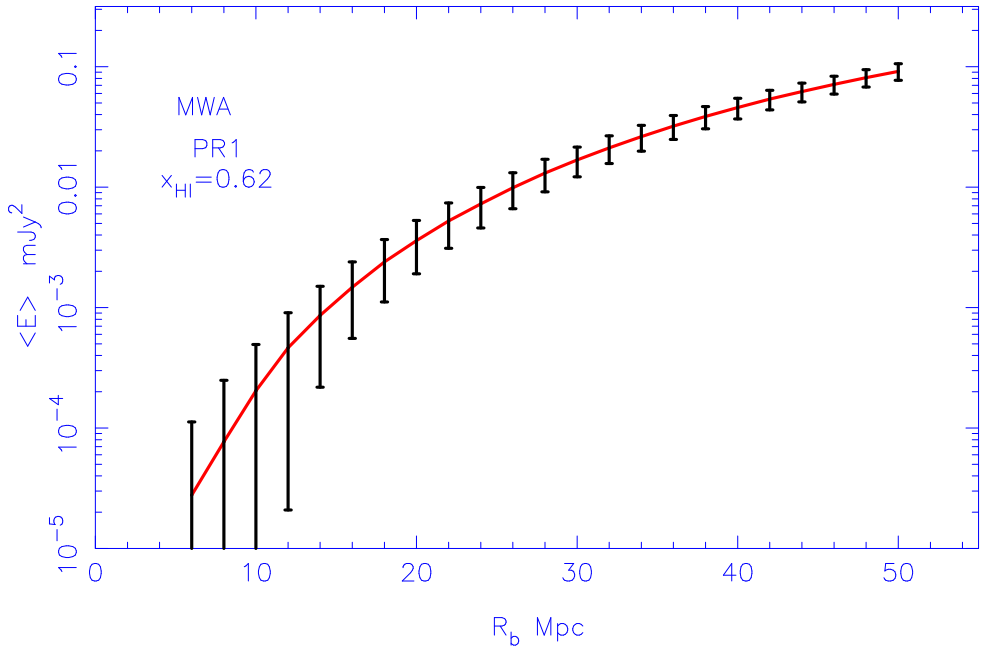}
\caption{Same as the Figure \ref{fig:R_Es_pr1} for the MWA.}
\label{fig:R_Es_pr1M}
\end{figure}

The SB scenario considered till now is the most optimistic scenario in
which the HI traces the dark matter. The presence of 
ionized  patches other than the one that we are trying to detect
is expected to increase the contribution from HI fluctuations. 
We first consider the PR1 scenario where there are several additional
ionized bubbles of radius  $6 \, {\rm Mpc}$ in the FoV. 
Figures \ref{fig:R_Es_pr1} \& \ref{fig:R_Es_pr1M} show the mean  
value of the estimator  $\langle\hat{E}\rangle$ and $3-\sigma$
error-bars  as a function of $R_f$ for the GMRT and the MWA 
respectively. These were estimated from $24$ different realizations of
the simulation, using a  filter  exactly
matched to the bubble. 


We find that the results are very similar to those for the SB scenario
except that  the signal is  down by $0.6$ due to the lower neutral fraction
($x_{\rm HI}=0.62$) in the PR scenarios. Ionized bubbles with radius 
$R_b=8\, {\rm Mpc}$ and  $=12\,{\rm Mpc}$ or smaller cannot be detected
by the GMRT and MWA respectively due to the HI fluctuations. These
limits are similar to  those obtained in  simulations of  the SB
scenario.  

In the PR2 scenario the FoV contains  other  ionized bubbles of the
same size as the bubble that we are trying to detect.  We find that
bubble detection 
is not possible in such a situation, the HI fluctuations always
overwhelm the  signal.   This result obviously depends on 
number of other  bubbles in the FoV,  and this is decided by 
$x_{\rm HI}$ which we take to be $0.62$. A detection may be possible
at higher $x_{\rm HI}$ where there would be fewer  bubbles in the FoV.

\begin{figure}
\includegraphics[width=100mm, angle=270]{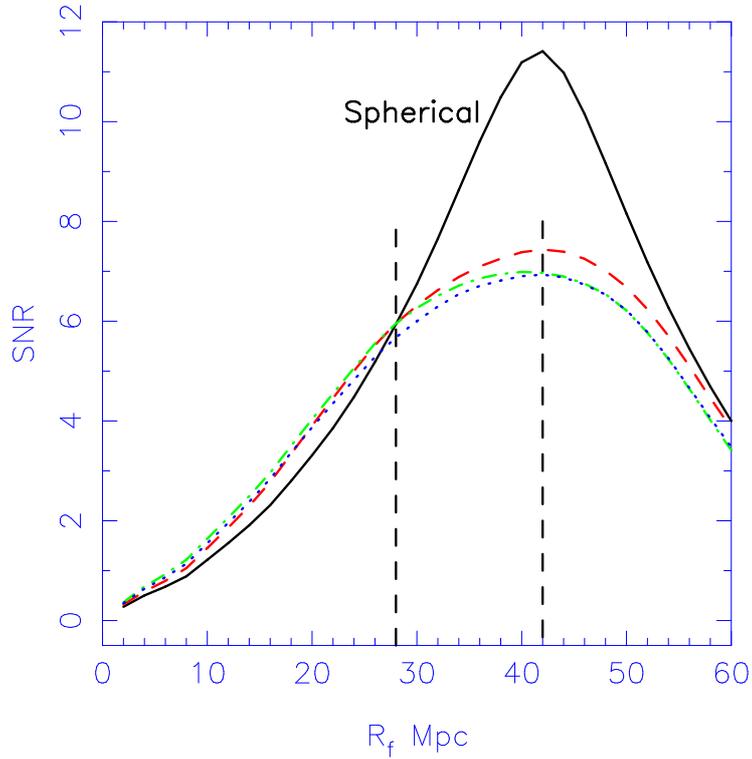}
\caption{Same as  Figure \ref{fig:find_R10} for the SM scenario for
  the GMRT.  The dotted, dashed dotted and dashed lines show
  results for three different realizations.  To show the 
  effect of non-sphericity, we compare these results
  with predictions for a  spherical bubble of sized $R_b=42\, {\rm
    Mpc}$ embedded in  uniform HI   with neutral fraction $x_{\rm
  HI}=0.5$(solid line).  The vertical line at $R_f=28\, {\rm 
    Mpc}$ shows the radius up to which the bubble is fully ionized and
  the SNR   follows the spherical predictions. The SNR peaks at
  $R_f=42\, {\rm     Mpc}$ marked by another vertical line.}  
\label{fig:find_R42}
\end{figure}

In the SM scenario, the very large computation time restricts us from
generating several realizations  with central ionized regions of
different sizes. Hence we are unable to study the  restriction on
bubble detection.    We have only three realizations all of which have  
the same ionized region located at  the center of the box.  Based on
these we find that the  mean estimator $\langle
\hat{E} \rangle$ is $\sim30$ times larger that the standard deviation
due to HI fluctuations. The  detection of a bubble  of the size
present in our simulation (Figure \ref{fig:image}) is not
restricted by the HI fluctuations. We present size determination
results in  Figure \ref{fig:find_R42}. We see that the SNR peaks at
$R_f=42\, {\rm Mpc}$ and not at $R_f=27\, {\rm  Mpc}$. Recall  that
in the 21-cm map (Figure \ref{fig:image}) we have visually identified
the  former  as the bubble's  outer radius which includes
several small patchy  ionized regions towards the periphery 
 and the latter is the  inner radius which encloses  
the  completely   ionized region. We see that the matched  filter
identifies the  bubble's outer radius. 
To study the effect of non-sphericity we compare our results in  
Figure \ref{fig:find_R42}  with the 
predictions for  a spherical bubble of radius $R_b=42\, {\rm Mpc}$
embedded in uniform HI with  the same neutral fraction $x_{\rm
  HI}=0.5$. We find that our  results for the SM scenario follow the
spherical bubble prediction up to a filter size $R_f=28\, {\rm
  Mpc}$  (marked with a vertical line in
Figure \ref{fig:find_R42}), corresponding to the bubble's inner
radius which encloses a perfectly 
 spherical ionized region. Beyond this, and upto the outer
radius of  $42 \, {\rm Mpc}$, the HI is not fully ionized. There are 
 neutral patches  which introduce deviations from
spherical symmetry and cause the SNR to fall below   the
predictions of a spherical bubble beyond  $R_f=28 \, {\rm Mpc}$. The 
deviations from sphericity also broadens the  peak in the SNR
relative to the predictions for a spherical bubble.

\begin{figure}
\includegraphics[width=100mm, angle=270]{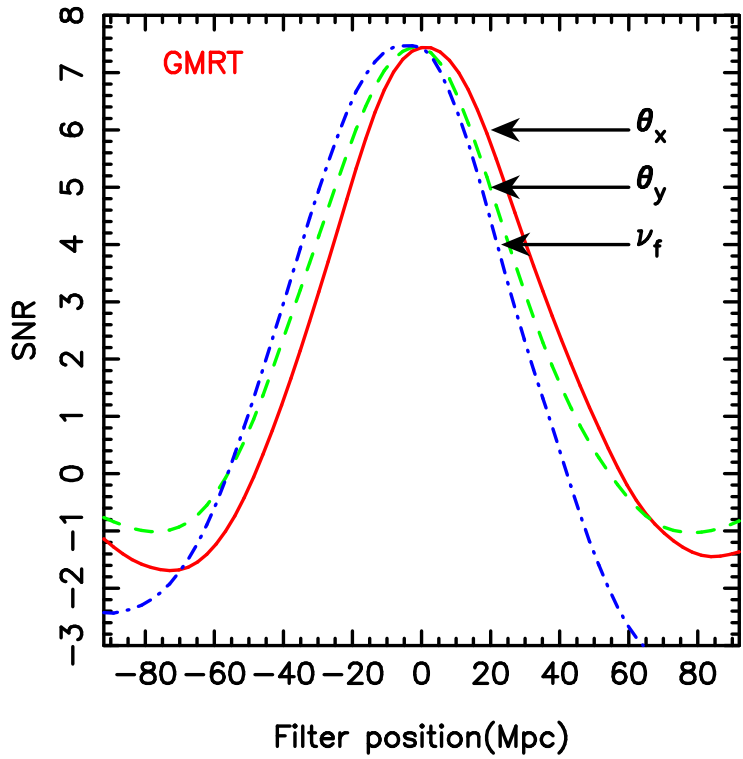}
\caption{Same as  Figure \ref{fig:find_theta_nu10} for the SM
  scenario for the GMRT. The x-axis shows the comoving distance of
  the filter position from the center of the box. The three curves 
respectively show  results for a  search along three $\theta_x$,
  $\theta_y$ and  $\nu$  axes.} 
\label{fig:find_theta_nu42}
\end{figure}

Our results based on the SM scenario show that the matched filter
technique works well for bubble detection and for determining the
bubble's size even when there are deviations from sphericity. We
obtain good estimates for the extents of  both,  the completely ionized
region and the partially ionized region.  For the SM scenario, 
 Figure \ref{fig:find_theta_nu42} shows   how well the bubble's 
position can be determined in a blind search. 
We have followed the  same  method as  described for the SB scenario
in  subsection 3.3. We  see that the SNR peaks at  the  expected
position. Further,  as  the bubble size is quite large $\gtrsim 
27$ Mpc there are no spurious peaks.

\section{Redshift Dependence}
Results shown so far are  all at  only one redshift $z=6$. It would
be  interesting and useful  to have  predictions for higher 
redshifts. However, addressing this issue through direct computations
at different redshifts would require considerable  computation 
beyond the scope of this work. Since we find that the analytic
predictions of the Chapter 3  are in good agreement with the simulations of
the SB scenario, we use the analytic formalism to predict how
different quantities are expected to scale with increasing $z$.

 The redshift dependence of some of the quantities 
like  the system noise, the background 21-cm brightness
$\bar{I_{\nu}}$,  and the angular and frequency extent of a bubble of
fixed comoving radius  causes the SNR to decrease with increasing
$z$. On the other hand the $z$ dependence of the 
neutral fraction, the  baseline distribution  function 
  and the effective antenna collecting area 
acts to increase the  SNR at higher redshifts.
 We find that with increasing $z$ both  $\langle \hat{E}\rangle$ and
 $\sqrt{\langle  (\Delta \hat E)^2 \rangle_{\rm  HF}}$  decrease  by
nearly  the same  factor  so that the   restriction on bubble
 detection does not change significantly at  higher redshift in the SB
 scenario. Assuming that the  neutral fraction does not change with
 $z$, the SNR for bubble detection decreases with increasing 
 redshift,  the change depending  on the bubble size. For  example, for
 the GMRT at $z=10$  the  SNR decreases by a factor $\sim 7$ and $6$
 for bubbles  of size   $R_b=10$ and $20\, {\rm Mpc} $
 respectively. For the  
 MWA this factor is $3$ for both these bubble sizes. 
We expect a similar  change in the SNR for the patchy reionization
scenarios.  The drop  in SNR is slower for the MWA relative to the GMRT 
because the effective antenna collecting area of the
 MWA increases at higher redshifts.   

The SNR is directly proportional to the global   neutral fraction $
x_{\rm HI}$  which increases  with $z$.  The details of how $x_{\rm
  HI}$ and the HI   fluctuations change with redshift depends on how 
reionization proceeds with  time, an issue beyond the scope of this
thesis.

\section{Summary}
We have used a visibility-based formalism, introduced in 
\cite{kkd2}, to simulate the detection of  spherical HII bubbles in
redshifted 
21 cm maps through a matched-filtering technique. The main aim
of this work is to use simulations to quantifying 
the limitations for bubble  detection  arising from the HI
fluctuations outside the bubble. We have computed 
the results for two instruments, namely, the  GMRT and the upcoming MWA.
Our main conclusions are as follows:

\begin{itemize}
\item In the case where the HI fluctuations outside the bubble
trace the dark matter distribution (SB scenario), 
we find that bubbles with radius 
$R_b=6\, {\rm Mpc}$ and  $=12\,{\rm Mpc}$ or smaller cannot be detected
by the GMRT and MWA respectively due to the HI fluctuations. Note that
this limitation is fundamental to the observations and cannot be
improved upon by increasing integration time.

\item For targeted observations of ionized bubbles, the bubble size
can be determined to an accuracy limited by the instrument's 
resolution; we find that HI fluctuations do not play any significant role.
However, the HI fluctuations can restrict the accuracy to which the 
neutral  fraction can be estimated. In addition, we find that
determining the position of the bubble in a blind search could be
quite difficult for small ($\sim 10 \, {\rm Mpc}$) bubbles as the HI
fluctuations introduce large fluctuations in the signal; for larger bubbles
the accuracy is determined by the instrument's resolution.

\item In a scenario of patchy reionization where the targeted HII
region is surrounded by many
small ionized regions of size $\sim 6 \, {\rm   Mpc}$ (PR1 scenario),
the lower  limit for bubble detection  is  similar to that in the SB
scenario.  
Thus the assumption that the HI traces the
dark matter gives a reasonable estimate of the contribution from HI
fluctuations if the background ionized bubbles are small 
 $\sim 6 \,{\rm Mpc}$. 
However, the situation is quite different when the surrounding bubbles  
as of similar size  as the targeted bubble (PR2 scenario). The large
HI fluctuations do not permit bubble detection for a neutral fraction
$x_{\rm HI}<0.6$.  Thus  for $x_{\rm HI}=0.6$  or lower,  bubble
detection is possible only if the other  ionized regions in the 
FoV  are much smaller than the bubble that we 
are trying to detect.

\item  The matched filter technique works well for 
more realistic cases based on the semi-numeric modelling of 
ionized regions \citep{choudhury08}. Here  the bubbles are substantially
non-spherical because of surrounding bubbles and inhomogeneous
recombination. Our method  gives a good estimate of the size of both
the  fully ionized and the partially ionized regions in the bubble.

\end{itemize}

To put our conclusions in an overall perspective, let us consider 
an ionized bubble around a luminous QSO at $z \gtrsim 6$. 
We expect $R_b \gtrsim 30 {\rm Mpc}$ 
from studies of QSO  absorption spectra (\citealt{wyithe,mesinger04}). 
It has also
been pointed out that these bubbles may survive as large ``gray fossils''
a long time after the source has shut down (\citealt{furlanetto08}).
It will be possible  to detect such bubbles only if 
the background bubbles are smaller, say, 
$< 30 \, {\rm Mpc}$. We find from  
models of \cite{mesinger07} 
that the typical sizes of ionized regions
when $x_{\rm HI} \sim 0.3 (0.1)$ is $\sim 20 (70) {\rm Mpc}$. Though 
these values could be highly model-dependent, it still gives 
us an idea that the bubbles around the  luminous
QSO would be detectable even in a highly ionized
IGM with, say, $x_{\rm HI} \sim 0.3$. If the size of the targeted
bubble is larger, then this constraint is less severe. This gives
a realistic hope of detecting these bubbles at $z \gtrsim 6$ with 
near-future facilities.

A caveat underlying a large part  of our analysis is  the
assumption that the bubbles  under consideration are perfectly
spherical. This is note the case in reality. For example,
non-isotropic emission from the sources (QSOs), density fluctuations
in the IGM and  radiative transfer effects would distort the 
shape of the bubble. The semi-numeric simulations (SM scenario)
incorporate some of these effects and give an  estimate  of the impact of
the deviations from sphericity on bubble detection.  This is
an important issue which we plan to address in more detail in future. 
In addition, the   finite light travel time gives rise to an 
apparent non-sphericity even if the physical shape is spherical
\citep{wyithe04a,yu05}.  
This effect can, in principle, be estimated analytically  and
incorporated in  the filter. We plan to address this effect
in future.

\newpage

%\clearpage{\pagestyle{empty}\cleardoublepage} %%%%%%%%%%%%%%%%%%%%

%\newpage
 \setcounter{section}{0}
 \setcounter{subsection}{0}
 \setcounter{subsubsection}{2}
 \setcounter{equation}{0}
 %\pagenumbering{arabic}

%-------------------------------------------
\chapter[Optimum Redshift for Detecting Ionized Bubbles]
{\bf \textbf {Optimum Redshift for Detecting Ionized
 Bubbles\footnote{{\bf \em { A sharpened, compact version of this
 chapter is presented in the paper ``The optimal redshift for detecting ionized
 bubbles in HI 21-cm maps''
 by \cite{kkd4}}}}}}

\vspace{1.5cm}
\section{Introduction}
In  Chapter  3 we have introduced a matched filter technique to detect
individual ionized bubbles in 21-cm maps. The technique optimally
combines the entire signal from an ionized bubble while minimizing the
noise and foreground. In Chapter  4 we use different simulated
21-cm maps to study the impact of the HI fluctuations outside the
bubble that we are trying to detect.      

Various redshift dependent parameters are important in bubble
detection. The background 21-cm brightness and the angular and
frequency extent of a bubble of a fixed comoving size increase from
higher to lower redshift. The system noise in low frequency radio
experiments is dominated by the sky temperature which becomes lower at
higher frequency. This makes the noise lower at lower redshifts. All
the above mentioned factors are expected to enhance the signal to noise
ratio (SNR) at lower redshifts. On the other hand, the neutral hydrogen
fraction decreases as the reionization process proceed and it becomes
almost zero around redshift $z=6$. We will also see later in the text
that the baseline distribution also changes significantly with the
frequency band of our interest. This change is 
also expected to enhance the SNR at higher redshifts. Again, for 
MWA like experiments the increase in effective collecting area of
the individual antenna is expected to reduce the system noise at
higher redshifts. Together these parameters will try to suppress the SNR
at lower redshifts with no signal at redshift $z=6$ where the Universe
is completely reionized. Because of these
two oppositely behaving set of parameters there is an
intermediate redshift where the SNR for a bubble of fixed comoving
size peaks. The detection of ionized bubbles is optimal at this
redshift. The prior knowledge of this optimum redshift is important
for any future attempt to detect ionized bubbles in 21-cm maps.

In this Chapter  we investigate the optimum redshift to detect ionized
bubbles considering different models of the neutral fraction evolution
with redshift. Predictions are made for the GMRT and the MWA. We also
establish  scaling relations for the SNR with redshift for a fixed
value of bubble size.

The Chapter is organized as follows. We begin with a brief description
on the 21-cm signal from an ionized bubbles and its redshift
dependence Section 5.2. In Section 5.3 we present a short note on our
matched filtering technique to detect ionized bubbles.  In Section 5.4
we establish scaling relations for the SNR with redshift considering
both uniform-frequency independent and nonuniform-frequency dependent
baseline distribution. In the same section we also briefly describe
the models of HI evolution that we adopt for our calculations. We
present our results in Section 5.5 and conclusions in Section 5.6

Throughout out this Chapter we adopt
cosmological parameters from Dunkley et al.(2008). For the GMRT we use
the antenna specifications from their website and for the MWA we use
the instrumental parameters from Bowman et. al. (2007).

\section{Ionized Bubbles in Redshifted 21-cm Observations}
The quantity measured in radio-interferometric observations is the
visibility $V(\u,\nu)$ which is measured in a number of frequency
channels $\nu$ across a frequency bandwidth $B$ for every pair of
antennas in the array. The visibility is related to the sky specific intensity 
$I_{\nu}(\th)$ as 
\be
V(\u,\nu)=\int d^2 \theta A(\th) I_{\nu}(\th)
e^{ 2\pi \th \cdot \u}
\label{eq:vis1}
\e
where the baseline $\u={\bf d}/\lambda$, ${\bf d}$ is physical separation between a pair of antennas projected on the plane perpendicular to the line of sight. $\lambda$ and $\th$ respectively are observing wavelength and two dimensional vector in the plane of the sky with
origin at the center of the field of view, and $A(\th)$ is the 
beam  pattern of the individual antenna. We have used $A(\th)$ described in Chapter 3.
The visibility  recorded in radio-interferometric
observations is a combination of four separate contributions 
\be
V(\u,\nu)=S(\u,\nu)+HF(\u,\nu)+N(\u,\nu)+F(\u,\nu)
\label{eq:vis2}
\e
where $S(\vec{U},\nu)$ is the HI signal that we are interested in, $ HF(\u,\nu)$
is contribution from fluctuating HI outside the target  bubble,
$N(\vec{U},\nu) $ is the system noise which is inherent to the 
measurement and $F(\vec{U},\nu)$ is the contribution from other
astrophysical sources  referred to as the foregrounds.

We consider a spherical ionized bubble of comoving radius $R_b$
centered at redshift $z_c$ at the center of the FoV. This bubble is
also assumed to be embedded in an uniform inter-galactic medium (IGM) with neutral hydrogen fraction $x_{\rm HI}$. The planar section
through the bubble at a comoving distance $\rn$ is a disk of comoving
radius $R_{\nu}=R_b \sqrt{1- (\Delta \nu/\Delta \nu_b)^2}$ where 
$\Delta \nu=\nu_c-\nu$ is the distance from the the bubble center $\nu_c$
in frequency space with $\nu_c=1420 \, {\rm MHz}/(1+z_c)$ and 
$\Delta \nu_b=R_b/r'_{\nu_c}$ is the bubble size in the 
frequency space. The expected observed visibility from this bubble can be written as (for details see Chapter 3) 

\be
S_{\rm center}(\u,\nu)=-\pi \bar{I_{\nu}} x_{\rm HI} \theta^2_\nu 
\left [ \frac{2 J_1(2 \pi U \theta_\nu
  )}{2 \pi U \theta_\nu}\right ] 
\Theta \left(1- \frac{\mid \nu -\nu_c \mid}{\Delta \nu_b} \right)
\label{eq:sig}
\e
where  $\bar{I_{\nu}}=2.5\times10^2\frac{Jy}{sr} \left (\frac{\Omega_b
  h^2}{0.02}\right )\left( \frac{0.7}{h} \right ) \left
(\frac{H_0}{H(z)} \right ) $  is the radiation background from the 
uniform HI distribution. $\theta_{\nu}=R_{\nu}/r_{\nu}$ is angular radius of the circular disc through the bubble at comoving distance $r_{\nu}$. $J_1(x)$ and $\Theta(x)$ are the first order Bessel function and the Heaviside step function respectively. 

The peak value of the signal is $S(0,\nu)=\pi
\bar{I}_{\nu} \theta_{\nu}^2 $. For a fixed value of bubble size the peak value changes with redshift. The Bessel function $J_1(x)$ has the first
zero crossing at $x=3.83$. As a result, 
in baseline the signal $S(\u,\nu)$ extends to $U_0=0.61
\theta_{\nu}^{-1}$ where it has the
first zero crossing. The signal extends over $\Delta \nu = \pm
\Delta \nu_b$ in frequency. Note that the angular size  $\theta_{\nu}$ and extent in frequency of a bubbles for a fixed value of comoving size change with redshift because of change in comoving distance $\rn$ and $r_{\nu}'$. We will discuss the effect of these change on bubble detection at various redshifts in Section 5.4. Detecting these ionized bubbles will  be a big challenge because  the
signal is  buried in noise and foregrounds which are both
considerably larger in amplitude. 
Whether we are able to detect the ionized bubbles
or not  depends critically on our ability to construct optimal filters
which  discriminate  the signal from other contributions.

\section{Matched Filtering Technique to Detect Ionized Bubbles}

For optimal signal detection we consider the  estimator (Chapter 3) 
\be
\hat{E}=  \left[ \sum_{a,b} S_{f}^{\ast}(\u_a,\nu_b)
\hat{V}(\u_a,\nu_b) \right]/\left[   \sum_{a,b} 1 \right]
\label{eq:estim1}
\e
where $S_f(\u,\nu)$ is a filter which has been constructed to detect
a particular ionized bubble, $\hat{V}(\u_a,\nu_b)$ respectively refer to the
observed visibilities and  $\u_a$ and $\nu_b$ refer to the
different baselines and frequency channels in the observation and in
eq. (\ref{eq:estim1}) we are to sum over all independent data points
(visibilities). The filter $S_f(\u,\nu)$  depends on $[R_f,z_c,\th_c]$
the  comoving  radius, redshift and angular position of the bubble
that we are trying to detect.

We now calculate $\langle \E \rangle$ the expectation value of the
estimator. Here the angular brackets  denote an average with respect
different realizations of the HI fluctuations, noise and foregrounds,
all  of which have 
been   assumed to be random variables with zero mean. This gives
$\langle \hat{V}(\u,\nu) \rangle =S(\u,\nu)$ in the continuum limit
and 
\be
\langle \E \rangle  =\int d^2U \, \int d\nu  \, \rho_N(\u,\nu) \, \, 
{S_f}^{\ast}(\u,\nu) S(\u,\nu)  
\label{eq:estim2}
\e  

The variance of the estimator which is the sum of
the contributions  from the noise (NS), the foregrounds(FG)  and the
HI fluctuations (HF) can be written as
\begin{eqnarray}
\langle (\Delta \E)^2 \rangle =\left <(\Delta \hat
E)^2 \right >_{{\rm NS}}+\left<(\Delta \hat E)^2 \right >_{{\rm FG
}} \,
 +\left<(\Delta \hat E)^2 \right >_{{\rm HF}} \,.\nonumber \\ 
\label{eq:16}
\end{eqnarray}

Assuming that the noise in different baselines and frequency channels are  
uncorrelated we have 
 
\be
\langle (\Delta \hat E)^2 \rangle_{\rm NS}
= \sigma^2 
\int d^2U \, \int d\nu  \, \rho_N(\u,\nu) \, \, 
\mid S_{f}(\u,\nu)\mid^2\,.
\label{eq:ns1}
\e

where $\sigma$ is the rms. noise in the image.

The variance due to HI fluctuations is given as

\begin{eqnarray}
\left <(\Delta \hat E)^2 \right >_{\rm{HF}}\!\!\!\!\!&=&\!\!\!\!\!
\int d^2 U  \int d \nu_1 \int d \nu_2 \left(\frac{d B}{d T}\right)_{\nu_1} \left(\frac{d B}{d
  T}\right)_{\nu_2} 
\nline
\!\!\!\!\!&\times&\!\!\!\!\!
\rho_N(\u,\nu_1) \rho_N(\u,\nu_2)
{S_f}^{\ast}(\u,\nu_1) {S_f}(\u,\nu_2)
\nline
\!\!\!\!\!&\times&\!\!\!\!\!
 C_{2 \pi U}(\nu_1,\nu_2)
\label{eq:fg1}
\end{eqnarray}

where $C_{2 \pi   U}(\nu_1, \nu_2)$  is the multi-frequency angular
power spectrum of the HI fluctuation  from \citet{kkd1} (Chapter 2).

\section{Redshift Dependence: Scaling Relations}

\subsection{Uniform and frequency independent baseline distribution}
The estimator $\langle E \rangle$ and its variance depend on the size
($R_b$), redshift ($z_c$) and position ($\th_c$) of the bubble. In
this Section we present a scaling relations for the estimator
describing its change with redshift. We assume a bubble of fixed size
at the center of the FoV and the filter is exactly matched with
bubble. In a situation where the baseline distribution
($\rho(\u,\nu)$) is (i) uniform in baseline over the signal
$S(\u,\nu)$ (ii) independent of frequency channels and (iii) the
frequency bandwidth is much larger than the bubble
($BW>>\Delta \nu_b$) the estimator scales exactly as
\be
\langle \E \rangle(z) \propto x^2_{\rm HI}(z){\bar I_{\nu}}^2
\theta_{\nu_c}^2\Delta\nu_b
\label{eq:scall_E1}
\e
where $x_{\rm HI}(z)$ quantifies the evolution of the neutral hydrogen
fraction. The redshifted 21-cm background ${\bar I_{\nu}}$ scales as
$(1+z)^{-1.5}$ at higher redshifts. The angular size $\theta_{\nu_c}$
which is $\propto 1/r_{\nu}$ approximately scales as $(1+z)^{-0.25}$
for a fixed value of bubble size. Similarly the frequency extent of
the bubble ($2\,\Delta\nu_b$) decreases with redshift as
$(1+z)^{-0.5}$. Considering all these quantities we get  
\be
\langle \E \rangle(z) \propto x^2_{\rm HI}(z)(1+z)^{-4}.
\label{eq:scall_E2}
\e

The system  noise is dominated by the sky temperature $T_{\rm sky}$
for the frequency range of interest. The sky temperature $T_{\rm sky}$
increases with observed wavelength and scales as $(1+z)^{\beta}$,
where $\beta$ is the spectral index. Considering this fact and under
the same assumptions mentioned above  the noise variance can be
written as  
\be
\langle (\Delta \hat E)^2 \rangle_{\rm
  NS}(z)\propto x^2_{\rm HI}(z)(1+z)^{2\,\beta-4}.
\label{eq:ns2}
\e
In this equation we have assumed that the individual antenna
collecting area  is independent of frequency (applicable to the
GMRT). For MWA like antenna the individual antenna collecting area is
expected to increase with observed wavelength as $\sim \lambda^2$ and
the noise variance scales as 
\be
\langle (\Delta \hat E)^2 \rangle_{\rm
  NS}(z)\propto(1+z)^{2\,\beta-8}.
\label{eq:ns3}
\e

The fluctuating HI outside the ionized bubble also contribute to the
variance of the estimator. This contribution  
can not be reduced by increasing observation time and put fundamental 
restrictions in detecting ionized bubbles. The HI distribution outside
the target bubble during reionizaintion is highly unknown and differs
drastically for different models of reionizations. For simplicity we
assume that the HI outside the bubble trace the dark matter
distribution. The presence of other ionized regions outside the bubble
are expected to increase this contribution. The assumption that the HI
outside the bubble trace the dark matter distribution will give the
minimum variance contribution to the estimator. Analytical derivation
of the scaling relation for the $\langle (\Delta \hat
E)^2 \rangle_{\rm HF}$ is not straight forward because of the term $C_{2 \pi
U}(\nu_1,\nu_2)$. Assuming that the HI to trace the dark matter we numerically calculate $\langle (\Delta \hat
E)^2 \rangle_{\rm HF}$ at various redshifts using
equation \ref{eq:fg1}. We then fit the data with a power law of $(1+z)$
which comes out to be  

\be
\langle (\Delta \hat E)^2 \rangle_{\rm HF}\propto x_{\rm HI}^2(z)(1+z)^{-10.5}.
\label{eq:hf1}
\e 
Detection of bubble is  possible when $ \langle \E \rangle(z) > \langle
(\Delta \hat E)^2 \rangle_{\rm HF}$. When this is satisfied we are
interested only in $\langle \E \rangle/ \sqrt{\langle (\Delta \hat
E)^2 \rangle_{\rm NS}}$ which we call SNR. Note that the term
$x_{\rm HI}(z)$  
in the equations \ref{eq:scall_E1}, \ref{eq:ns2} is almost constant
along the line of 
sight of the ionized bubble. This factor does not affect  the value of the
quantity SNR. One can drop $x_{\rm HI}(z)$ and similarly ${\bar I_{\nu}}$
from the filter
$S_f(\u,\nu)$ without losing the effectiveness of the method. We calculate the SNR at different redshifts
ranging from $z=6$ to $16$ for a fixed value of bubble size. Using
equations \ref{eq:scall_E2} and \ref{eq:ns2} for a GMRT like antenna we have 

\be
SNR\propto x_{\rm HI}(z)(1+z)^{-\beta-2}
\label{eq:snr1}
\e
For a MWA like antenna this is 
\be
SNR\propto x_{\rm HI}(z)(1+z)^{-\beta}
\label{eq:snr2}
\e
Note that in the above two equations $x_{\rm HI}(z)$ is a increasing 
function of  redshift $z$ and  for a  typical value
of $\beta \approx 2.5-2.8$, we expect the SNR to be peaked at an
intermediate redshift  
during reionization. As we discussed earlier this is due to three reasons, 
firstly the sky temperature increases at low frequency ie, at higher
redshifts. Second the background specific intensity ${\bar I_{\nu}}$ also
decreases at higher redshift. Finally the angular size and 
frequency extent of the bubble also become smaller at higher redshift
due to the increase in $r_{\nu}$ and $r'_{\nu}$ with redshift . We see 
that the two factors are opposite in nature- one increases with redshift 
whereas the other decrease. There is an intermediate redshift $z$ where the 
SNR is maximum.

\subsection{Non uniform and frequency dependent baseline distribution}
The baseline distribution in general is not uniform in $\u$ (see
figure 3.5 in Chapter 3) and also frequency dependent. This non
uniformity changes the above scaling relations. We introduce a
parameter $n$ to take into account this effect and now modified SNR
can be written as 

\be
SNR\propto x_{\rm HI}(z)(1+z)^{n/2-\beta-2}
\label{eq:snr3}
\e
for the GMRT like antenna. For MWA like antenna this is

\be
SNR\propto x_{\rm HI}(z)(1+z)^{n/2-\beta}
\label{eq:snr4}
\e
The normalized baseline distribution function $\rho_N(\u, \nu)$ is
frequency dependent and the parameter $n$ takes care of that. Note
that $n=0$ is situation for uniform baseline coverage.  The value of
$n$ depends on the bubble size.  For large bubbles the signal remains
confined within small baselines whereas for smaller bubbles a
significant amount of signal 
spreads to larger baselines. Thus the normalized baseline distribution
function $\rho_N(\u, \nu)$ acts in different manner on bubbles of
different sizes. For a situation when the $\rho_N(\u, \nu)$ is
constant in $\u$ over the signal we have $n=2$. In a situation where the
signal is constant in $\u$  over  the whole baseline range
$\rho_N(\u,\nu)$  does not play significant role and 
$n=0$. In general we have $0<n<2$. We will
discuss more about the exact values of $n$ for the GMRT and the MWA
later.

The GMRT baseline distribution is well described by the sum of a Gaussian 
(at small baselines) and an exponential (at large baseline)(Figure 3.5
in Chapter 3). Here $\rho_N(\u, \nu)$ remains almost constant up to a
baseline $\u=100$ for the frequency range of interest and then it
decays. For the larger bubbles of radius $R_b\ge50 {\rm Mpc}$, the
signal is confined with a maximum baseline of
$\u=100$ and we have $n=2$ for which $SNR\propto x_{\rm HI}(z)(1+z)^{-3.6}$ 
where we assume $\beta=2.6$. For smaller bubbles the signal extends to
larger baselines where the 
baseline distribution function $\rho_N(\u, \nu)$ decays with $\u$ and
the $n$ value also becomes smaller. For example, for the bubbles of
sizes $R_b=30, 20$ and $10 {\rm Mpc}$ the values of $n$ are $1.27,
1.07$ and  $0.5$ respectively. We have assumed that the MWA antenna
distribution is $\rho_{ant}\sim r^{-2}$, where $r$ is physical
distance from the center of the array. Here $\rho_N(\u,\nu)$ also
decays monotonically. For the bubble sizes $R_b=50, 30, 20$ and $10\,
{\rm Mpc}$ the values of $n$ are $0.45, 0.18$ and $0$ respectively.

The behavior of the  HI fluctuations contribution also changes with
  redshift when we calculate for non-uniform and frequency dependent
  baseline distribution. As we go to higher redshifts the HI
  fluctuations decreases but
  slowly than uniform baseline case. This is due to the same reason
  as discussed in the 1st paragraph of this subsection.  Using
  equation \ref{eq:fg1} we calculate $\langle (\Delta \hat
  E)^2 \rangle_{\rm HF}$ numerically at different redshifts for 
  the bubble sizes $R_b=10, 20, 30$ and $50\, {\rm Mpc}$. For the
  GMRT we find that $\sqrt{\langle (\Delta \hat E)^2 \rangle_{\rm HF}}\sim
   x_{\rm HI}(z)(1+z)^{-3.67},  x_{\rm HI}(z)(1+z)^{-3.37}, x_{\rm
  HI}(z)(1+z)^{-3.25}$ and $x_{\rm HI}(z)(1+z)^{-3.16}$ 
  respectively. For the MWA these are $x_{\rm HI}(z)(1+z)^{-4.26}$,
  $x_{\rm HI}(z)(1+z)^{-4.18}$, $x_{\rm HI}(z)(1+z)^{-4}$ and $x_{\rm
  HI}(z)(1+z)^{-3.85}$ respectively.   

\subsection{Evolution of neutral fraction with redshift}
\begin{figure}
\includegraphics[width=1.0\textwidth]{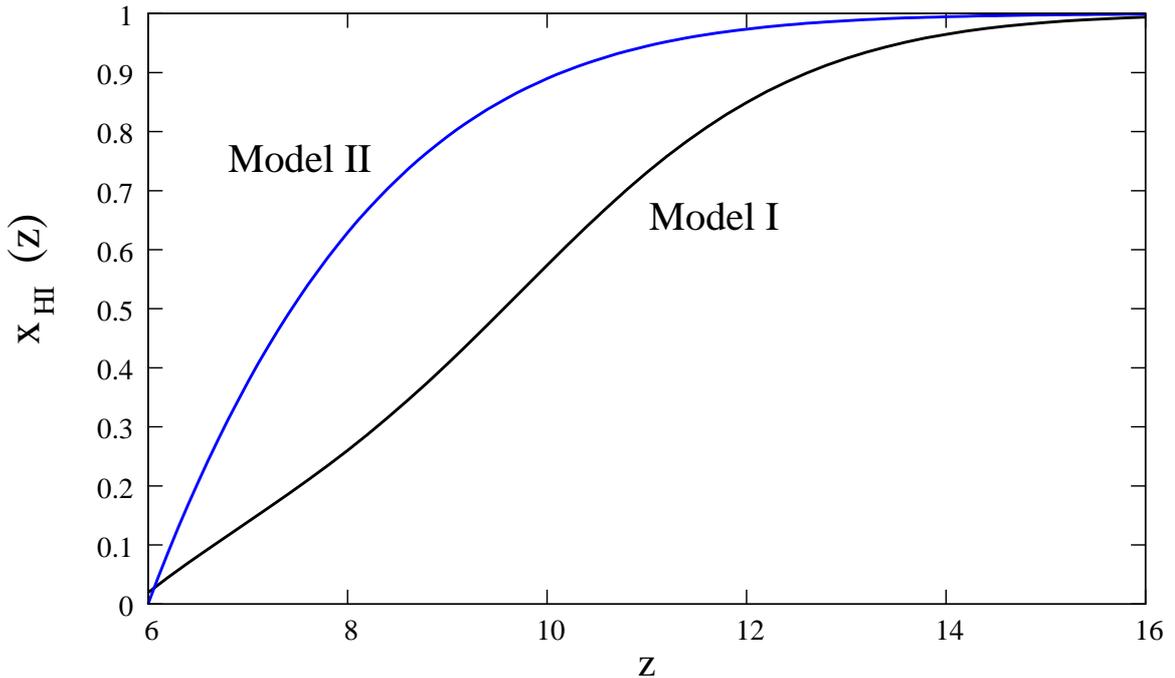}
\caption{This shows the evolution of the mean neutral fraction
  $x_{HI}$ with redshift for the two different
  reionization models discussed in the text.}
\label{fig:zvsh1}
\end{figure}
In this work, we consider two physically motivated models of reionization,
namely, the early reionization model and the late reionization model
which we call Model I and Model II respectively.
These models are constructed using the semi-analytical formalism 
\citep{chou05,cf06a} which implements most of the relevant
physics governing the thermal and ionization history of the IGM, such as
the inhomogeneous IGM density distribution, three different classes of
ionizing photon sources (massive PopIII stars, PopII stars and QSOs),
radiative feedback inhibiting star formation in low-mass galaxies and
chemical feedback for transition from PopIII to PopII stars. The
models are consistent with various observational data, namely,
the redshift evolution of Lyman-limit absorption systems,
the Gunn-Peterson effect, electron scattering optical depths,
temperature of the IGM and cosmic star formation history. 

In Model I, hydrogen reionization starts around $z
\approx 16$ driven by metal-free (PopIII) stars, and it is $50 \%$
complete by $z \approx 10$. The contribution of PopIII stars decrease below
this redshift because of the combined action of radiative and chemical
feedback. As a result, reionization is extended considerably completing
only at $z \approx 6$ (Figure \ref{fig:zvsh1}).

In Model II, the contribution from the metal-free stars
is ignored, which makes reionization start much later and is only $50$ per
cent complete only around $z \approx 7.5 $. The main difference between
this model with the previous one is in their predictions for the electron
scattering optical depth (which is $0.15$ for the early reionization model
and $0.06$ for the late reionization model). Recent measurements of
the electron scattering optical depth
$\tau_e=0.087\pm0.017$ (\citealt{dunkley08}) suggests that these two models are
possibly at two extreme ends of different possible reionization scenarios.

\section{Optimum Redshift to Detect Ionized Bubbles}

\begin{figure}
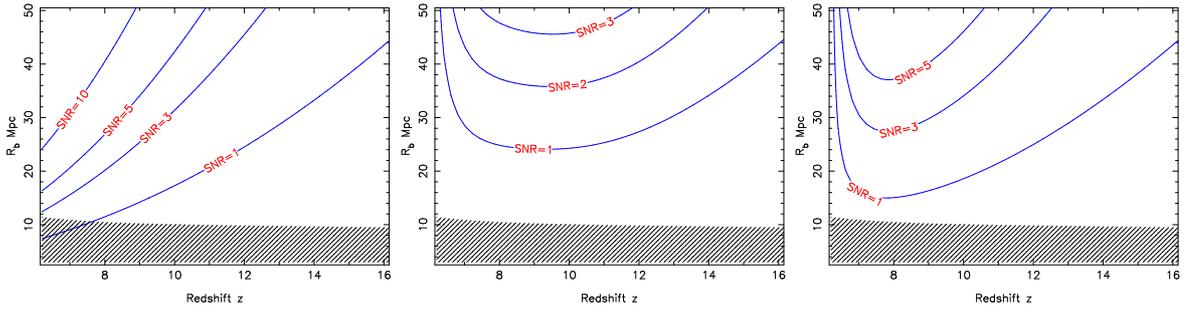

\includegraphics[width=.26\textwidth, angle=270]{chap5fig1.ps}
\includegraphics[width=.26\textwidth, angle=270]{chap5fig2.ps}
\includegraphics[width=.26\textwidth, angle=270]{chap5fig3.ps}
\caption{This shows the SNR contours for $1000 \, {\rm hrs}$
  observations with the GMRT. The left panel shows results for a constant
  neutral fraction $x_{\rm HI}=1$. Middle and right panel show results for the
  Model I and the Model II respectively. The HI fluctuations dominate over the
  signal in the shaded region and bubble detection is not possible.}
\label{fig:z_Rf_snr}
\end{figure}

We consider bubbles of comoving size ranging from $2\, {\rm Mpc}$ to
$50\, {\rm Mpc}$ at redshifts  from $z=6$
to $16$. Figure \ref{fig:z_Rf_snr} shows the SNR contours for $1000\,
{\rm hrs}$ of observations with the GMRT.  The left panel shows results
for a constant neutral fraction $x_{\rm HI}=1$. This shows the joint
effect of factors such  
as the background 21-cm brightness ${\bar I_{\nu}}$, angular 
and frequency extent of the bubble, sky temperature, normalized
baseline distribution function and the effective
area of individual antenna. We see that the SNR is maximum at redshift the 
$z=6$ for any bubble size. As we have discussed
in  Section 5.4 (equations \ref{eq:snr3} and \ref{eq:snr4}) this is
because of the lower  noise  at lower redshift. The region to the left of a
 line is allowed for the detection. For example a $5\,\sigma$
 detection is possible for a bubble size $R_b>16\, {\rm
 Mpc}$ at redshift $z=6$. At $z\ge 11$ a $5\,\sigma$ is not possible even for
 bubbles of size $50\, {\rm Mpc}$. The middle  shows the
 same results for Model I.  We see that the redshift range $z\sim 9 $
 to $10$ where the SNR is maximum is the optimum redshift range for
 bubble detection. We 
 also see that a  $3\, \sigma$ detection is possible  for bubbles of size
 $\ge 46\, {\rm 
 Mpc}$ at the optimum redshift. A $3\, \sigma$ detection is also
 possible for a bubble size $R_b\ge50\, {\rm Mpc}$ in the redshift
 range $z\sim 7$ to $11$ . Results for the Model II are presented in
 the right panel. Here we see that redshift $z=8$
 is optimum for bubble detection. In this model
the  reionization occurred at lower redshift and hence the SNR is comparatively
 higher.  Bubble detection is relatively easier
 in Model II. For example a $5\sigma$ detection
 is possible for $R_b\ge38\, {\rm Mpc}$ at $z=8$ which is not possible
 in the Model I. In all the panels bubble detection is not possible on
 the shaded region due to the HI fluctuations.

\begin{figure}
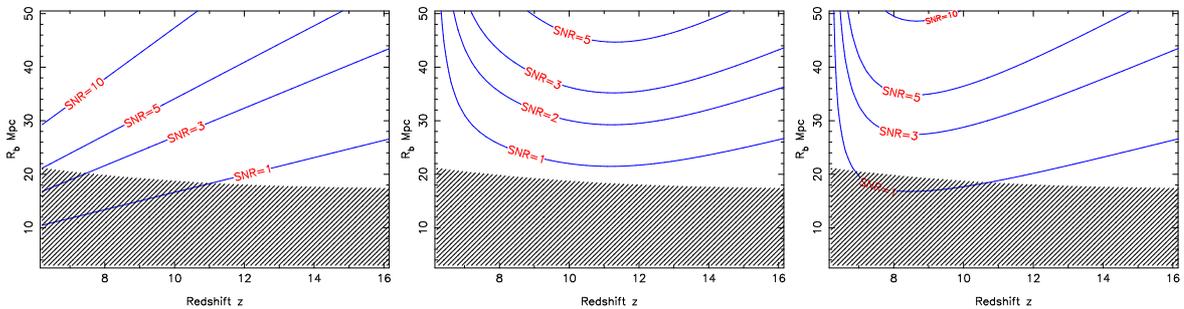

\includegraphics[width=.26\textwidth, angle=270]{chap5fig4.ps}
\includegraphics[width=.26\textwidth, angle=270]{chap5fig5.ps}
\includegraphics[width=.26\textwidth, angle=270]{chap5fig6.ps}
\caption{Same as the figure \ref{fig:z_Rf_snr} for the MWA.}
\label{fig:z_Rf_snrM}
\end{figure}

Figure \ref{fig:z_Rf_snrM} shows the same results for the MWA. The effective  
collecting area of
the individual MWA antenna is expected to increase with 
wavelength i.e, with redshift as $\sim(1+z)^2$. This will reduce the noise
rms at higher redshifts and make  bubble detection 
easier at higher redshifts than the GMRT. For
example a $5\sigma$ detection is possible in Model I for bubble
$R_b=50\, {\rm Mpc}$ in the redshift range $z=9$ to $15$ which seems
impossible with the GMRT. The SNR is $3$ times
higher than the GMRT at redshift $z=16$ for a  $R_b=46\,
{\rm Mpc}$ bubble. The
optimum redshift in  Model I for the MWA is $z\sim 11$ which is
slightly higher than the GMRT. Detection of bubbles in model II is
more optimistic in the MWA where $5\sigma$ is possible for bubbles of
$R_b\ge38\, {\rm Mpc}$ at redshift $z\sim8.5$ which is the optimum
redshift of detection in this model.

\begin{figure}
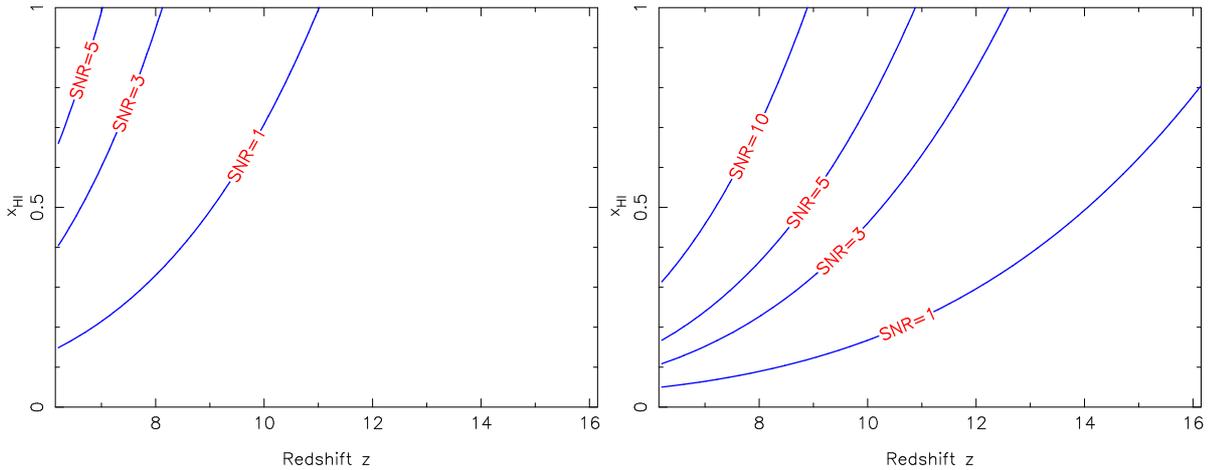

\includegraphics[width=.40\textwidth, angle=270]{chap5fig7.ps}
\includegraphics[width=.40\textwidth, angle=270]{chap5fig8.ps}
\caption{This shows the SNR contours for a fixed value of bubble
size with the GMRT. Left panel shows results for a 
$R_b=20\,{\rm Mpc}$ bubble and $4000\, {\rm hrs}$ of observations. Right panel
shows results for $R_b=50\,{\rm Mpc}$ bubble size and $1000\, {\rm
hrs}$ of observations} 
\label{fig:z_xhiG}
\end{figure}

\begin{figure}
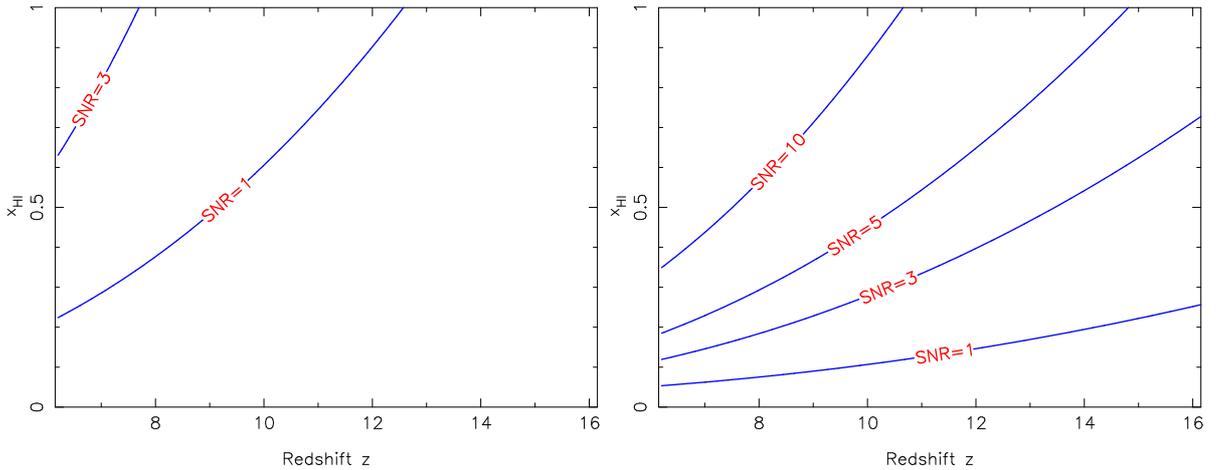

\includegraphics[width=.40\textwidth, angle=270]{chap5fig9.ps}
\includegraphics[width=.40\textwidth, angle=270]{chap5fig10.ps}
\caption{Same as the figure \ref{fig:z_xhiG} for the MWA.}
\label{fig:z_xhiM}
\end{figure}

In the above discussion we consider different models of the mean
HI evolution. The HI evolution surrounding the ionized bubbles is 
highly unknown and model dependent. Keeping this in mind we consider a
bubble of a fixed size and calculate the SNR for neutral fraction
$x_{\rm HI}=0$ to $1$ at various redshift ranging from $z=6$ to
$16$. We present results in Figures \ref{fig:z_xhiG}
and \ref{fig:z_xhiM} for the GMRT and the MWA respectively for bubble
size $R_b=20\,{\rm Mpc}$ (left panel) and $50\,{\rm Mpc}$ (right
panel). The left panel in the figure \ref{fig:z_xhiG} shows results for
$4000\, {\rm hrs}$ of observation. The Figure demonstrates the
detectibility of a
$R_b=20 (50)\,{\rm Mpc}$ bubble embedded in the HI of some neutral
fraction. We see that a higher neutral fraction is required to detect
bubbles at higher redshifts. For example  we
see that a $3\sigma$ detection for a bubble 
$R_b=50\,{\rm Mpc}$ is possible for neutral fraction $\sim 0.1$ at
redshift $z=6$ with the GMRT (Figure \ref{fig:z_xhiG}) but the neutral
fraction $\sim1$ is required at $z=12$ to detect the same
bubble. Bubble detection is found to be easier at higher redshifts for
the MWA.

\section{Conclusions}

We investigate the optimum redshift  
for bubble detection.  We find that
for early reionization the redshift $z=9$ and $11$ are
respectively the optimum redshifts for the GMRT
and the MWA.  For late reionization redshift $z=8$ is
found to be the optimum redshift for both the experiments. Bubble
detection will be  easier if the reionization occurred
late. This  is because of less
noise and higher 21-cm background intensity. The MWA is suitable for
bubble detection at higher redshifts ($z>8.5$) than the GMRT. We also
find that at redshift $z=6$ if surrounding mean neutral fraction 
is $ x_{\rm HI}\ge0.1$ a bubble of size 
$R_b>50\, {\rm Mpc}$  is possible to detect with $1000\, {\rm
hrs}$ of observation for both the experiments.

\newpage

%\clearpage{\pagestyle{empty}\cleardoublepage} %%%%%%%%%%%%%%%%%%%%

%\newpage
\clearpage{\pagestyle{empty}\cleardoublepage} %%%%%%%%%%%%%%%%%%%%
\chaptermark{Appendix}
\addcontentsline{toc}{chapter}{Appendix } 
\appendix
\thispagestyle{empty}
%\setcounter{section}{0}
%\setcounter{subsection}{0}
%\setcounter{subsubsection}{2}
%\setcounter{equation}{0}
%\pagenumbering{arabic}
\hspace{-.8cm}{\huge{\bf Appendix}}
\section{Calculation of the angular power spectrum $C_{\lowercase{l}}(\nu_1,\nu_2)$}\label{sec:cl}

In this section, we present the details of the calculation for
the 21 cm angular power spectrum $C_l(\nu_1,\nu_2)$. The first
step would be to calculate the spherical harmonic component $a_{lm}$
of the brightness temperature $T(\nu, {\bf \hat{n}})$. 
Using the expression (\ref{eq:eta_tilde}) 
for $\tilde{\eta}_{\rm HI}\left({\bf k}\right)$ in 
equation (\ref{eq:a_lm}), the expression for
$a_{lm}$ can be written as
\bear
a_{lm}(\nu) 
&=& 
\bar{T} ~ \bar{x}_{\rm HI}
\int \de \Omega ~ Y^*_{lm}({\bf \hat{n}}) ~ 
\int \f{\de^3 k}{(2 \pi)^3}
\nline
&\times&
\left[\Delta_{\rm HI}({\bf k}) 
+ ({\bf \hat{n} \cdot \hat{k}})^2 \Delta({\bf k})\right]
{\rm e}^{-{\rm i} k r_{\nu} ({\bf \hat{k} \cdot \hat{n}})}
\ear
which then essentially involves solving angular integrals of the forms
$\int \de \Omega ~ Y^*_{lm} ~ 
{\rm e}^{-{\rm i} k r_{\nu} ({\bf \hat{k} \cdot \hat{n}})}$
and $\int \de \Omega ~ Y^*_{lm} ~ ({\bf \hat{n} \cdot \hat{k}})^2
{\rm e}^{-{\rm i} k r_{\nu} ({\bf \hat{k} \cdot \hat{n}})}$ respectively.
Expanding the term 
${\rm e}^{-{\rm i} k r_{\nu} ({\bf \hat{k} \cdot \hat{n}})}$
in terms of spherical Bessel functions $j_l(kr_{\nu})$, one can show that
\be
\int \de \Omega ~ Y^*_{lm}({\bf \hat{n}}) ~ 
{\rm e}^{-{\rm i} k r_{\nu} ({\bf \hat{k} \cdot \hat{n}})}
=
4 \pi (-{\rm i})^l j_l(k r_{\nu}) Y^*_{lm}({\bf \hat{k}})
\e
Differentiating the above equation with respect to $k r_{\nu}$ twice
\be
\int \de \Omega ~ ({\bf \hat{k} \cdot \hat{n}})^2 ~ Y^*_{lm}({\bf \hat{n}}) ~ 
{\rm e}^{-{\rm i} k r_{\nu} ({\bf \hat{k} \cdot \hat{n}})}
= -4\pi
(-{\rm i})^l ~ j''_l(k r_{\nu}) Y^*_{lm}({\bf \hat{k}})
\e
where $j''_l(x)$ is the second derivative of $j_l(x)$ 
with respect to its
argument, and can be obtained through the recursion relation
\bear
(2 l + 1) j''_l(x) \!\!\!\!\! &=& \!\!\!\!\!\f{l(l-1)}{2 l -1} j_{l-2}(x)
-\left[\f{l^2}{2 l - 1} + \f{(l+1)^2}{2 l +3}\right] j_l(x)
\nline
\!\!\!\!\!&+&\!\!\!\!\!
\f{(l+1)(l+2)}{2 l +3} j_{l+2}(x)
\ear
So the final expression of $a_{lm}$ is given by
\bear
a_{lm}(\nu) 
&=&
4 \pi \bar{T}~ \bar{x}_{\rm HI}
(-{\rm i})^l~\int \f{\de^3 k}{(2 \pi)^3} Y^*_{lm}({\bf \hat{k}})
\nline
&\times&
\left[\Delta_{\rm HI}({\bf k}) j_l(k r_{\nu})
- \Delta({\bf k}) j''_l(k r_{\nu}) \right]
\ear

The next step is to calculate the 
the power spectrum
$C_l(\nu_1, \nu_2) \equiv \langle a_{lm}(\nu_1) ~ a^*_{lm}(\nu_2) \rangle$.
The corresponding expression is then given by
\bear
C_l(\nu_1, \nu_2) 
\!\!\!\!\!&=& \!\!\!\!\!
(4 \pi)^2 \bar{T}(z_1)\bar{T}(z_2) ~ \bar{x}_{\rm HI}(z_1)\bar{x}_{\rm HI}(z_2)
\nline
\!\!\!\!\!&\times&\!\!\!\!\!
\int \f{\de^3 k_1}{(2 \pi)^3} \int \f{\de^3 k_2}{(2 \pi)^3}
Y^*_{lm}({\bf \hat{k}_1}) Y_{lm}({\bf \hat{k}_2}) 
\nline
\!\!\!\!\!&\times&\!\!\!\!\!
\left\langle
\left[\Delta_{\rm HI}(z_1, {\bf k_1}) j_l(k_1 r_{\nu_1})
- \Delta(z_1, {\bf k_1}) j''_l(k_1 r_{\nu_1}) \right]
\right.
\nline
\!\!\!\!\!&\times& \!\!\!\!\!
\left.
\left[\Delta^*_{\rm HI}(z_2, {\bf k_2}) j_l(k_2 r_{\nu_2})
- \Delta^*(z_2, {\bf k_2}) j''_l(k_2 r_{\nu_2}) \right]
\right\rangle
\nline
\ear
where we have put back the redshift-dependence into the expressions
for clarity.
Now note that we would mostly be interested in cases where 
$\nu_2 - \nu_1 \equiv \Delta \nu \ll \nu_1$. In such cases, one can
safely assume $\bar{T}(z_2) \approx \bar{T}(z_1)$ and
$\bar{x}_{\rm HI}(z_2) \approx \bar{x}_{\rm HI}(z_1)$.
Furthermore, the terms involving the ensemble averages
of the form
$\langle \Delta \Delta^* \rangle$ can be approximated
as $\langle\Delta(z_1, {\bf k_1}) \Delta^*(z_2, {\bf k_2})\rangle
\approx (2 \pi)^3 \delta_D({\bf k_1 - k_2}) P(z_1, k_1)$
and similarly for terms involving $\Delta_{\rm HI}$.
We can then use the Dirac delta function 
$\delta_D({\bf k_1 - k_2})$ to compute the 
${\bf k_2}$-integral, and thus can write the 
angular power spectrum as
\bear
C_l(\Delta \nu) \!\!\!\!\!&\equiv& \!\!\!\!\! C_l(\nu, \nu + \Delta \nu)
\nline
\!\!\!\!\!&=& \!\!\!\!\!
(4 \pi)^2 ~ \bar{T}^2 ~ \bar{x}^2_{\rm HI}
\int \f{\de^3 k_1}{(2 \pi)^3} 
Y^*_{lm}({\bf \hat{k}}) Y_{lm}({\bf \hat{k}}) 
\nline
\!\!\!\!\!&\times&\!\!\!\!\!
\left[j_l(k r_{\nu}) j_l(k r_{\nu_2})
P_{\Delta^2_{\rm HI}}(k)
\right.
\nline
\!\!\!\!\!&-&\!\!\!\!\!
\{j_l(k r_{\nu}) j''_l(k r_{\nu_2}) 
+ j_l(k r_{\nu_2}) j''_l(k r_{\nu})\}
P_{\Delta_{\rm HI}}(k)
\nline
\!\!\!\!\!&+&\!\!\!\!\! 
\left. j''_l(k r_{\nu}) j''_l(k r_{\nu_2}) P(k)
\right]
\ear
Using the normalization property of the spherical harmonics
$\int \de {\bf {\hat n}} |Y_{lm}({\bf \hat{n}})|^2 = 1$, one can 
carry out the angular integrals in the above expression, 
and hence obtain the final result (\ref{eq:cl_delta_nu}) 
as quoted in the
main text.

\section{Correspondence between all-sky and flat-sky power spectra}
\label{sec:flat-all}

As discussed in section \ref{sec:flat-sky}, 
we shall mostly be interested in very small
angular scales, which corresponds to $l \gg 1$. 
For high values of $l$, it is most useful to
work in the flat-sky approximation, where 
a small portion of the sky can be
approximated by a plane.
Then the unit vector 
${\bf \hat{n}}$ towards the direction of observation
can be decomposed into ${\bf \hat{n}} = {\bf m} + {\bf \theta}$,
where ${\bf m}$ is a vector towards the center of the 
field of view and ${\bf \theta}$ is a two-dimensional
vector in the plane of the sky.

Without loss of generality, let us now
consider a small region around the pole $\theta \to 0$.
In that case the vector ${\bf \theta}$ can 
be treated as a Cartesian vector with components 
$\{\theta \cos \phi, \theta \sin \phi\}$.
This holds true for any two-dimensional vector on the sky, 
in particular
${\bf U} = \{U \cos \phi_U, U \sin \phi_U\}$.
Then the spherical harmonic components of $T(\nu,{\bf \hat{n}})$ 
[defined in equation (\ref{eq:a_lm})] can be written as
\be
a_{lm}(\nu) \approx \int \de {\bf \theta} ~ Y^*_{lm}(\theta, \phi) ~ 
T(\nu,{\bf \hat{n}}) 
\label{eq:a_lm_flat}
\e
where we have replaced $\int \de \Omega \rightarrow \int \de {\bf \theta}$.
Now use the expansion 
\be
{\rm e}^{-2 \pi {\rm i} {\bf U \cdot} {\bf \theta}}
= \sum_m (-{\rm i})^m J_m(2 \pi U \theta) 
{\rm e}^{{\rm i} m (\phi_U-\phi)}
\label{eq:exp_bessel_2d}
\e
where $J_m(x)$ is the ordinary Bessel function.
Further, we use the approximation for spherical harmonics
\be
Y_{lm}(\theta, \phi) ~ ^{\approx}_{\theta \to 0} ~ J_m(l\theta) 
\sqrt{\f{l}{2 \pi}} ~ {\rm e}^{{\rm i} m \phi}
\e
to write
\be
{\rm e}^{-2 \pi {\rm i} {\bf U \cdot} {\bf \theta}}
\approx \sqrt{\f{1}{U}} ~ 
\sum_m (-{\rm i})^m Y^*_{2 \pi U, m}(\theta, \phi) 
{\rm e}^{-{\rm i} m \phi_U}
\label{eq:exp_ylm_2d}
\e
Then the two-dimensional Fourier transform of the 
brightness temperature [defined in equation (\ref{eq:T_four})]
will be
\bear
\tilde{T}(\nu, {\bf U}) 
\!\!\!\!\!&=&\!\!\!\!\! 
\int \de {\bf \theta} ~  
{\rm e}^{-2 \pi {\rm i} {\bf U \cdot} {\bf \theta}} ~ 
T(\nu,{\bf \hat{n}}) 
\nline
\!\!\!\!\!&\approx&\!\!\!\!\!
\sqrt{\f{1}{U}} ~ 
\sum_m (-{\rm i})^m
{\rm e}^{-{\rm i} m \phi_U}~ 
\int \de {\bf \theta} ~  Y^*_{2 \pi U,m}(\theta, \phi)~  T(\nu,{\bf \hat{n}}) 
\nline
\!\!\!\!\!&=&\!\!\!\!\!
\sqrt{\f{1}{U}} ~ 
\sum_m (-{\rm i})^m
{\rm e}^{-{\rm i} m \phi_U}~ 
a_{2 \pi U, m}(\nu)
\ear
where we have used the expression (\ref{eq:a_lm_flat}) for $a_{lm}$
in the last part.
This gives a relation between the flat-sky Fourier transform
$\tilde{T}(\nu, {\bf U})$ and and its the full-sky equivalent
$a_{lm}(\nu)$.

Using the above relation, we can
calculate the power spectrum
\bear
\langle \tilde{T}(\nu_1, {\bf U}) \tilde{T}^*(\nu_2, {\bf U'}) \rangle
\!\!\!\!\!&\approx&\!\!\!\!\!
\sqrt{\f{1}{U U'}} ~ 
\sum_{m m'} (-{\rm i})^{m-m'}
{\rm e}^{-{\rm i} m \phi_U}~ {\rm e}^{{\rm i} m' \phi_{U'}}
\nline
\!\!\!\!\!&\times&\!\!\!\!\!
\langle a_{2 \pi U, m}(\nu_1) a^*_{2 \pi U', m'}(\nu_2) \rangle
\ear
Use the definition $\langle a_{lm}(\nu_1) a^*_{l' m'}(\nu_2) \rangle
= C_l \delta_{ll'} \delta_{mm'}$
and the property 
\be
\sum_m{\rm e}^{-{\rm i} m (\phi_U - \phi_{U'})} = 
2 \pi \delta^{(1)}_D(\phi_U - \phi_{U'})
\label{eq:exp_delta_1d}
\e
to obtain
\be
\langle \tilde{T}(\nu_1, {\bf U}) \tilde{T}^*(\nu_2, {\bf U'}) \rangle
= 2 \pi ~ C_{2 \pi U}(\nu_1, \nu_2) ~ \f{\delta_{U U'}}{U} ~ 
\delta^{(1)}_D(\phi_U - \phi_{U'})
\label{eq:T_tilde_sf}
\e
The last step involves writing the right hand side of the above equation
in terms of the two-dimensional Dirac delta function, which
follows from the expansion
\be
\delta^{(2)}_D({\bf U - U'}) =
\int \de {\bf \theta} ~ 
{\rm e}^{-2 \pi {\rm i} {\bf (U - U') \cdot} {\bf \theta}}
\e
The exponentials can be written in terms of the 
spherical harmonics using equation (\ref{eq:exp_ylm_2d}):
\bear
\delta^{(2)}_D({\bf U - U'})
\!\!\!\!\!&\approx&\!\!\!\!\!
\int \de {\bf \theta} ~ 
\sqrt{\f{1}{U U'}} \sum_{m m'} 
(-{\rm i})^{m-m'}
\nline
\!\!\!\!\!&\times& \!\!\!\!\!
Y^*_{2 \pi U, m}(\theta, \phi) Y_{2 \pi U', m'}(\theta, \phi) 
{\rm e}^{-{\rm i} m \phi_U} {\rm e}^{{\rm i} m' \phi_{U'}}
\nline
\ear
Finally use the orthonormality 
property of spherical harmonics 
$\int \de {\bf \theta} Y^*_{lm}(\theta, \phi) 
Y_{l'm'}(\theta, \phi) = \delta_{ll'} \delta_{mm'}$
and the relation (\ref{eq:exp_delta_1d}) to obtain
\be
\delta^{(2)}_D({\bf U - U'})=
2 \pi ~ \f{\delta_{U U'}}{U} 
\delta_D(\phi_U - \phi_{U'})
\e
Putting the above relation into (\ref{eq:T_tilde_sf}), we obtain
equation (\ref{eq:T_Cl}) used in the final text.

\section{Relation between visibility-visibility correlation and MAPS}
\label{sec:vis}
In this appendix we give the calculations for expressing the 
two visibility correlation in terms of the Multi-frequency 
angular power spectrum (MAPS). We can write the visibility 
$V(\u,\nu)$ as a two-dimensional Fourier transform 
of the brightness temperature $T(\th, \nu)$ [see equation (\ref{eq:1})]
\be
V(\u,\nu)=\left(\frac{\del B}{\del T}\right)_{\nu}
\int d^2 \theta A(\th, \nu) T(\th, \nu)
e^{ 2\pi \imath \th \cdot \u}
\e
where $(\del B/\del T)_{\nu}$ is the conversion factor from 
temperature to specific intensity and $A(\th, \nu)$ is the 
beam pattern of the individual antenna. The visibility-visibility correlation
is then given by
\begin{eqnarray}
\langle V(\u_1,\nu_1) V(\u_2 ,\nu_2) \rangle \!\!\!\!\!&=& \!\!\!\!\!
\left(\frac{\del B}{\del T}\right)_{\nu_1}
\left(\frac{\del B}{\del T}\right)_{\nu_2}
\nonumber\\
\!\!\!\!\!&\times&\!\!\!\!\!
\int d^2 \theta \int d^2 \theta'
A(\th,\nu_1) A(\th',\nu_2)
\nonumber\\
\!\!\!\!\!&\times&\!\!\!\!\!
\langle T(\th, \nu_1) T(\th', \nu_2) \rangle
e^{2\pi \imath (\th \cdot \u_1 + \th' \cdot \u_2)}
\nonumber\\
\label{eq:vis-vis}
\end{eqnarray}
The correlation function for the temperature fluctuations on the sky
would simply be the two-dimensional Fourier transform of the MAPS 
$C_{2 \pi U}(\nu_1,\nu_2)$
\be
\langle T(\th, \nu_1) T(\th', \nu_2) \rangle = 
\int d^2 U ~ C_{2 \pi U}(\nu_1,\nu_2) 
e^{-2\pi \imath (\th - \th') \cdot \u}
\e
Using the above equation in equation in (\ref{eq:vis-vis}), we obtain
\begin{eqnarray}
\langle V(\u_1,\nu_1) V(\u_2 ,\nu_2) \rangle \!\!\!\!\!&=& \!\!\!\!\!
\left(\frac{\del B}{\del T}\right)_{\nu_1}
\left(\frac{\del B}{\del T}\right)_{\nu_2}
\nline
\!\!\!\!\!&\times&\!\!\!\!\!
\int d^2 U ~ C_{2 \pi U}(\nu_1,\nu_2) 
\nonumber\\
\!\!\!\!\!&\times&\!\!\!\!\!
\tilde{A}(\u_1-\u, \nu_1) \tilde{A}(\u_2+\u, \nu_2) 
\end{eqnarray}
where $\tilde{A}(\u, \nu)$ is the Fourier transform of the 
beam pattern $A(\th, \nu)$. If the beam pattern is assumed to be Gaussian
$A(\th, \nu) = e^{-\theta^2/\theta_0^2}$, the 
Fourier transform too is given by a Gaussian function
\be
\tilde{A}(\u, \nu) = \pi \theta_0^2 e^{-\pi^2 U^2 \theta_0^2}
\e
Hence, the visibility correlation becomes
\begin{eqnarray}
\langle V(\u_1,\nu_1) V(\u_2 ,\nu_2) \rangle \!\!\!\!\!&=& \!\!\!\!\!
\left(\frac{\del B}{\del T}\right)_{\nu_1}
\left(\frac{\del B}{\del T}\right)_{\nu_2}
\pi^2 \theta_1^2 \theta_2^2
\nline
\!\!\!\!\!&\times&\!\!\!\!\!
\int d^2 U ~ C_{2 \pi U}(\nu_1,\nu_2) 
\nonumber\\
\!\!\!\!\!&\times&\!\!\!\!\!
e^{-\pi^2[(\u_1 - \u)^2 \theta_1^2 + (\u_2 + \u)^2 \theta_2^2]}
\end{eqnarray}
where $\theta_1$ and
$\theta_2$ are the values of $\theta_0$ at $\nu_1$ and $\nu_2$
respectively. Now, since the two Gaussian functions in the above
equation is peaked around different values of $\u$, the 
integrand will have a non-zero contribution only
when $|\u_1 + \u_2| < (\pi ~ {\rm max}[\theta_1, \theta_2])^{-1}$.
In case the typical baselines are much larger than the quantity
$(\pi ~ {\rm max}[\theta_1, \theta_2])^{-1}$, 
the integral above can be well approximated as  being  non-zero only
when $\u_1 = -\u_2$. Then 
\begin{eqnarray}
\langle V(\u_1,\nu_1) V(\u_2 ,\nu_2) \rangle \!\!\!\!\!&\approx& \!\!\!\!\!
\delta_{\u_1,-\u_2} 
\left(\frac{\del B}{\del T}\right)_{\nu_1}
\left(\frac{\del B}{\del T}\right)_{\nu_2}
\nline
\!\!\!\!\!&\times&\!\!\!\!\!
\pi^2 \theta_1^2 \theta_2^2 C_{2 \pi U_1}(\nu_1,\nu_2)
\nline
\!\!\!\!\!&\times&\!\!\!\!\! 
\int d^2 U
e^{-\pi^2[(\u_1 - \u)^2 (\theta_1^2 + \theta_2^2)]}
\nonumber\\
\!\!\!\!\!&=& \!\!\!\!\!
\delta_{\u_1,-\u_2} \pi \left(\frac{\theta_1^2 \theta_2^2}{\theta_1^2
  + \theta_2^2}\right)
 \left(\frac{\del B}{\del T}\right)_{\nu_1}\left(\frac{\del B}{\del
   T}\right)_{\nu_2}
\nline
\!\!\!\!\!&\times&\!\!\!\!\! 
C_{2 \pi U_1}(\nu_1,\nu_2) 
\end{eqnarray}
which is what has been used in equation (\ref{eq:11}).

In the continuum limit, the Gaussian $\tilde{A}(\u, \nu)$ can be approximated
by a delta function, i.e., $\tilde{A}(\u, \nu) \approx \delta^{(2)}_D(\u)$
(which corresponds to the limit $\theta_0 \to \infty$);
the visibility-visibility correlation is then given as 
\begin{eqnarray}
\langle V(\u_1,\nu_1) V(\u_2 ,\nu_2) \rangle
= \delta^{(2)}_D(\u_1 + \u_2) \left(\frac{\del B}{\del
  T}\right)_{\nu_1} \left(\frac{\del B}{\del T}\right)_{\nu_2}
\nline 
\times C_{2 \pi U_1}(\nu_1,\nu_2) 
\end{eqnarray}
which corresponds to equation (\ref{eq:21}) in the main text.

\newpage

%\clearpage{\pagestyle{empty}\cleardoublepage} %%%%%%%%%%%%%%%%%%%%
\newpage
\chaptermark{Publications}
\addcontentsline{toc}{chapter}{List of Publications} 
\newpage

\setcounter{section}{0}
\setcounter{subsection}{0}
\setcounter{subsubsection}{2}
\setcounter{equation}{0}

\begin{center}
{\Large{\bf List of Publications}}
\end{center}

 \vspace{0.5cm}
\noindent(a) {\bf Research Papers in Refereed Journals:}

\begin{enumerate}

\item{Datta, K.~K., Majumdar, S., Bharadwaj, S., \& Choudhury, T.,~R., 
(2008), Simulating the impact of HI fluctuations on matched filter search for ionized bubbles in redshifted 21-cm maps, \mnras, Vol. 391, Issue 4, pp. 1900-1912}  

\item{Datta, K.~K., Bharadwaj, S., \& Choudhury, T.,~R. (2007), Detecting ionized bubbles in redshifted 21-cm maps, \mnras, Vol. 382, Issue 2, pp. 809-818 }

\item{Datta, K.~K., Choudhury, T.,~R., \& Bharadwaj, S. \
(2007), The multifrequency angular power spectrum of the epoch of
  reionization 21-cm signal, \mnras, Vol. 378, Issue 1, pp. 119-128}

\end{enumerate}

\noindent (b) {\bf Submitted manuscripts:}\\
\begin{enumerate}

\item{Guha Sarkar, T., Datta, K.~K., \& Bharadwaj, S. (2008), The CMBR
  ISW and HI 21-cm Cross-correlation Angular Power Spectrum, arXiv:0810.3649 }

\item{Datta, K.~K., Bharadwaj, S., \& Choudhury, T.,~R., The optimum
redshift for detecting ionized bubbles in hi 21-cm maps, arxiv:0906.0360}

\end{enumerate}

\noindent(c) {\bf Research Paper in Conference Proceedings:}
\begin{enumerate}

\item{Datta K.~K. (2006), Frequency decorrelation properties of the
epoch of reionization 21 cm signal, Proceedings of XVII DAE- BRNS High
Energy Physics 
Symposium, December 11- 15, 2006, IIT Kharagpur, pp. 217-220}

\newpage

\item{Datta, K.~K., Majumdar, S., Bharadwaj, S., \& Choudhury, T.,~R., 
(2009), Searching for Ionized bubbles in 21-cm maps, To appear in ASP Conference Series, Vol. 407, The Low-Frequency Radio Universe, eds D. J. Saikia, D. A. Green, Y. Gupta and T. Venturi (Conference held at NCRA, Pune, India from 8th to 12th December 2008)}
\end{enumerate}

\end{document}